\documentclass[12pt]{article}
\usepackage[matrix,arrow,ps,all,cmtip]{xy}
\usepackage{amsmath}
\usepackage{amsfonts}
\usepackage[noend]{algcompatible}
\usepackage[section]{algorithm}
\usepackage{ulem}

\makeatletter
\def\BState{\State\hskip-\ALG@thistlm}
\makeatother

\usepackage{multibib}
\newcites{appendix}{References}
\newcites{supplement}{References}

\makeatletter
\let\OldStatex\Statex
\renewcommand{\Statex}[1][3]{%
	\setlength\@tempdima{\algorithmicindent}%
	\OldStatex\hskip\dimexpr#1\@tempdima\relax}
\makeatother

 \usepackage{lscape}
\usepackage{comment}
\usepackage{amssymb,amsthm,natbib} 
\usepackage{setspace} 
\usepackage[margin=1.1in]{geometry}
\usepackage{verbatim}  
\usepackage{graphicx,eepic} 
\usepackage[usenames,dvipsnames]{color}
\usepackage{hyperref} 
\usepackage{xcolor}
\usepackage{colortbl}
\usepackage{multirow}
\definecolor{lightgray}{rgb}{0.7421875,0.7421875,0.7421875}
\hypersetup{pdfstartview={FitH}}
\usepackage[capitalize]{cleveref}
	\crefname{claim}{Claim}{Claims}
	\Crefname{claim}{Claim}{Claims}
	\crefname{figure}{Figure}{Figures} 

\usepackage{pdflscape}
\usepackage{booktabs}
\usepackage{array}
\usepackage{multicol}
\usepackage{longtable} 
\usepackage{rotating}
\newcolumntype{K}{>{\arraybackslash}m{8.5cm}}
\newcolumntype{N}{>{\arraybackslash}m{2.2cm}}
\newcolumntype{n}{>{\arraybackslash}m{1.8cm}}
	
\usepackage{caption}
\usepackage{subcaption}

\theoremstyle{definition}

\begin{document}

\onehalfspacing

\title{Meritocracy and Its Discontents:\\
\Large{Long-run Effects of Repeated School Admission Reforms}\\
}
\author{
\and Chiaki Moriguchi
\and Yusuke Narita
\and Mari Tanaka\thanks{
We thank Zach Bleemer, David Card, Yeon-Koo Che, John Friedman, Caroline Hoxby, Pat Kline, Joe Price, Al Roth, Chris Walters, Seth Zimmerman, and seminar participants at AEA, Berkeley, Columbia, Harvard, Opportunity Insights, Penn State, Princeton, Stanford, Rice, Yale, and the NBER (education group and market design group) for their helpful feedback, 
Hidehiko Ichimura and Yasuyuki Sawada for sharing a part of the data, and an extensive list of students for research assistance. 
We gratefully acknowledge financial support from JSPS Grant JP19K13719 and 24H00145, Yale Council on East Asian Studies, Hitotsubashi University IER Joint Usage and Research Center Grant (FY 2018). 
Chiaki Moriguchi: Institute of Economic Research, Hitotsubashi University, email: \textsf{chiaki.m@r.hit-u.ac.jp}. 
Yusuke Narita (corresponding author): Department of Economics, Cowles Foundation, and Y-RISE, Yale University, email: \textsf{yusuke.narita@yale.edu}, address: Room 328, 87 Trumbull Street, New Haven CT 06511.  
Mari Tanaka: Faculty of Economics, University of Tokyo, and Institute of Economic Research, Hitotsubashi University, email: \textsf{m.tanaka@e.u-tokyo.ac.jp}.
}
}
\date{\today}
\maketitle

\begin{abstract}
What happens if selective colleges change their admission policies? We
study this question by analyzing the world’s first implementation of
nationally centralized meritocratic admissions in the early twentieth
century. We find a persistent meritocracy-equity tradeoff. Compared to
the decentralized system, the centralized system admitted more high-achievers and produced more occupational elites (such as top income earners) decades later in the
labor market. This gain came at a distributional cost, however. Meritocratic centralization also increased the number of urban-born
elites relative to rural-born ones, undermining equal access to higher
education and career advancement.\\

\end{abstract}


\maketitle
\newpage


\setcounter{page}{1}
\section{Introduction}

One major trend in college admissions around the world is a growing degree of centralization. 
Today, over thirty countries use regionally- or nationally-integrated, single-application and single-offer college admissions (Appendix Figure \ref{tab:world}). 
These systems have well-specified admission criteria, mixing meritocratic achievement elements (such as GPA and entrance exams) and other priority considerations. 
Before the turn of the 20th century, however, no country used such a centralized system. How does the centralization of admission affect students' life trajectories?
What are their impacts on the national production of highly skilled individuals and their composition?
A key challenge in studying these questions is the lack of clear policy changes and data about students' long-run outcomes.

In this paper, we study the impacts of centralized and meritocratic college admissions.
Our investigations reveal their pros and cons, especially a tension between meritocratic centralization and equal 
access to higher education and career achievements in the long run. 
We reach these findings by combining a series of natural experiments in history and newly assembled historical data that trace students over decades.

Our empirical setting is the world's first known transition from decentralized to nationally-centralized school admissions. 
At the end of the 19th century, to modernize its higher education system, the Japanese government set up National Higher Schools (roughly equivalent to today's liberal arts colleges) as an exclusive entry point to the most prestigious tertiary education.\footnote{After WWII, National Higher Schools were transformed into today's national
universities. For example, the First National Higher School in Tokyo became the first two years of the University of Tokyo.}  
We study these selective colleges, as they later produced many of the most influential leaders of society, including several Prime Ministers, Nobel Laureates, and founders of global companies like Toyota. 
Their prestige and social influences were similar to Oxbridge in the UK, Grandes \'Ecoles in France, and the Ivy League in the US.

Acceptance into these schools was based on annual entrance exams. 
Initially, the government let each school run its own exam and admissions based on exam scores, similar to 
many of today's decentralized K-12 and college admissions. 
Under this decentralized system, the schools typically held exams on the same date so that each applicant could apply for only one school.\footnote{Similar restrictions on the number of applications exist today in the college admission systems of Italy, Nigeria, and the UK.}
As a result, many high-achieving applicants applied to and were rejected by the most competitive school, failing to enter any school. 

This problem motivated the government to introduce a centralized system in 1902. 
The new system prioritized the admission of the highest-scoring students. Specifically, applicants were asked to submit preference rankings over multiple schools and to take a single unified exam.
\footnote{As shown later, the defining feature of the centralized system was to allow applicants to list multiple schools, not the use of a single unified exam.}
Given their preference rankings and exam scores, each applicant was assigned to a school (or none if unsuccessful) based on a computational algorithm.
The algorithm was a mix of the Deferred Acceptance and Immediate Acceptance (Boston) algorithms combined with a meritocracy principle that assigns only the highest-scoring applicants to any school.  
To our knowledge, this instance is the first nationwide use of any matching algorithm.\footnote{The earliest known large-scale use of the Boston algorithm is the assignment of medical residents to hospitals in New York City in the 1920s \citep{roth:90}. 
The oldest known national use of the Deferred Acceptance algorithm is the National Resident Matching Program (NRMP) in the 1950s \citep{roth:84a}. See \cite{abdulkadiroglu/sonmez:03} for the details of these algorithms in school admission contexts.} 
The reform has similarities to contemporary admission reforms that adopt variants of the Deferred Acceptance and Immediate Acceptance algorithms.

Furthermore, the government later re-decentralized and re-centralized the system several times, producing multiple natural experiments for studying the consequences of the different systems.\footnote{During 1900--1930, there were three periods of centralization in 1902--1907, 1917--1918, and 1926--1927.} 
We exploit these bidirectional institutional changes to identify the impacts of meritocratic centralization. 
For the short-run analysis, we newly collect and digitize application and enrollment data from administrative school records and government documents, including Government Gazettes and Ministry of Education Yearbooks.

We first find that meritocratic centralization had large effects on application behavior and enrollment outcomes. In particular, centralization caused applicants in all areas to be more risk-taking and more often rank the most selective school first. 
Applicants also made longer-distance applications, leading students to be enrolled at schools further away from their regional origin. 
The reform thus made the selective higher education market more meritocratic, competitive, and regionally integrated.

The effect of centralization was heterogeneous, however. 
Because high-achieving students were disproportionately located in urban areas (mainly Tokyo), 
the centralized system caused more urban applicants to be admitted to schools in rural areas, typically after being rejected by their first-choice schools.\footnote{We use ``urban areas'' to refer to a region surrounding Tokyo (as defined in Section \ref{section:short}). In the nomenclature of ``urban'' and ```rural'' schools,  urban schools were located in extended metropolitan areas around Tokyo while rural schools were located in provincial cities. As shown later, the urban areas were characterized by a greater population, higher income, and better educational infrastructure.}
As a result, urban high-achievers crowded out rural applicants: the number of urban-born entrants to any national higher school increased by about 10\% during centralization.\footnote{It is also empirically true that the centralized system made more rural applicants apply to and enter urban schools. 
The centralized system thus increased regional mobility across the country. 
But their net effects are such that urban high-achievers crowded out rural applicants.} 
Historical documents suggest that this distributional consequence upset rural schools and their local communities. Such rural discontents were a reason why the government oscillated between the centralized and decentralized systems.
This distributional effect echoes the widespread concerns about regional inequality in access to higher education.

Our main results are the long-run effects of the centralized system on the national production and distribution of future occupational elites. 
To study the distributional consequence, we compare urban vs rural-born individuals' long-term career outcomes by each cohort’s exposure to the centralized system. 
The career outcome data is based on the Personnel Inquiry Records (PIR) published in 1939 (more than thirty years after the first period of meritocratic centralization). 
The data provides a list of socially distinguished individuals---encompassing economic, political, and cultural elites---and their biographical information.\footnote{We provide extensive investigations about the quality of the data (such as the coverage and sampling bias) in Section \ref{section:long_run_analysis1}.}

The distributional effects of meritocratic centralization turn out to be persistent. 
Almost four decades after the reform, relative to the decentralized system, the centralized system produced a greater number of occupational elites (such as top income earners and top politicians and bureaucrats) who came from urban areas compared to rural areas. 
The number of urban-born elites increased by 10--20\% for the cohorts exposed to centralization. 
We also provide suggestive evidence that the centralized system increased the number of elites living in urban areas in their adulthood. Meritocratic centralization thus impacted both the origins and destinations of highly skilled individuals.

We conclude our analysis by examining the impact of meritocratic centralization on the national production of occupational elites. 
We first focus on a specific subgroup, i.e., top bureaucrats. 
Top bureaucrats were an essential category of elites in our empirical context, where higher civil service was one of the most prestigious and coveted jobs due to their political and legislative influence in building the state. 
Using comprehensive administrative data of all individuals who passed the national exams to become higher civil officials, we compare the total number of bureaucrats promoted to top ranks between cohorts exposed to the centralized admissions and cohorts exposed to the decentralized admissions. We find that the number of top-ranking bureaucrats was 15\% greater in the centralization cohorts than in the decentralization cohorts, suggesting that centralization cohorts won promotion competitions over decentralization cohorts.

Moreover, meritocratic centralization produced more elites across fields and occupations, not just government officials. 
We find that meritocratic centralization produced a greater number of all occupational elites in the PIR data, including top-income earners. 
Thus, the increasing production of top-ranking government officials is not at the expense of other occupations. 
Meritocratic centralization boosted the production of elites for the whole country.


Overall, our findings show that the design of admission rules affects the production and distribution of highly educated and skilled individuals, which is an important determinant of economic growth and inequality \citep{glaeser2011triumph, moretti2012new}. 
On one hand, meritocratic centralization achieved the goal of recruiting applicants with higher academic performance. This meritocracy then resulted in producing a greater number of occupational elites in the long run. The policymaker may, therefore, endorse meritocratic centralization on the grounds that it produces more high-quality leaders for society.

On the other hand, this gain came at the equity cost of urban-born high-achievers overwhelming rural-born ones both in the short and long run. Meritocratic centralization expanded the urban-rural gap in the ability to produce highly skilled individuals. 
Meritocracy may thus dampen the society's capability to diversify the backgrounds of individuals in leadership positions. This meritocracy-equity tradeoff would be relevant for various decentralized admissions where costly applications make applicants strategically self-select into a small number of schools. 

\textbf{Literature.} Our analysis contributes to the empirical literature that uses policy reforms to study the impacts of the design of admission systems. Existing literature finds the short-run effects on strategic responses by applicants \citep{abdulkadiroglu2006changing, chen2020empirical} and enrollment outcomes. Some studies emphasize the equalizing effects of introducing or expanding centralized school choice \citep{campos2024impact}, while others highlight its distributional effects \citep{machado2021,terrier2021immediate}. Moreover, 
\citet{abdulkadiroglu/agarwal/pathak:13} and \citet{kapor2022aftermarket} find positive effects of centralized school admissions on academic achievement, graduation, and welfare. These results jointly suggest the importance of both efficiency and equity effects.\footnote{Other studies measure the effects of selective schools conditional on particular admission systems \citep{dale2002estimating, 
hastings/neilson/zimmerman:13, pop-eleches/urquiola:13, deming/hastings/kane/staiger:13, 
kirkeboen/leuven/mogstad:15, 
clark2016,
abdulkadiroglu2017regression, zimmerman2019elite, abdulkadiroglu2019breaking, 
narita2021theory, 
michelman2021boysclub, 
chetty2023diversifying}.}

In contrast to these prior studies on the short-run effects, to our knowledge, this paper provides the first evidence about the long-run impacts of centralized admissions on career outcomes.\footnote{With its interest in long-run effects, this paper also relates to studies of the long-term effects of educational reforms (e.g., \citealp{duflo2001schooling, 
meghir/palme:05,ichimura}). 
These studies focus on the effects of expanding resources (such as school constructions and compulsory education extensions), while we investigate the effects of changing resource allocation mechanisms given a fixed amount of resources.}
We find not only the distributional effect but also the overall talent-production effect for the whole country.\footnote{The resulting meritocracy vs equity tension has some similarities to ongoing policy discussions on affirmative actions, which are surveyed in \cite{arcidiacono2016affirmative}. Recent important contributions in this area include \cite{bleemer2018top, bleemer2020affirmative}, \cite{kapor} and \cite{otero2021affirmative}. \cite{kamada2015efficient} and \cite{agarwal2017policy} study regional equity in other matching markets.} 
We identify the impacts by taking advantage of bidirectional, repeated policy changes in history, echoing other studies with similar identification strategies \citep{niederle2003unraveling}.


From a broader historical perspective, this study relates to the literature 
that uses historical natural experiments \citep{Cantoni_Yuchtman2021} and
investigates the long-term effects of institutions \citep{nunn2009survey}, particularly the effects of resource allocation mechanisms \citep{bleakley2014land}. 
Our analysis is also related to \cite{bai2016elite}, who examine the political consequences of the abolition of a meritocratic elite recruitment system (civil service exam) in early twentieth-century China. 
Our study also builds on the historical and sociological literature on higher education and elite formation in Japan, which is summarized in Appendix Section \ref{appendix:historical_literature}.

The next section provides the historical and institutional background. 
Section \ref{section:short} examines the short-term effects of meritocratic centralization, while Section \ref{section:long} analyzes the long-term impacts. 
Section \ref{section:direction} summarizes our findings and outlines future directions.

\section{Background}\label{section:background}




\subsection{Bidirectional Admission Reforms}
\label{section:admission_reforms}

To analyze the impact of centralized admissions, we take advantage of unique historical episodes in early twentieth-century Japan. 
Since the opening of the country in the mid-19th century, 
education reforms became a central part of modernization efforts by the Japanese government. 
In 1894, the government set up a national higher education system consisting of one Imperial University (three-year program) and five National Higher Schools (three-year program).\footnote{The curriculum of National Higher Schools roughly corresponds to that of the first two years of today's universities. It was divided into fields such as law, social sciences, engineering, and medicine, and taught by academic scholars who graduated from Imperial Universities.} 
By 1908, the system was expanded to four Imperial Universities and eight National Higher Schools located across the country.\footnote{First in Tokyo, Second in Sendai, Third in Kyoto, Fourth in Kanazawa, Fifth in Kumamoto, Sixth in Okayama, Seventh in Kagoshima, and Eighth in Nagoya, named after the order of establishment; see Appendix Figure \ref{fig:map} for their locations.} 
We focus on these eight National Higher Schools and refer to them as Schools 1--8 for short.\footnote{The number of higher education institutions increased after 1918, as the government permitted not only national but also local public and private higher schools and universities.
See Appendix Section \ref{appendix:admission_system} and Table \ref{tab:chronological_table2} for details. In our empirical analyses, we control for the number of National Higher Schools as well as other characteristics of higher education institutions.} 

Entering Schools 1--8 was considered a passport to the elite class. 
Students must attend National Higher Schools to be admitted to Imperial Universities, the most prestigious tertiary education.
Virtually all graduates of National Higher Schools were admitted to an Imperial University without further selection.
Due to its rigorous curriculum, Imperial University graduates were wholly or partially exempted from highly selective national qualification exams for physicians and higher government officials (including judges and diplomats).\footnote{In the period we study, Imperial University graduates were only partially exempted.}

Schools 1--8 produced distinguished and influential individuals, including Prime Ministers, Nobel Laureates, world-leading mathematicians and novelists, and founders of global companies like Toyota.
To apply to these schools, one must be male aged 17 or older and have completed a five-year middle school.\footnote{The eligibility was changed in 1919 to males aged 16 or older who have completed the fourth year of middle school.}
Schools 1--8 admitted less than 0.5\% of the cohort of males aged 17 throughout 1900--1930. Descriptive statistics are presented in Appendix Table \ref{tab:statistics}.


There was a clear selectivity hierarchy among Schools 1--8. School 1 in Tokyo was considered the most prestigious due to its location in the capital and its geographical proximity to Tokyo Imperial University (today's University of Tokyo).
The next most prestigious was School 3 in Kyoto. 
By contrast, Schools 5 and 7 were considered the least prestigious among all schools. 
Consequently, the schools differed substantially in their popularity and selectiveness.\footnote{Tuition was mostly uniform across Schools 1--8.} 

The admission to Schools 1--8 was based on annual entrance exams. 
Initially, the government let each school administer its own exam and admissions. 
Schools 1--8 coordinated to hold their exams on the same date so that each applicant could only apply to one school (see Appendix Table \ref{tab:chronological_table2} for details). 
There was no restriction on which school an applicant could apply to.
We call this system ``decentralized admissions" (or ``decentralization'' for short). 
The single choice aspect captures an essential feature of decentralization, which incentivizes each applicant to self-select into an appropriate school by comparing the selectivity of schools with his own standing \citep{che2016decentralized}. 
We focus on this particular decentralized system, but as we detail below, the lessons from our results would hold for other types of decentralized admissions as long as costly applications make applicants self-select into a limited number of schools.


Under decentralization, however, many high-achieving students were rejected by popular schools, while lower-achieving students entered less popular schools.
For the government, which wanted to send the most promising students to selective higher education, the decentralized system appeared inefficient.\footnote{Applicants who failed to enter Schools 1--8 chose either (1) to retake the exam in the following year, (2) to enter local public or private higher educational institutions, or (3) to give up receiving higher education.}
The Education Minister criticized the decentralized system as follows:

\begin{quote}
``\textit{[Under the decentralized system] many applicants are rejected by Schools 1 and 3 [in Tokyo and Kyoto], which attract a large number of high-ability applicants, despite the fact that these applicants have superior academic performance to that of applicants admitted to other more rural schools. (...) As a result, hundreds of applicants with sufficiently high academic ability to enter rural schools are idly wasting another year [to retake the exam]. This is not only a pity for them, but also a loss for the country.}"\footnote{A quote from 
\textit{Educational Review} No.1146, p.21, published in February 15, 1917.}
\end{quote} 

\noindent To solve this problem, in 1901, the government first asked all schools to unify their entrance exams to a single one, while maintaining decentralized admission decisions. In 1902, the government launched a centralized system. Under the new system, applicants were allowed to apply for multiple schools by submitting their preference rankings over schools before taking the unified exam. 
Based on their exam scores and school preferences, applicants were then assigned to a school (or none if unsuccessful) by an assignment algorithm announced ex ante. 
The assignment algorithm was specified as follows.\footnote{This is a simplified version of the assignment algorithm in 1917 (see \citet{moriguchi2021higher} for a detailed discussion of original algorithms). A reprint of the original assignment algorithm published in the Government Gazette is shown in Appendix Figure \ref{fig:algo}. } 

\begin{quote}
\begin{enumerate}
\item In the order of exam scores, select the same number of applicants as the sum of all schools' capacities. In the case of a tie, decide by lottery.
\item For applicants selected in (1), in the order of exam scores, assign each applicant to the school of his first choice. For an applicant whose first-choice school is already filled, hold his assignment until the next step. In the case of a tie, decide by lottery. 
\item For those applicants whose first-choice schools are already filled at the end of (2), in the order of exam scores, assign each applicant to the school of his second choice. For an applicant whose second-choice school is already filled, hold his assignment until the next step. In the case of a tie, decide by lottery. 
\item For those applicants whose second-choice schools are already filled at the end of (3), assign each applicant to the school of his third choice or below, repeating the same procedure as (3). 
\item If all the schools in an applicant's preference list are filled, then the applicant is not admitted to any school.
\end{enumerate}
\end{quote}

\noindent 
The above algorithm imposes meritocracy up front in which only top-scoring applicants were considered for admission regardless of preferences (Step (1)). The centralized system is meritocratic in that the test score distribution among the assigned students first order stochastically dominates that under any other mechanisms, including the original decentralized system. 
Step (1) also selects the same applicants as those who would be admitted by any school under the Deferred Acceptance algorithm, one of the most widely used algorithms in today's college and K-12 admissions.\footnote{See Supplementary Materials Section A for details.}  
These applicants are assigned to one of Schools 1--8 using the Immediate Acceptance (Boston) algorithm (Steps (2) to (4)).
This algorithm is therefore a variant of the Immediate Acceptance algorithm with a meritocracy constraint, making it closer to the Deferred Acceptance algorithm. 
To our knowledge, this is the world's first nationwide use of any assignment algorithm. 
We call this system centralized admissions (or centralization for short).\footnote{This centralized system differs from decentralized systems with multiple choices, where each college independently makes its own admissions based on its own exam but each student is allowed to apply to multiple colleges. Theoretically, \citet{che2016decentralized} show that with uncertain student preferences, such decentralized systems with multiple choices produce different equilibrium allocations from centralized systems such as the one we analyze. Empirically, in the context of the NYC high school match, \citet{abdulkadiroglu/agarwal/pathak:13} find that the old decentralized system with multiple choices results in different allocations compared to the new centralized system based on the deferred acceptance algorithm.} 

\subsection{Political Economy of the Admission Reforms}\label{section:political_economy}

This institutional innovation was short-lived, however.
Due to the opposition detailed below, the government switched back to decentralization (with a unified exam) in 1908.
The government then continued to oscillate between the two regimes, reintroducing centralization in 1917, moving back to decentralization in 1919, reinstituting centralization in 1926, and finally settling down to decentralization in 1928.
In the space of thirty years, there were three periods of centralized admissions: first in 1902--1907, next in 1917--1918, and finally in 1926--1927.\footnote{For details of the evolution of the admission system, see Appendix Table \ref{tab:chronological_table2}.} 
We exploit the series of bidirectional policy changes to examine the impacts of centralization. 

Historical documents suggest that the repeated policy changes resulted from annual bargaining between the Ministry of Education (representing the central government) and the Council of School Principals (representing the local interests of Schools 1--8).\footnote{Detailed accounts are provided in \cite{moriguchi2021higher}, pp.195-199.} 
The admission
policy, as a result of bargaining, was decided only a few months before the exam, making
it difficult for applicants to anticipate the exact timing of the reforms. 
Indeed, we confirm that the timing of the reforms is not associated with the size and composition of applicants, school capacities, and other potential confounding factors (see Section \ref{section:robustness}).

Throughout the bargaining process, the Ministry of Education insisted on centralized admissions to select the best and brightest,  while the Council of School Principals was opposed to centralization, pointing out its adverse impacts on rural schools.\footnote{Other reasons against centralization included a loss of school autonomy and independence as well as the high administrative cost of implementing centralized admissions.} 
They claimed that these rural schools lost the most talented students in their local areas (who would have entered local schools under decentralization) to urban schools, especially School 1. 
They also complained that rural schools instead received many reluctant and unmotivated students who only came to these schools as fallback options.

Moreover, a review of admission results revealed that the number of applicants from rural areas (areas where rural schools were located) admitted to any of Schools 1--8 decreased considerably under centralization.\footnote{Ministry of Education (1917), \textit{Report on National Higher School Entrance Examination of 1917: Extra Issue}.} 
This was upsetting to rural schools and their local communities.
A noted historian summarizes the situation as follows (\citealp{takeuchi}, p.121): 

\begin{quote}
``\textit{Urban applicants `overwhelmed' rural applicants by applying for rural schools as fallback options. Urban applicants robbed rural applicants of opportunities that were once open to them. This ruined the meaning of building national higher schools across the nation.}''
\end{quote} 

\noindent These observations highlight a potential meritocracy-equity tradeoff of centralization. On the one hand, the centralized admissions made the school seat allocation more meritocratic, enabling high-achieving students to enter one of Schools 1--8 even if they failed at the most prestigious one. On the other hand, this meritocracy might have come at the expense of equal regional access to national higher education, whereby high-achieving urban applicants dominated rural applicants.

\section{Short-run Impacts}\label{section:short}

\subsection{Predicted Impacts}
Motivated by the above historical observations, we first make predictions on the short-run effects of centralization. (In Supplementary Materials Section A, we formally derive the following predictions using a simple model.)


First, we predict changes in application behavior. Specifically, the centralized system would encourage applicants all over the country to rank the most preferred and selective urban school as their first choice. 
This is because the centralized system provides applicants with an option to apply for second and lower-choice schools as insurance, incentivizing students to be more risk-taking at 1st choice selection. 

Second, if high-achieving applicants are disproportionately located in a specific geographical area, such as an urban area, then a larger fraction of all school seats in Schools 1--8 would be assigned to students coming from that area under the centralized system. 
This is because of the meritocratic constraint (Step 1) in the centralized assignment algorithm, which assigns only the top-scoring applicants to any school.

Finally, a smaller fraction of applicants would be assigned to their local school under centralized admissions. 
Intuitively, meritocratic centralization enables high-scoring applicants from urban areas to enter rural schools even if they fail to enter the most competitive urban school. At the same time, the centralized system would make a greater number of high-scoring rural applicants apply to and enter urban schools. Under the centralized system, therefore, rural students would more often move to urban areas, while urban students would more often move to rural areas. 
These effects would jointly increase the spatial mobility of students.

\subsection{Data}\label{section:data}

To empirically analyze the short-run effects of the centralized admission system, we newly assemble data on application and enrollment.
First, we collect data on the number of applicants by school of their first choice from 1900 to 1930, using the \textit{Ministry of Education Yearbook} and other sources. 
For two specific years (1916 and 1917), we have additional data on the number of applicants by school of their first choice and by prefecture of their middle school (the country was divided into 47 prefectures). 

Second, we compile data on the number of entrants (first-year students) by school and by their birth prefecture from 1900 to 1930, using \textit{Student Registers} of Schools 1--8. 
Finally, to control for the size of potential applicants and the number of competing schools, we collect data on the number of middle-school graduates by prefecture of their middle school, as well as the number of other Higher Schools (established in addition to Schools 1--8 starting in 1919) by prefecture from 1900 to 1930. 
Supplementary Materials Section B.1 provides more detailed explanations for data sources, samples, and definitions. 
Descriptive statistics of main variables are reported in Appendix Table \ref{tab:statistics}.



\subsection{Strategic Responses by Applicants}\label{section:application}

As an immediate effect, switching back and forth between the centralized and the decentralized admission systems causes stark strategic responses in application behavior. 
Figure \ref{fig:application} shows that the three periods of centralization are associated with sharp increases in the share of applicants who chose the most selective School 1 in Tokyo as their first choice, as predicted. 
This finding suggests that applicants understood the nature of strategic incentives they face, which complements existing evidence about strategic behavior \citep{abdulkadiroglu2006changing, chen2020empirical}. 

To observe regional variations, we examine how the propensity of applicants to rank School 1 as their first choice changed between 1916 (under decentralization) and 1917 (under centralization).
Appendix Table \ref{tab:application} shows that meritocratic centralization induced applicants in all regions to rank the most prestigious school first and to make more long-distance applications.\footnote{The effect is largest for applicants from School 1's area, presumably because School 1's area has a disproportionate fraction of applicants who prefer School 1 due to local preferences. If such applicants would often not apply to School 1 under decentralization, School 1's area could experience a particularly big increase in the fraction of applicants who rank School 1 first under centralization.}
The competition to enter School 1 thus became even more intense under centralization (Supplementary Materials Figure D.1). 
%

\subsection{Regional Mobility in Enrollment}\label{section:mobility}

The changes in the assignment algorithm and application behavior influence enrollment outcomes.
To measure the geographical mobility of entrants, we compute the ``enrollment distance," defined as the direct distance between an entrant's birth prefecture and the school he entered.
As Figure \ref{fig:mobility} shows, the centralized system is associated with a sharp and discontinuous increase in enrollment distance.
\footnote{The effect is especially stark in the first two periods of centralization. The third period of centralization in 1926--27 was qualitatively different from that in the first and second periods. In the third period of centralization, schools were divided into two groups and applicants were allowed to choose and rank at most two schools (one school per group) in 1926--27.}
This increase in regional mobility is driven by a sharp reduction in the number of ``local entrants" defined as entrants who entered a school in (or near) their birth prefecture. To show this, we estimate the following regression for each school $s$ separately: 
\begin{equation} \begin{split} 
Y_{pt}^s =& \ \beta_1^s \times Centralized_{t}\times 1\{\text{school }s\text{ is located in prefecture }p\} \\ 
&+\beta_2^s \times Centralized_{t}\times 1\{\text{school }s\text{ is 1--100 km away from prefecture }p\} \\ 
& +\alpha^s X_{pt}+\gamma_{t}^s +\gamma_{p}^s+\epsilon_{pt}^s, \end{split}  \label{eq:local_monopoly}
\end{equation} 
\noindent where $Y_{pt}^s$ is the number of entrants born in prefecture $p$ who entered school $s$ in year $t$.
$Centralized_t$ is the indicator that the admission system was centralized in year $t$. 
$1\{$school $s$ is located in prefecture $p\}$ is the indicator that school $s$ was located in prefecture $p$. 
$1\{$school $s$ is 1--100 km away from prefecture $p\}$ is the indicator that school $s$ was located 1--100 km away from prefecture $p$, which roughly corresponds to the definition of school regions (defined by a set of prefectures for which school $s$ is the nearest school).
$X_{pt}$ controls for observable characteristics of prefecture $p$ and year $t$, including the number of middle-school graduates from prefecture $p$ in year $t$ and the number of higher schools other than Schools 1--8 in prefecture $p$ in year $t$. 
$\gamma_{t}^s$ and $\gamma_{p}^s$ are year and prefecture fixed effects, respectively. 

As shown in Panel (a) of Table \ref{tab:mobility1}, centralization reduces the number of local entrants born in the school's prefecture across the country. The coefficients 
$\beta_1^s$ are significantly negative for all schools. 
Column (1) shows that the number of School 1 entrants born in Tokyo prefecture declined by about 26\% relative to its mean under decentralization.
Schools 2--7 experienced reductions in the number of entrants born not only from the school's prefecture but also from surrounding prefectures.
In other words, centralization weakened the local monopoly power of each school by integrating local markets into a national market, which is consistent with our prediction. 
This evidence aligns with previous findings on the effects of centralized matching systems on regional mobility \citep{niederle2003unraveling, machado2021}. 


\subsection{Meritocracy vs Equal Access}\label{section:equality}

\begin{figure}[h!]
        \caption{Which Regions Win from Centralization?}\label{fig:equality}
        \begin{center} \vspace{-0.1in} 
        \begin{subfigure}[b]{0.7\textwidth}
        \begin{center}
                \caption{Change in No. of Local Entrants to Schools 1-8 from Each Prefecture under Centralization}\label{fig:equality2}
                \includegraphics[width=0.7\textwidth]{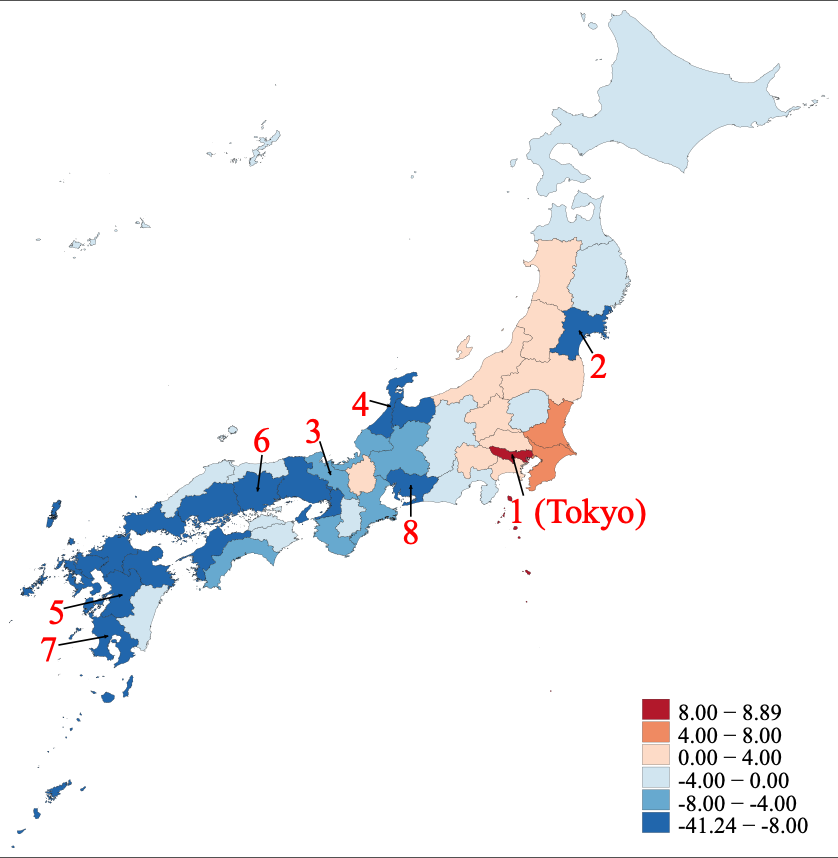}
                \end{center}
        \end{subfigure}
        \end{center}
                \begin{center}
        \begin{subfigure}[b]{0.7\textwidth}
                \begin{center}\caption{Centralization Increased Urban-born Entrants} 
 \label{fig:equality1}
                \includegraphics[width=100mm]{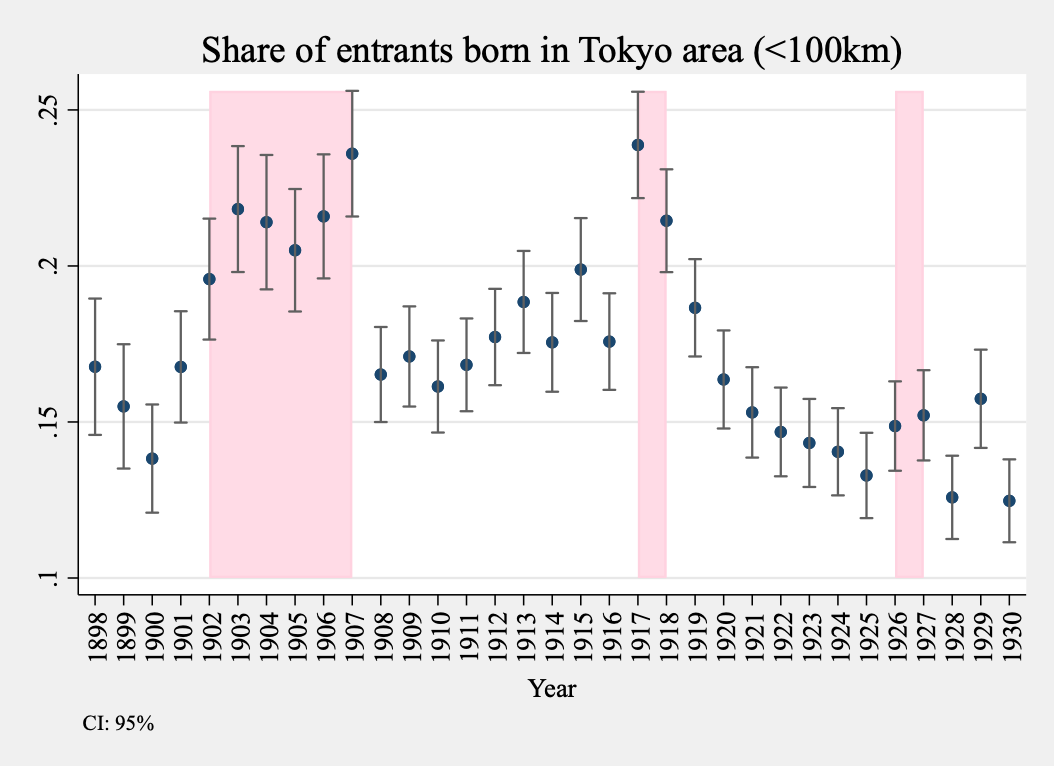}      
                \end{center}      
        \end{subfigure}
        \end{center}
        {\scriptsize \textit{Notes}: 
Panel (a) plots the estimated prefecture-specific coefficient $\beta_p$ in $\#entrants_{pt} = \beta_p Centralized_t +  \alpha_p X_{pt} + e_{pt}$, using the 1900-1930 data for each prefecture $p$, where $\#entrants_{pt}$ is the number of entrants in year $t$ who were born in prefecture $p$ and $X_{pt}$ is the number of schools other than Schools 1--8 in prefecture $p$ in year $t$. 
Panel (b) uses the entrant-level data from 1898 to 1930 to show the time evolution of the fraction of entrants to Schools 1--8 who were born in the Tokyo area (defined as the set of prefectures within 100 km from Tokyo; see Appendix Figure \ref{fig:map} for a map). 
Bars show the 95 percent confidence intervals. 
See Section \ref{section:equality} for discussions about this figure. \par}
\end{figure}

We now investigate the distributional effect of the centralized system on the allocation of school seats at Schools 1--8.
Figure \ref{fig:equality2} plots the change in the number of entrants to Schools 1--8 from decentralization to centralization by their birth prefecture, where the darker blue color indicates the greater decline and the darker red color indicates the greater increase.
The figure shows that Tokyo prefecture (where School 1 was located) and its surrounding area gained school seats under centralization, while most of the western and northern prefectures lost school seats.


%

To see this quantitatively, we regress 
the number of entrants to Schools 1--8 in year $t$ born in prefecture $p$ on the interactions between $Centralized_t$ and each of the following two indicator variables: the indicator that takes 1 if prefecture $p$ was Tokyo and the indicator that takes 1 if prefecture $p$ was 1--100 km away from Tokyo. 
To control for the decline of local entrants, we add the interactions between $Centralized_t$ and each of the following two indicator variables: the indicator that takes 1 if any of Schools 1--8 was located in prefecture $p$ and the indicator that takes 1 if any of Schools 1--8 was located 1--100 km away from prefecture $p$. 
As in Equation (\ref{eq:local_monopoly}), we control for characteristics of prefecture $p$ and year $t$ and include year and prefecture fixed effects.

Column (1) in Table \ref{tab:mobility1} Panel (b) shows that the number of entrants born in Tokyo prefecture increased by 12\% per year under centralization relative to its mean under decentralization. The increase in the number of entrants is even higher for its surrounding area; the number of entrants born in prefectures 1--100 km away from Tokyo increased by 45\% per year under centralization relative to its mean under decentralization.\footnote{One possible reason for this concentration of the effect in surrounding prefectures is that applicants from this area might be marginal applicants who would not be admitted under decentralization but would be admitted under centralization. By contrast, many high-scoring applicants from the Tokyo prefecture might be admitted even under decentralization.} 
To see which schools received more entrants born in or near Tokyo under centralization, we regress the number of entrants to each school $s$ in year $t$ born in prefecture $p$ on the interaction terms between $Centralized_t$ and the variables indicating whether entrants' birth prefecture $p$ is Tokyo or near Tokyo.\footnote{For School 1, the specification and the result are the same as Column (1) in Panel (a)  because School 1 is in Tokyo prefecture. } We include the same variables used in Equation (\ref{eq:local_monopoly}) to control for school locations and other prefecture and year characteristics. 
Under centralization, more students born in or near Tokyo entered less selective rural schools, as shown in Columns (2)--(8) in Table \ref{tab:mobility1} Panel (b).\footnote{The results remain almost the same when we additionally control for observable prefecture characteristics (see Supplementary Materials Table D.1).}

To capture the areas that benefited most under centralization, we define the ``Tokyo area" as the set of prefectures located within 100 km from Tokyo (see the map in Appendix Table \ref{tab:application} for its location).  
Figure \ref{fig:equality1} depicts the time evolution of the share of Schools 1--8 entrants born in the Tokyo area. 
Consistent with the above results, the share of Tokyo-area-born entrants rose during the periods of centralization.

Such regional inequality in access to the selective higher education becomes a concern if it creates urban-rural inequality in the geographical distribution of future elites.
Indeed, the Tokyo area exhibited characteristics that are generally considered urban, i.e., a greater population, higher income per capita, and better educational infrastructure (proxied by the number of middle-school graduates), as shown in Appendix Table \ref{tab:tokyo_area}. 
In Appendix Table \ref{tab:whytokyoarea}, we regress the number of entrants to Schools 1--8 on the interaction between $Centralized_t$ and each of these urban characteristics, instead of the Tokyo area indicator (using the same specification as in Table \ref{tab:mobility1} Panel (b) Column (1)). These urban characteristics, especially educational infrastructure, positively predict the areas that produced a greater number of entrants under centralization. 


An additional reason can be that the most competitive School 1 was located in Tokyo.  
Since students had a preference for a local school, it is likely that under decentralization, students born in the Tokyo area were more inclined to apply to School 1 and failed to enter any school, even though they might have been able to enter Schools 2--8 had they applied.
If so, the location of School 1 in the Tokyo area might have contributed to the increased share of Tokyo-born-areas entrants in Schools 1-8 under centralization. 
The concentration of the most selective universities in the capital, however, is widely observed in many countries.\footnote{Countries that have the most selective universities in the capital include France (\'Ecoles Normales Sup\'erieures, Sorbonne, and several others in Paris), China (Peking and Tsinghua Universities in Beijing), and South Korea (Seoul National University in Seoul).} 
In general, the presence of selective schools can be viewed as an urban characteristics shared by major metropolitan areas.

To summarize the short-run effects, under centralization, although School 1 lost a sizable number of urban students, Schools 2--8 admitted an even greater number of urban students.
As a result, a net outcome was such that urban-born high achievers crowded out rural students from selective higher education. 
In the next section, we explore long-run persistence of the regional inequality.

\subsection{Other Institutional Changes}\label{section:robustness}

We finish the short-run analysis by discussing potential threats. In particular, 
if changes in other institutional and policy factors were correlated with the admission reforms, it could influence application behavior and enrollment outcomes, explaining our finding that more applicants applied for School 1 during centralization. 
Such other institutional factors include simultaneous reforms in middle schools, the total number of applicants, capacities at Schools 1--8, and the capacity of School 1 relative to the capacities of other schools. 

We investigate these concerns and confirm that centralization periods are not correlated with time-series changes in the following variables (Columns (1)--(6) in Appendix Table \ref{tab:other}): the number of middle school graduates, the national number of applicants as well as the level of competitiveness (measured by the number of entrants divided by the number of applicants), the total number of entrants to Schools 1--8, the share of entrants to School 1 in all entrants, the probability of unsuccessful applicants retaking the exam in subsequent years, the average age of entrants, and government expenditure on higher education. 

This is consistent with historical documents about how one of the two admission systems was chosen and announced. 
Each year, the government announced that year's admission policy in April, three months before the exam in July, as a public notice in the Government Gazette. 
This timeline suggests that it was difficult for applicants to anticipate the exact timing of the reforms. 

A potential concern with the above time-series analysis is that the insignificant results in Appendix Table \ref{tab:other} may be due to a small sample size (around 30). 
Yet, using the same empirical specification, we find that centralization is significantly correlated with our main outcome variables (the share of applicants to School 1, the enrollment distance, and the share of entrants who were born in the Tokyo area), as shown in Columns (8)--(10) of Appendix Table \ref{tab:other}. 
Taken together, these results suggest that our findings are unlikely to be driven by institutional changes other than the school admission reforms.

\begin{figure}[h!]
        \caption{Short-run Effects of Centralization: First Look}
        \begin{center}
        \begin{subfigure}[b]{0.65\textwidth} 
        \caption{Centralization Caused Applicants to Apply Aggressively} \label{fig:application}
\includegraphics[width=95mm]{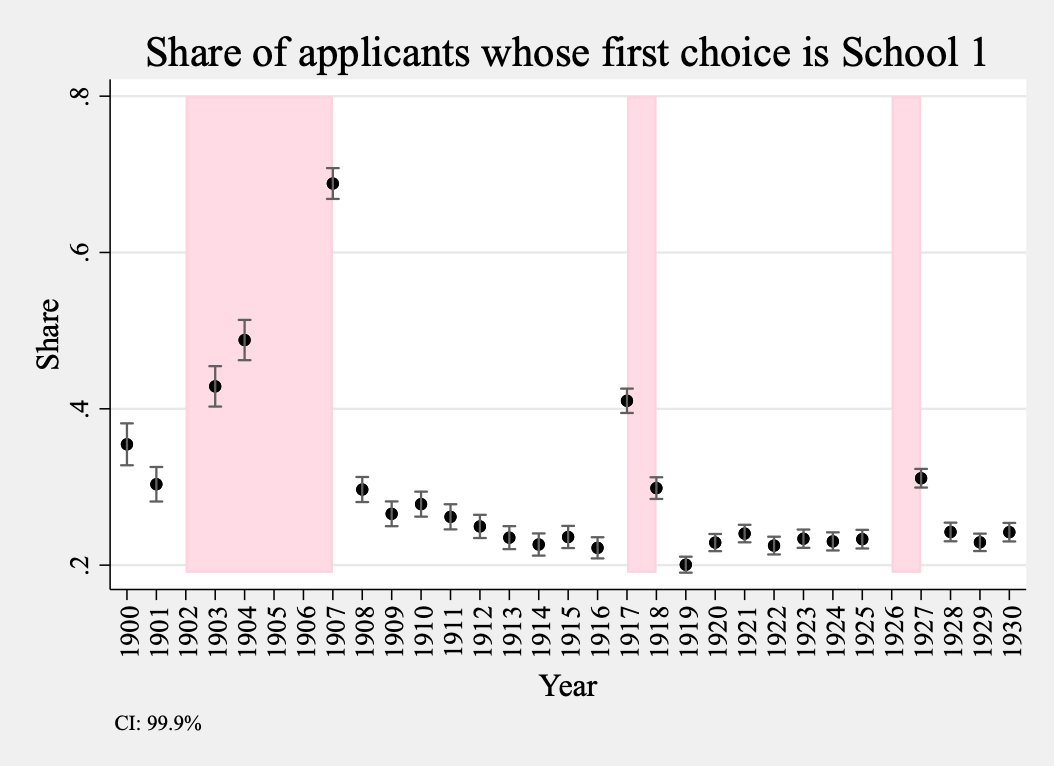} \vspace{0.1in}
        \end{subfigure} 
        \begin{subfigure}[b]{0.65\textwidth} 
        \caption{Centralization Increased Regional Mobility} \label{fig:mobility}
\includegraphics[width=95mm]{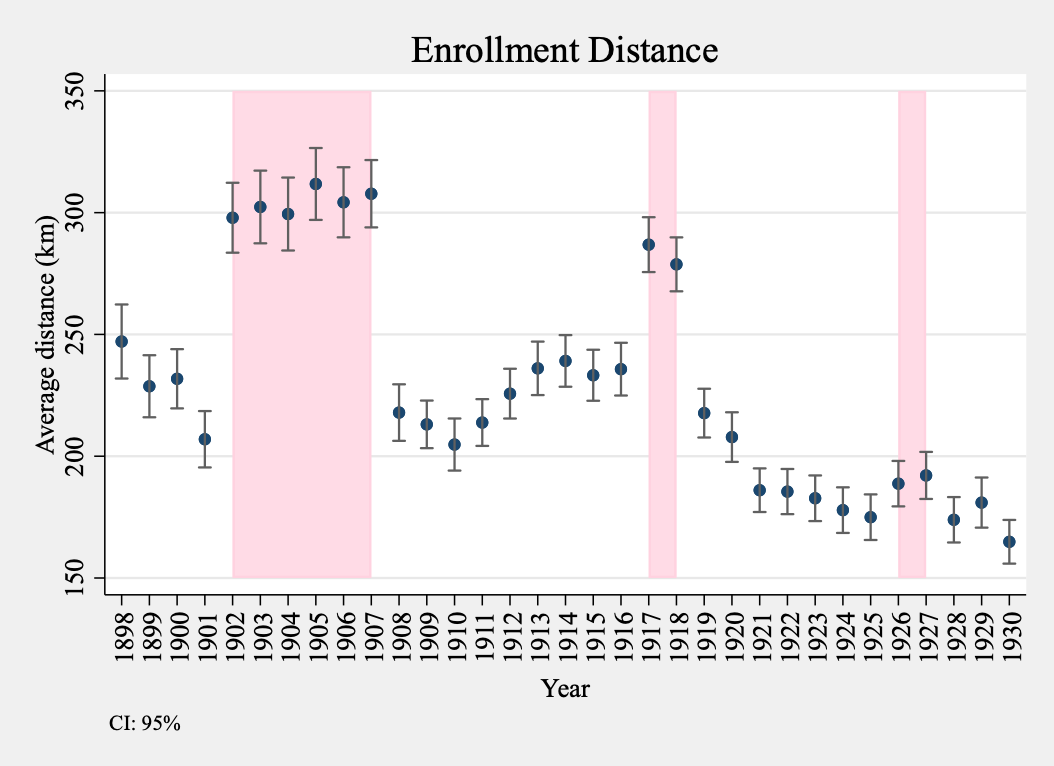}
        \end{subfigure} 
        \end{center}
        {\scriptsize \textit{Notes}: 
Panel (a) shows the time evolution of the share of applicants who selected the most prestigious School 1 (in Tokyo) as their first choice in all applicants. No data is available for 1902, 1905, 1906, and 1926.
Colored years (1902--07, 1917--18, and 1926--27) indicate the three periods of the centralized admission system.  	
Bars show the 99.9 percent confidence intervals. 
See Section \ref{section:application} for discussions about this figure. 
Panel (b) shows the time evolution of the average enrollment distance (defined by the distance between an entrant's birth prefecture and the prefecture of the school he entered).
Colored years indicate the three periods of the centralized admission system. 
Bars show the 95 percent confidence intervals.
See Section \ref{section:mobility} for discussions about this figure. \par}
\end{figure}

Finally, the centralization reform introduced not only the meritocratic assignment algorithm, but also the unified entrance exam that applicants could take at any school location (see Appendix Table \ref{tab:chronological_table2}).  As such, the estimated impacts of centralization may be confounded by the unification of entrance exams and more flexible exam location choices. To investigate this issue, we analyze how key outcomes changed from 1900 to 1901, during which the government also introduced a single entrance exam that applicants were allowed to take at any school while the assignment method remained decentralized. Figures \ref{fig:application} and \ref{fig:mobility} show that this institutional change from 1900 to 1901 induced little changes in application and enrollment patterns. The estimated impacts of centralization are therefore likely due to the meritocratic assignment algorithm rather than the changes in exam contents and locations.

\section{Long-run Impacts}\label{section:long}
To assess the long-run effects of meritocratic centralization, we provide two sets of empirical analyses. 
First, we investigate the persistence of the distributional effect of centralization, which we document in the short-run analysis. 
We do so by a difference-in-differences strategy comparing labor-market outcomes of urban- and rural-born individuals across birth cohorts with differential exposure to centralized admissions.
Our analysis shows that centralization increased both (a) the number of occupational elites coming from urban areas and (b) the number of occupational elites living in urban areas as adults. The admission reforms thus influence the regional origins and destinations of elites.

Second, we compare the national production of elites between the two admission systems. 
We find that cohorts more intensely exposed to the centralized admissions produced a larger national number of occupational elites, such as top-income earners and civil officials who were promoted to top ranks. This finding indicates that centralization boosted the national production of individuals reaching leadership positions.

\subsection{Regional Distribution of Elites}\label{section:long_run_analysis1}

\subsubsection*{Personnel Inquiry Records Data}

To analyze the long-run effects of centralization on students' career outcomes, we merge data from two editions of the \textit{Personnel Inquiry Records} (PIR) published in 1934 and 1939.\footnote{The 1939 edition of the PIR was based mostly on the information in 1938. Although Japan entered the military regime after the National General Mobility Act of 1938 and economic interventions began in the same year, their effect was relatively small in 1938--1939 \citep{moriguchi2008evolution}.}
The PIR is an equivalent of \textit{Who's Who} and compiles a selective list of socially distinguished individuals with biographical information. 
The numbers of listed individuals correspond to 0.07\% (26,117) and 0.15\% (55,742) of the adult Japanese population in 1934 and 1939, respectively.\footnote{Among 26,117 and 55,742 individuals listed in the 1934 and 1939 editions, 18,792 individuals appear in both editions. We examine the coverage of the PIR below and describe its sources in Supplementary Materials Section B.2.1.}
As shown in Appendix Table \ref{tab:occupation_PIR} Column (1), about 60\% of these individuals are corporate executives, 10\% are politicians or bureaucrats, and the rest consist mainly of professionals such as scholars and engineers.
Throughout this section, we refer to the individuals listed in the PIR as ``elites.'' 

We use this data to capture the effects of the first period of the centralized admissions in 1902--1907. We do so by focusing on the cohorts born in 1880--1894, who turned age 17 (the age eligible for application) in 1897--1911. 
These cohorts were 40 to 54 years old in 1934 and 45 to 59 years old in 1939.\footnote{Since the cohorts who turned age 17 around the second period of centralization (1917-1918) were too young to reach top positions and be listed in the PIR published in 1934 or 1939, we restrict our analysis to the first period of centralization.}
We have the following information for each individual: full name, birth year, birth prefecture,
prefecture of residence, final education, 
occupation titles and organization names, the medal for merit and the court rank awarded (if any), and the amount of national income tax and business tax paid (if any). 
Among all the elites in our dataset, 23\% graduated from Imperial Universities. 

We define the following subgroups of elites (as a subset of PIR-listed individuals): (1) the top 0.01\% and 0.05\% income earners according to the national income distribution,
(2) prestigious medal recipients (civilians who received high-ranking imperial medals for their exceptional service or merit in various fields),\footnote{These prestigious medal recipients held a broad range of occupations, including scholars, executives, engineers, and bureaucrats as shown in Appendix Table \ref{tab:occupation_PIR} Column (4).}
(3) corporate executives (individuals who hold an executive position in a corporation and pay a positive amount of income or business tax), (4) top politicians and bureaucrats (individuals whose occupation is either Imperial Diet member, Cabinet member, or high-ranking central government official), 
(5) Imperial University professors or associate professors,
(6) all occupational elites (all individuals listed in the PIR except for hereditary elites defined as individuals whose sole occupation is peerage or landlord).\footnote{Supplementary Materials Section B.2.1 and Table D.2 provide definitions and descriptions of the elite subgroups.} 
These categories encompass economic, political, and cultural definitions of occupational elites.
In addition, we also look at PIR-listed Imperial University graduates as a proxy for Schools 1--8 graduates.\footnote{Recall that those who graduated from Schools 1--8 were automatically accepted to Imperial Universities during this period.}   

We first count the number of individuals in each subgroup defined above by birth prefecture by birth cohort. 
We then divide these counts by male birth population estimates in each prefecture to adjust for the differences in prefectural populations.\footnote{As described in Supplementary Materials Section B.2.1, we first estimate cohort-prefecture-level male birth population and then use the average of this estimate over the cohorts born in 1880--1894 to reduce the influence of measurement errors in population estimates. The regression results are similar when we use cohort-prefecture-level male birth population estimates instead (Supplementary Materials Table D.3).} 
This procedure allows us to conduct a difference-in-differences analysis that compares long-term elite density among urban- and rural-born individuals by each cohort's exposure to the centralized admission system.
Descriptive statistics are summarized in Appendix Table \ref{tab:statistics}.

\subsubsection*{Assessing the Coverage and Bias of PIR Data}

Since our PIR data is not exhaustive administrative data, we are concerned about potential sample selection bias. 
For the top income earners and Imperial University professors, we can compute the exact sampling rates by comparing the number of individuals in our data against complete counts reported in government statistics.  
We find that the sampling rates are decent even by modern standards: 47\% (or 70\%) for Imperial University professors, 50\% (or 53\%) for the top 0.01\% income earners, and 40\% (or 40\%) for the top 0.05\% income earners in the 1934 (or 1939) edition of the PIR. 
Consistent with the nature of the PIR that lists only highly distinguished individuals, the sampling rates increase with the income level (see Appendix Figure \ref{fig:sampling_rate1}). 

Sample selection bias threatens our difference-in-differences analysis only if the difference in sampling rates between urban and rural areas changes with the cohort's exposure to the centralized admission system.
To assess this possibility, we examine the prefecture-level sampling rates for the top income earners. 
As Appendix Figure \ref{fig:sampling_rate2} shows, the prefecture-level sampling rate (defined by the number of high income earners in our PIR data divided by the complete count from tax statistics) is not systematically correlated with the prefecture-level complete count of high income earners. In particular, while Tokyo prefecture has the largest number of high income earners in the tax statistics, the sampling rate is close to the average of all prefectures.

Even so, one additional concern is that Schools 1-8 and Imperial University graduates might have a higher likelihood of being sampled by the PIR even after controlling for the income level. 
However, we find no positive correlation between the sampling rates of top income earners and the numbers of entrants to Schools 1--8 per birth population across prefectures (see Appendix Figure \ref{fig:sampling_rate2}). 
We further find that our main results remain similar even when we multiply the number of top income earners in the PIR by the inverse of the sampling rates or restrict the sample to prefectures with relatively high sampling rates (as discussed later).
This series of findings suggests that possible sample selection bias in our PIR data is unlikely to drive our empirical results.

\subsubsection*{Urban-Rural Disparity in Producing Elites}\label{section:long_run_results1} 

We estimate the long-run impacts of the centralized admission system by conducting a difference-in-differences analysis by birth cohorts and birth areas. 
The idea behind our empirical strategy is that urban-born applicants experienced a greater gain in entering Schools 1--8 under centralization relative to decentralization, as shown in the short-run analysis in Figure \ref{fig:equality} and Table \ref{tab:mobility1}.   
We exploit this differential gain in school access to compare the career outcomes of individuals born inside and outside the Tokyo area by the cohort's exposure to centralization. 
If admission to Schools 1--8 increases one's chance of becoming an elite, we should observe more elites born inside the Tokyo area for the cohorts exposed to centralization. 
We estimate a difference-in-differences specification as follows:
\begin{equation} 
Y_{pt} = \beta \times Centralized_t\times Tokyo\_area_p +\gamma_p+\gamma_t +\epsilon_{pt},
\label{eq:pir}
\end{equation}

\noindent where $Y_{pt}$ is the number of elites (per 10,000 male births) born in cohort $t$ and prefecture $p$. 
$Centralized_t$ is a measure of cohort $t$'s exposure to centralization, which is the binary indicator that takes 1 if cohort $t$ turns age 17 during centralization in 1902--1907.  $Tokyo\_area_p$ is the indicator that takes 1 if prefecture $p$ is in the Tokyo area.  
$\gamma_p$ and $\gamma_t$ are prefecture and cohort fixed effects. 
To allow for serial correlation of $\epsilon_{pt}$ within prefecture over time, we cluster the standard errors at the prefecture level in our baseline specification.\footnote{\cite{bertrand2004much} evaluate approaches to deal with serial correlation within each cross-sectional unit in panel data. They suggest that clustering the standard errors on each cross-section unit performs well in settings with 50 or more cross-section units, as in our setting.} In addition, we report the results of clustering at the cohort level, which are estimated by wild cluster bootstrap \citep{cameron2015practitioner, roodman2019fast} due to the small number of clusters (15 cohorts). 

The above specification defines $Centralized_t$ to be the binary indicator as the simplest proxy for the cohort's exposure to centralization.
In reality, however, a nontrivial number of unsuccessful applicants retook the exam at age 18. 
We thus drop the cohort who turned age 17 in 1901 (under decentralization) and age 18 in 1902 (under centralization) from the sample in the baseline specification. As an alternative specification, we also use the granular intensity of the cohort's exposure to centralization (defined as the share of the number of exams taken under centralization in the total number of exams taken, estimated for each cohort).\footnote{
See Supplementary Materials Section B.2.2 for the data and method to estimate the share of exams taken under centralization for each cohort.}

\begin{figure}[hbp] \vspace{-0.4in}
        \caption{Long-run Impacts of Centralization: Geographical Origins of Elites}\label{fig:long-run1} 
         \begin{center} \vspace{-0.2in}
        \begin{subfigure}[b]{0.42\textwidth}
                \includegraphics[width=\textwidth]{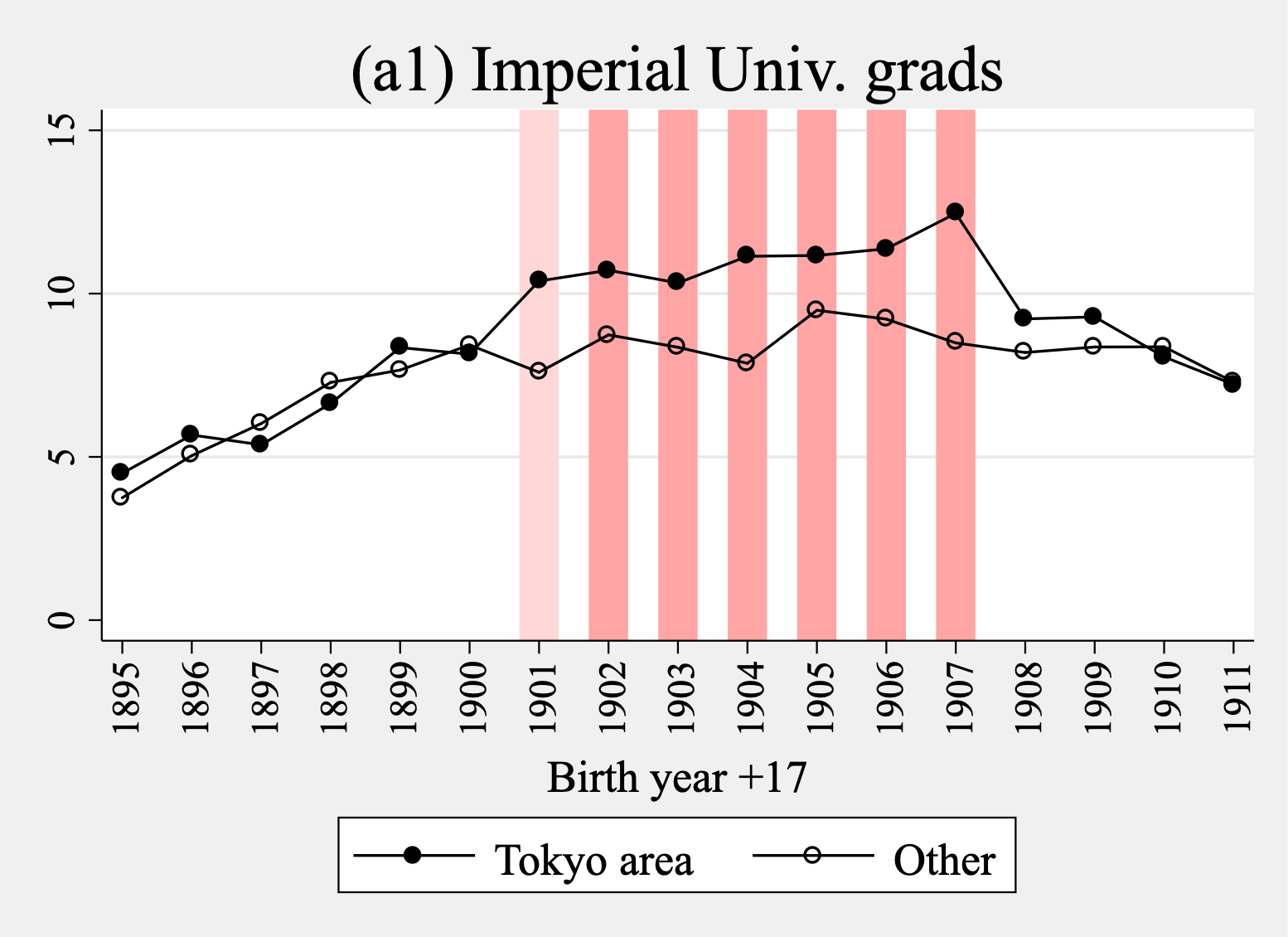}
        \end{subfigure}
                ~ 
        \begin{subfigure}[b]{0.42\textwidth}
                \includegraphics[width=\textwidth]{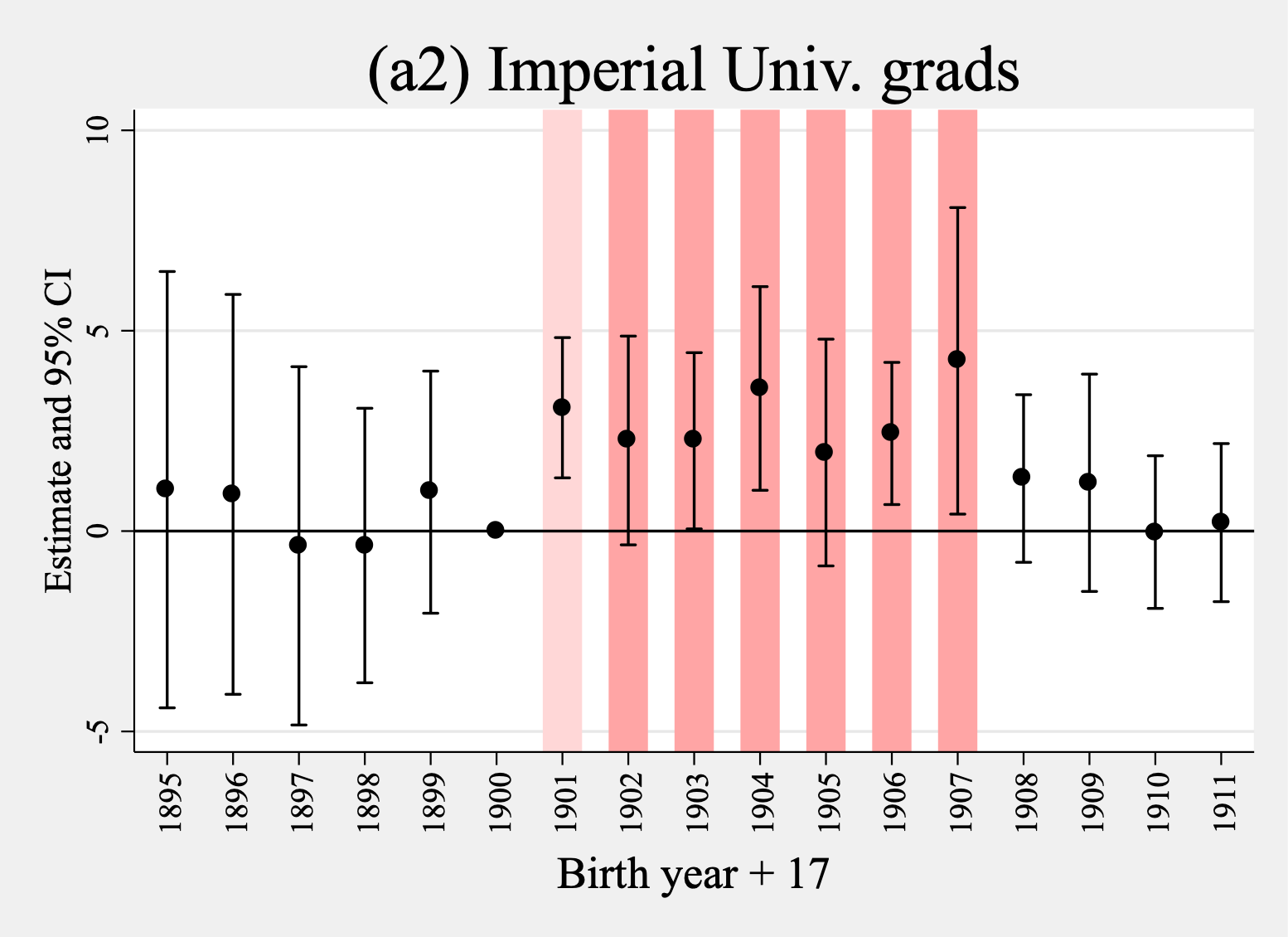}            
        \end{subfigure}
                ~
        \begin{subfigure}[b]{0.42\textwidth}
                \includegraphics[width=\textwidth]{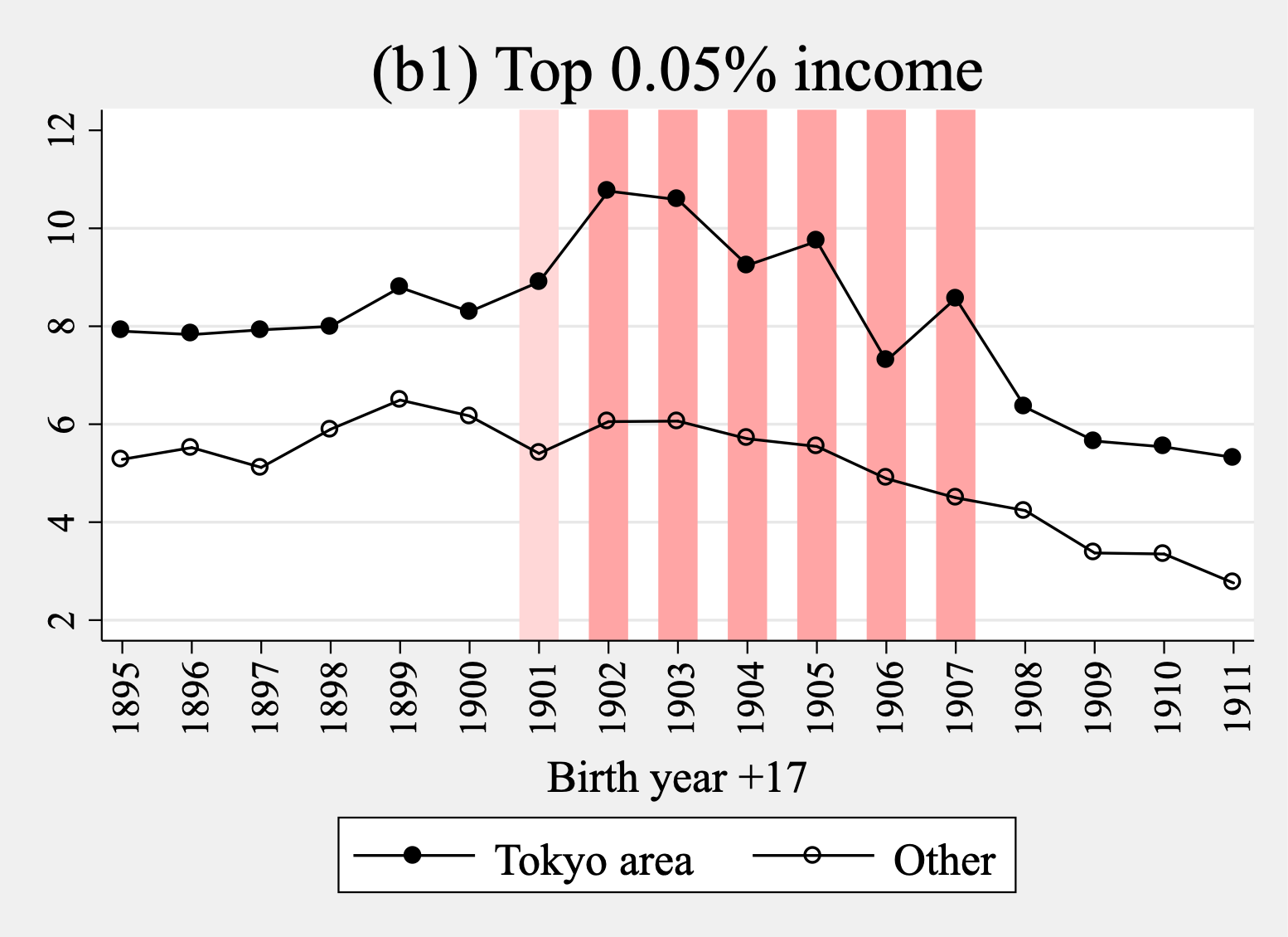}
        \end{subfigure}
                ~ 
        \begin{subfigure}[b]{0.42\textwidth}
                \includegraphics[width=\textwidth]{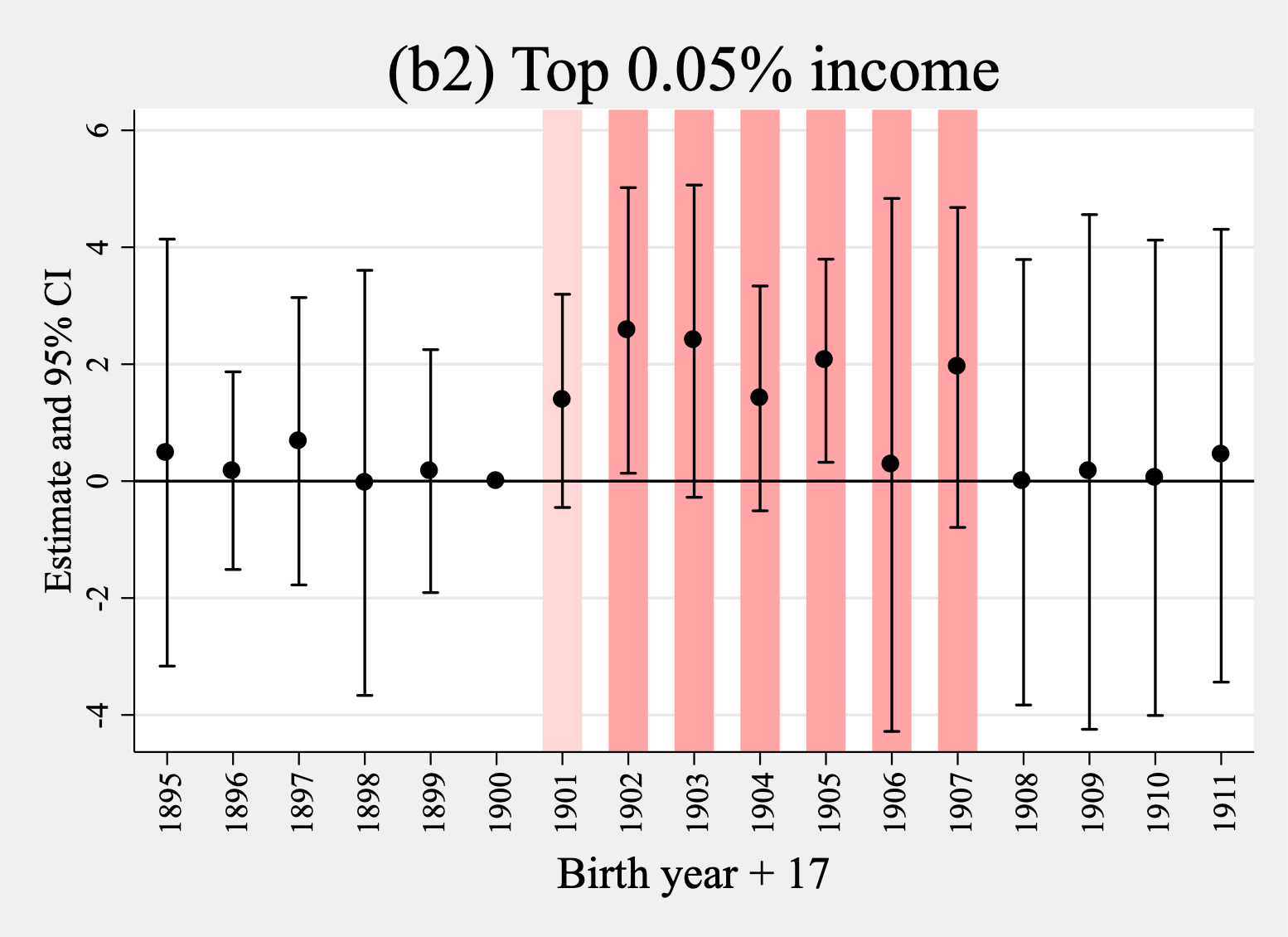}         
        \end{subfigure}  
                        ~ 
        \begin{subfigure}[b]{0.42\textwidth}
                \includegraphics[width=\textwidth]{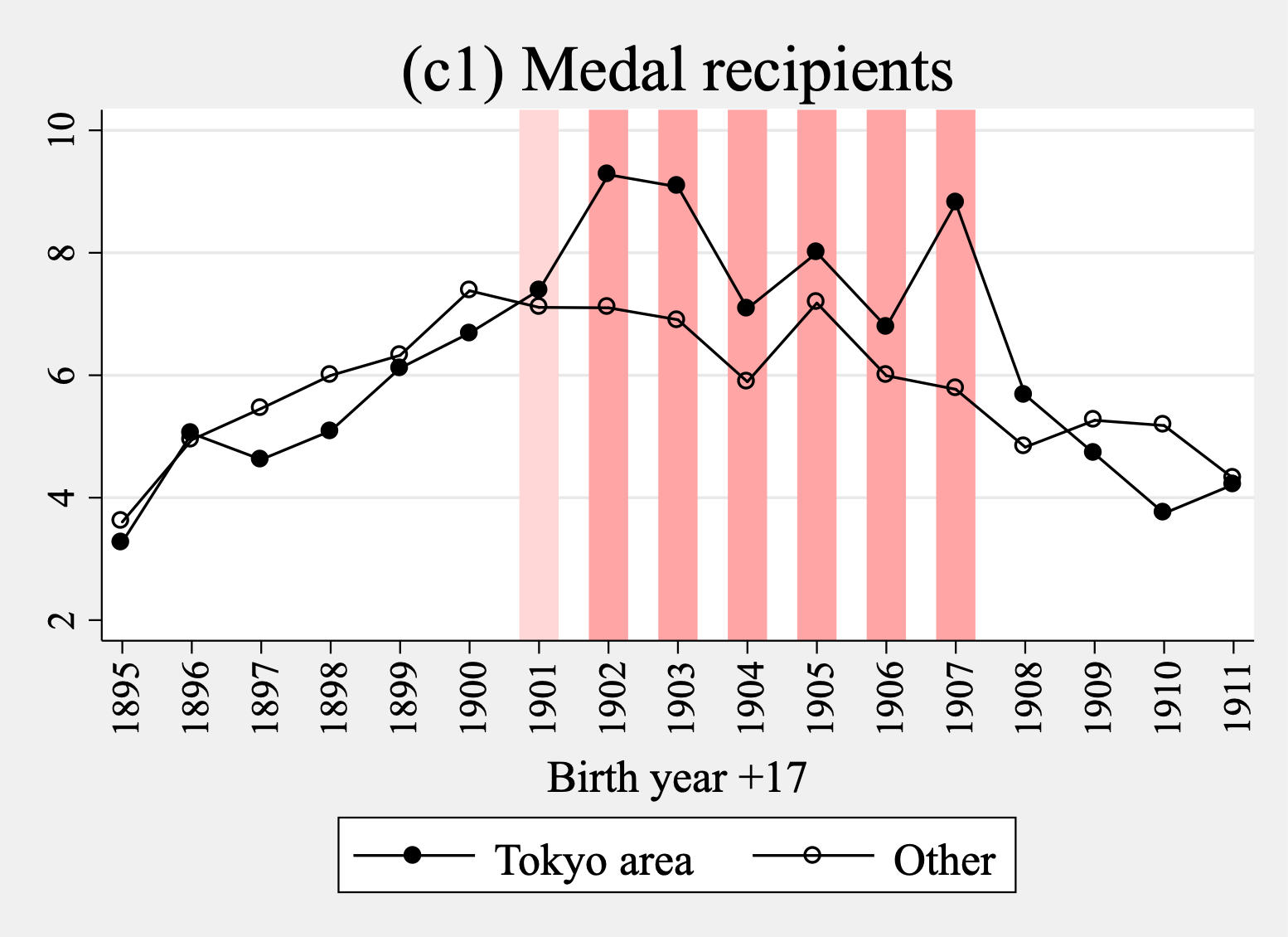}            
        \end{subfigure}   
                ~ 
        \begin{subfigure}[b]{0.42\textwidth}
                \includegraphics[width=\textwidth]{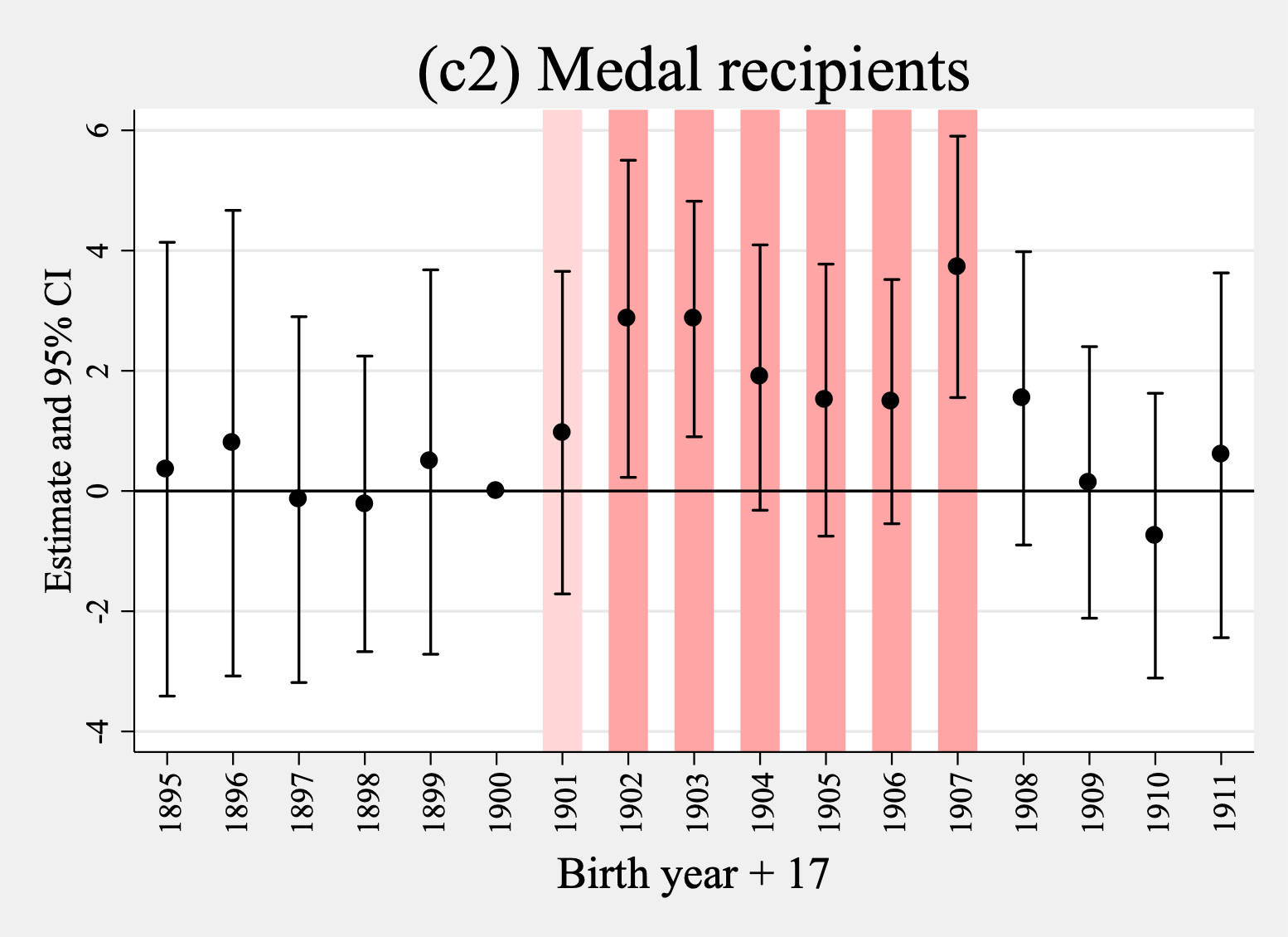}         
        \end{subfigure}
                                ~ 
        \begin{subfigure}[b]{0.42\textwidth}
                \includegraphics[width=\textwidth]{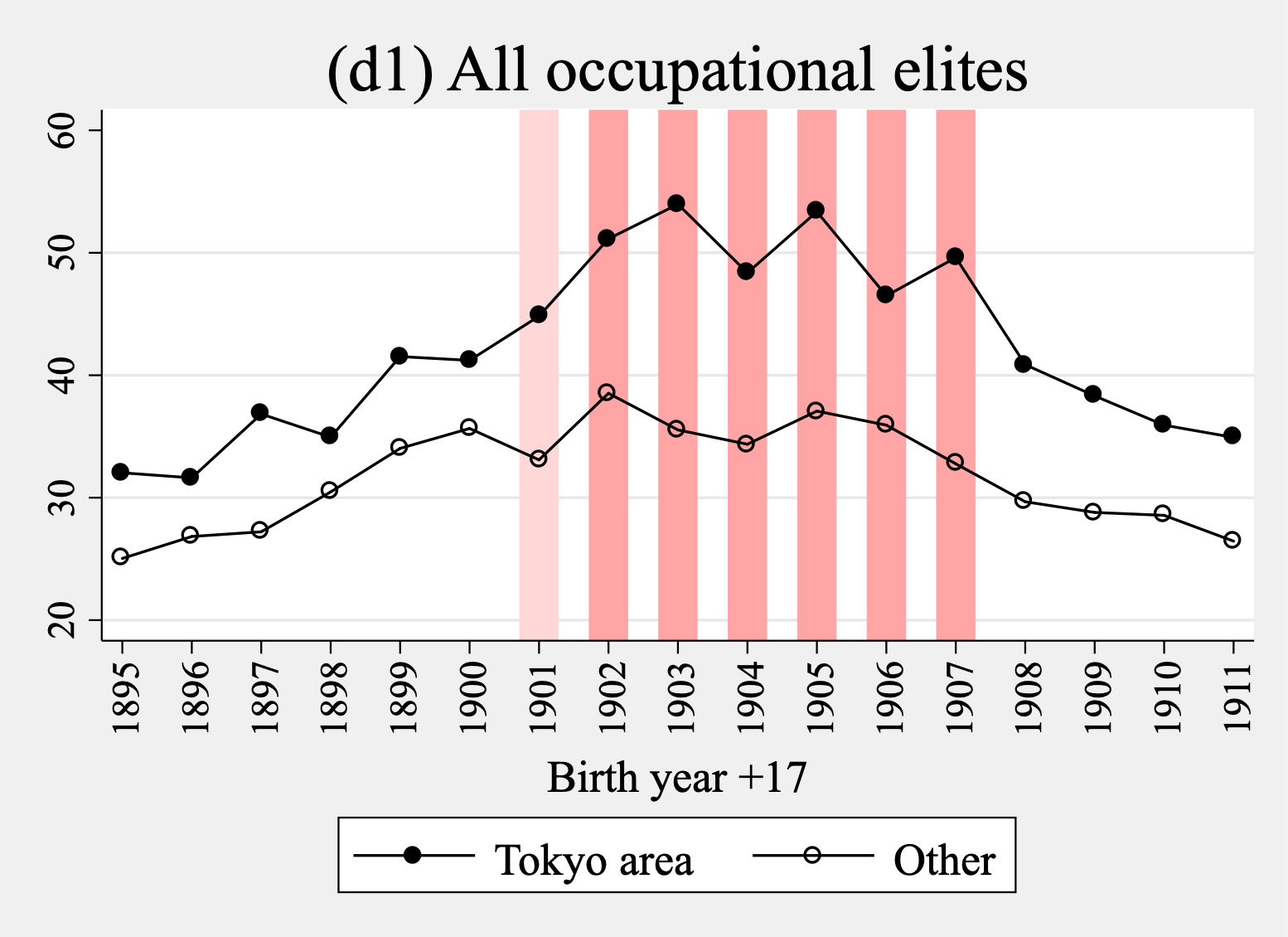}            
        \end{subfigure}   
                ~ 
        \begin{subfigure}[b]{0.42\textwidth}
                \includegraphics[width=\textwidth]{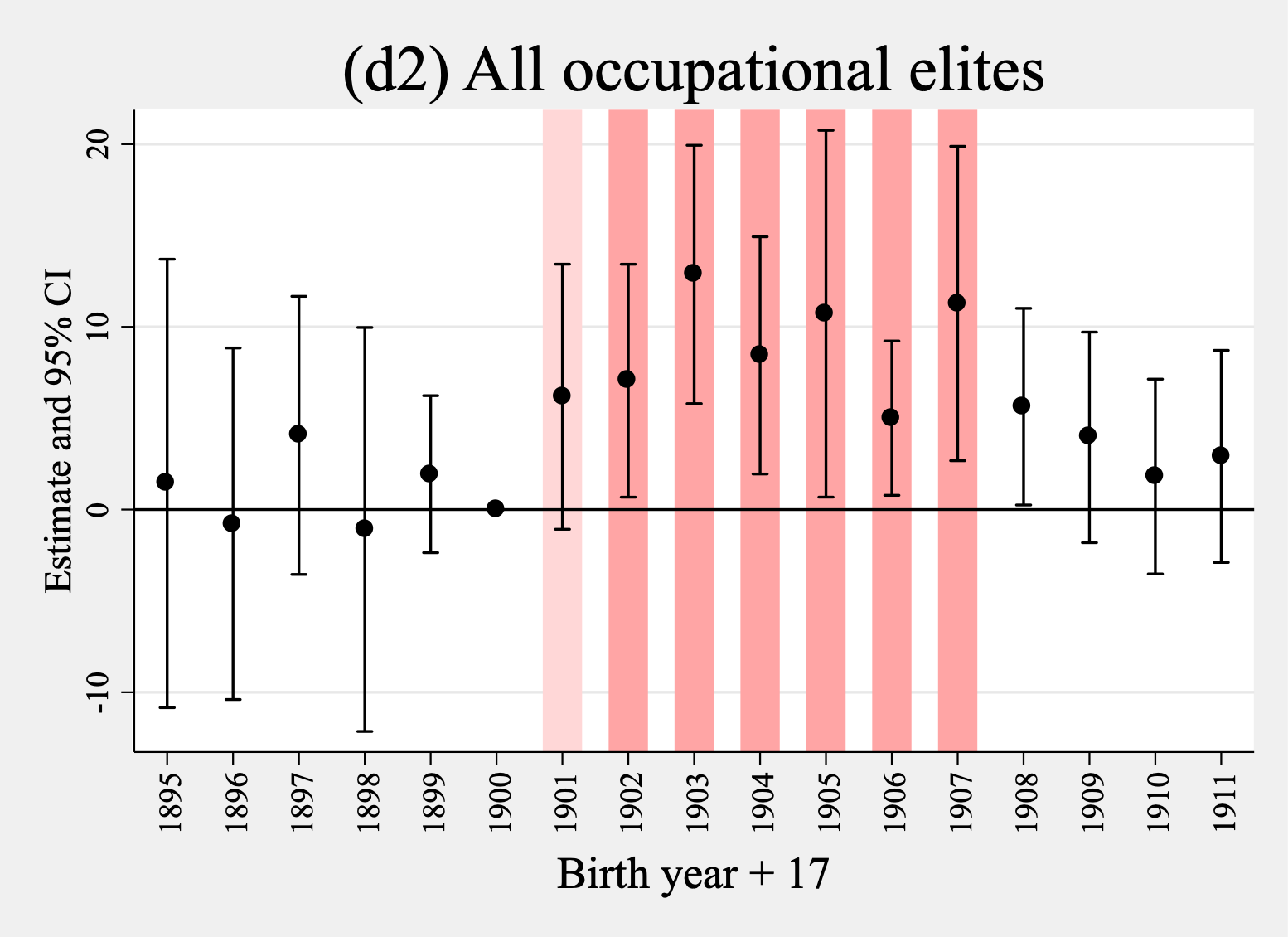}         
        \end{subfigure}
        \end{center} \vspace{-0.1in}
        {\scriptsize \textit{Notes}: 
In each panel, the horizontal axis shows cohorts defined by the year in which they turn age 17 (the minimum application age).
The cohorts turning age 17 in 1902--1907 (under centralization) are colored in dark pink. 
The cohort turning age 17 in 1901 (under decentralization) and age 18 in 1902 (under centralization) is colored in pale pink.
Left panels take the prefecture-cohort-level number of specified elites (per 10,000 male births) and compare their average inside and outside the Tokyo area for each cohort. 
Right panels are event-study plots that show the coefficients of the Tokyo area indicator interacted with the cohort indicators (controlling for cohort and prefecture fixed effects), where the whisker around the coefficient for each cohort indicates the 95\% confidence interval based on the standard errors clustered at the prefecture level (except for the 1900 cohort which is the baseline cohort). 
The data is from the two editions of the \textit{Personnel Inquiry Records} published in 1934 and 1939.
See Section \ref{section:long_run_analysis1} for discussions about this figure. \par}

\end{figure}

We first present visual results in Figure \ref{fig:long-run1}. 
In the horizontal axis, cohorts are indicated by the year in which they turn age 17. The cohorts turning age 17 in 1902--1907 under centralization are colored in dark pink, and the cohort turning age 17 in 1901 and age 18 in 1902 is colored in pale pink.
In the left panels, using the prefecture-cohort-level elite counts (per 10,000 male births) defined above, we compare their average inside and outside the Tokyo area.
In the right panels, we plot event-study estimates of the coefficients of the Tokyo area indicator interacted with binary cohort indicators (controlling for prefecture and cohort fixed effects).

We first check in Panel (a) if the area that produced more Schools 1--8 entrants under centralization generated more Imperial University graduates. We then present the results for the top 0.05\% income earners in Panel (b), medal recipients in Panel (c), and all occupational elites in Panel (d).
In all of the right panels, the urban-rural difference in the number of elites stays close to zero in pre-centralization cohorts, supporting parallel pre-trends. The urban-rural difference then becomes positive for most of the centralization cohorts and falls when centralization ends. 
We obtain similar results for the top 0.01\% income earners, corporate executives, top politicians and bureaucrats, and Imperial University professors, as shown in Appendix Figure \ref{fig:long-run_additional}.

In Table \ref{tab:long_run}, we report the regression estimates of $\beta$ in Equation \ref{eq:pir}, confirming that the long-run effects of centralization are economically and statistically significant. In Panel A, we control only for cohort and prefecture fixed effects. In Panel B, we also control for time- and cohort-varying prefecture characteristics,\footnote{We control for the number of primary schools, the number of middle school graduates, and GDP, all at the cohort-prefecture level. See Supplementary Materials Section B.2.1 for data and definitions.}  where the coefficients remain sizable.
For centralization cohorts, the number of elites born inside the Tokyo area (compared to those born outside the Tokyo area) increases by 31\% for the top 0.01\% income earners, 21\% for the top 0.05\% income earners, 34\% for medal recipients, 13\% for corporate executives, 53\% for top politicians and bureaucrats, 42\% for Imperial University professors, and 13\% for all occupational elites (see Panel B).\footnote{In Supplementary Materials Table D.4, we also examine other types of occupational elites and obtain similar results for scholars and physicians.}

Panel C of Table \ref{tab:long_run} shows that the effects are symmetric with respect to the direction of the admission reforms, i.e., the change from decentralization to centralization and the change from centralization to decentralization produce similar effects of the opposite sign. 
In Panel D, instead of the binary indicator, we use the continuous measure of the cohort's intensity of exposure to centralization (and include the cohort who turned age 17 in 1901 in the sample). The results remain qualitatively the same as the baseline results. 

Almost four decades after its implementation, meritocratic centralization had lasting impacts on students' career trajectories. 
These results are robust to alternative specifications.  
First, we conduct additional tests for the parallel trends assumption and its sensitivity.
Appendix Table \ref{tab:pretrend} verifies that differences in pre-event trends between the areas of comparison are small and mostly insignificant in non-parametric (joint significance test), linear, and non-linear specifications. 
We also examine the sensitivity of the estimates to potential violations of the parallel trend assumption. We use the method of \citet{Rambachan_Roth2023}, which provides robust confidence intervals for difference-in-differences estimates in the presence of bounded differential trends. Appendix Figure \ref{fig:long-run_sensitivity} shows that our estimates for all occupational elites are robust to the presence of differential trends of up to 0.15--0.2 percentage points per year. 

Second, another potential threat to our identification strategy is that there may be some age-specific trends in the number of elites that covary with the cohort-area variation we use. 
For instance, the number of elites listed in the 1939 edition of the PIR peaks at around the cohort who were 51 years old in 1939 (who were age 17 in 1905) and gradually falls for younger and older cohorts. This suggests that there are certain ages at which individuals are more likely to be listed in the PIR. 

Such age effects may generate different trends in the number of elites between the Tokyo area and other areas. 
To address this concern, we use PIR (1934) and PIR (1939) separately and conduct the same regression analysis using each edition of the PIR. 
Appendix Table \ref{tab:JPIR1934} shows that our key results are qualitatively the same across the two editions, even though they differ in the age at which centralization cohorts are observed. 
This also confirms that our main results are not driven by potential differences in the selection criteria of elites between the two editions. 

Third, to examine the potential influence of differences in sampling rates across prefectures, we multiply the number of top income earners by the inverse of the prefecture-level sampling rate. In Appendix Table \ref{tab:long_run_inversesample}, we use the adjusted number of top income earners as alternative outcomes and show that the results are qualitatively the same. Furthermore, as shown in Appendix Table \ref{tab:long_run_high_sample_rate}, restricting the sample to prefectures with relatively high sampling rates of top income earners (equal to or higher than that of Tokyo prefecture) yields coefficients of similar magnitude to the baseline estimates by PIR edition in Appendix Table \ref{tab:JPIR1934}.

Finally, we conduct placebo tests to examine if the results are driven by other factors, such as the sample selection of the PIR or changes in the cohort population. The urban-rural difference in the cohort's birth populations does not change significantly with the cohort's exposure to centralization, as shown in Appendix Table \ref{tab:long_run_placebo}. 
As an additional placebo test, we also look at unrelated career outcomes. 
Among PIR-listed individuals, we expect that landlords are least likely to be affected by centralized admissions as receiving higher education was not a typical pathway to become a landlord.\footnote{For landlord definition, see Supplementary Materials Table D.2.}
As shown in Appendix Table \ref{tab:long_run_placebo}, the effect of centralization on the number of landlords is small and statistically insignificant. These findings further support our conclusion that meritocratic centralization enlarged the urban-rural gap in the capacity to produce individuals in leadership positions.



\subsubsection*{Geographical Destinations of Elites}

Having established that centralization affects the geographical \textit{origins} of occupational elites, we now ask how it affects their \textit{destinations}. 
While the former concerns regional inequality in educational opportunities, the latter concerns regional inequality in the supply of highly skilled human capital. 

We first examine if the centralization-induced increase in inter-regional mobility in the short run boosted the geographical mobility of elites in the long run. 
It did not: The urban-rural difference in the fraction of elites whose prefectures of residence differ from their birth prefectures did not increase for centralization cohorts (Appendix Table \ref{tab:long_run_placebo}). 
We find similar results when we use the distance between an elite's birth prefecture and his prefecture of residence as an alternative measure of long-run mobility. 
Even though a greater number of Tokyo-area-born students entered rural schools under centralization, most of them returned to the Tokyo area when pursuing their careers. 

If the increased number of Tokyo-area-born students admitted to rural schools under centralization returned to the Tokyo area eventually, then we expect a greater number of elites to be living in the Tokyo area for the cohorts exposed to centralization.
To test this hypothesis, we redefine the outcome variables by changing the prefecture ($p$) from birth prefecture to prefecture of residence and estimate the same equation. 

Table \ref{tab:long_run_destination} indicates large positive effects of centralization on the urban-rural gap in the number of elite residents (per 10,000 male births in the prefecture).
For the cohorts exposed to centralization (relative to decentralization), the number of elites living in the Tokyo area in their middle age increases (compared to those living outside the Tokyo area). The magnitude is 31\% for the top 0.01\% income earners, 24\% for the top 0.05\% income earners, 29\% for medal recipients, 21\% for corporate executives, 22\% for top politicians and bureaucrats,  40\% for Imperial University professors and 19\% for all occupational elites (Panel B). 
These results suggest that meritocratic centralization intensified the concentration of elites in urban areas in the long run.

\subsection{National Production of Elites}\label{section:long_run_analysis2}

The above analysis examines the distributional consequence of centralized admissions. A complementary problem is its overall impact on the whole country. To study this, we explore whether meritocratic centralization increased the national production of occupational elites in the long run. 

We first focus on a specific subgroup of elites---higher civil officials---for whom we have complete count data from administrative records. 
They are also a key group of occupational elites in our empirical context, as higher civil service was considered to be one of the most prestigious jobs due to their political and legislative influence. 
As a result, high-achieving students at the top universities competed to enter the Ministry of Finance and other selective ministries \citep{shimizu}.
We investigate whether cohorts exposed to the centralized admissions produced a greater number of top-ranking civil officials compared to cohorts exposed to the decentralized admissions. 

Finally, using PIR data, we analyze the impact of centralization on the national production of all occupational elites as well as major subgroups of elites (top income earners, medal recipients, and corporate executives). 


\subsubsection*{Higher Civil Officials Data}

Our main data is the list of individuals who passed the Higher Civil Service Examinations (HCSE) and their biographical information \citep{hata}. The HCSE were selective national qualification exams held annually from 1894 to 1947. 
The 1893 ordinance required all individuals, including Imperial University graduates, to pass the HCSE for appointment in the administrative division of higher civil service (\citealp{spaulding}, Chapter 25; \citealp{shimizu}, Chapter 5).\footnote{See Supplementary Materials Section B.2.2 for exam details.} We digitized the information of all individuals who passed the HCSE in 1894--1941, including their full name, education (both university and higher school), year of university graduation, year of passing the exam, starting and final positions, year of retirement, and other notable positions held. 

We define ``top-ranking officials'' as higher civil officials who were internally promoted to one of the top three ranks by the end of their career.
In the prewar Japanese bureaucracy system, the higher civil service refers to the top ten ranks of national government offices. 
Within the higher civil service, the top three ranks were distinctively called ``imperial appointees.'' The first rank consisted of minister-level positions, while the second and third ranks consisted of vice-minister-level positions (such as vice minister, director general, bureau chief, and prefectural governor). 
Supplementary Materials Section B.2.2 provides more details. 

We count the number of top-ranking officials and the number of individuals who passed the administrative division of the HCSE (hereafter ``exam passers'') by cohort. 
In this section, cohort is defined by ``the year of entering a higher school or its equivalent.'' 
Since we only observe the year of university graduation in the HCSE data, we collect Student Registers data and impute this variable partially (see Supplementary Materials Section B.2.2 for the data and method).
In this analysis, we focus on the cohorts who entered a higher school or its equivalent between 1898 and 1930, which cover the three periods of centralization.

Out of 6,255 exam passers, 55.8\% are Schools 1--8 graduates, and 15.7\% are top-ranking officials. 
Among 982 top-ranking officials, 71.4\% are Schools 1--8 graduates. 
More descriptive statistics are shown in Appendix Table \ref{tab:statistics}. 

\subsubsection*{Effect on Top-Ranking Higher Civil Officials}
Our main outcome variable is the number of higher civil officials who were internally promoted to the top three ranks by the end of their career.
We also examine two subgroups of these top-ranking officials: (a) those who graduated from School 1--8 and (b) those who did not. For all top-ranking officials and each subgroup, we estimate the following equation: 
$$Y_t = \theta Centralized_t + \xi_1  X_t +  f(Trend_t) +\omega_t,$$
where $Y_t$ is the number of top-ranking officials in a given group in cohort $t$ (defined by the year of entering a higher school or its equivalent). $Centralized_t$ is the indicator that cohort $t$ entered a higher school or its equivalent during centralization.
We control for the number of exam passers in cohort $t$ (denoted by $X_t$) and a quadratic or 6th-order polynomial time trend where $Trend_t$ is the number of years since 1897. The results remain similar for any of the 1st to 6th-order polynomial time trend controls.

Meritocratic centralization produced a larger number of high-quality bureaucratic elites for
the whole country. Table \ref{tab:long_run_toprankgov} Columns (1)-(2) show that 
the total number of top-ranking officials increased for centralization cohorts by 15--17\%. 
This result is consistent with the idea that the centralized admission system (which assigned top-scoring students to Schools 1--8) increased the quality of Schools 1--8 entrants, which in turn improved the likelihood of Schools 1--8 graduates to win promotion competitions over decentralization cohorts around them. 
Indeed, Columns (3)-(4) indicates that centralization had a large, positive, and significant effect on the number of top-ranking officials who graduated from Schools 1--8. 

In fact, centralization's positive effect on Schools 1--8 graduates was so large (Columns (3)-(4)) that it dominated centralization's small and negative effect on those who did not graduate from Schools 1--8 (Columns (5)-(6)), resulting in a positive effect on the total number of top-ranking officials (Columns (1)-(2)). This finding suggests that the meritocratic assignment of high achievers to Schools 1--8 produced a net positive effect on the national production of top-ranking officials. Notably, this result is inconsistent with the pure selection hypothesis that the role of centralized admissions is simply to select and send a fixed number of high-achieving students to national higher education and that the value added of national higher education on subsequent career outcomes is homogeneous across all students. 

This particularly large value-added at Schools 1--8 for high-achieving students can be due to multiple factors. 
First, human capital returns to education by these schools may be higher for high-achieving students (e.g., due to demanding curriculum). Gathering high-achieving students in these schools may also produce greater peer effects. 
Second, high-achieving students may benefit more from their signaling value or from gaining connections with powerful alumni. 
Our data and variation do not permit clear differentiation of these factors. Yet, our analysis suggests that the effect of meritocratic centralization remains similar even after controlling for the size of same-cohort colleagues from the same schools or focusing on those officials with the highest educational qualification (Supplementary Material Section C.1). Much of Schools 1--8's effects therefore likely come from the first human capital factor. 

The human capital hypothesis is also consistent with additional evidence from an intermediate outcome of passing the HCSE. 
Appendix Table \ref{tab:long_run_toprankgov_add} Panel (a) shows that meritocratic centralization increased the number of exam passers who were Schools 1--8 graduates by 12--18\%, while reducing that of non-Schools 1--8 graduates by the same magnitude. 
Namely, Schools 1--8 graduates in centralization cohorts were more likely to pass the competitive national exams.\footnote{
When we control for the number of exam passers who were Schools 1--8 graduates, the effect of centralization on the number of top-ranking officials who were Schools 1--8 graduates remains significant, while its magnitude declines by half (Appendix Table \ref{tab:long_run_toprankgov_add} Panel (b)). 
This implies that the greater number of top-ranking officials in centralization cohorts who were Schools 1--8 graduates is due not only to their larger size of the exam passers, but also to their higher likelihood of being promoted to the top ranks.
}

One potential threat to our identification is that the number of available top-ranking positions might have increased during the centralization periods. 
To investigate this concern, we first confirm that centralization did not have a significant effect on the total number of exam passers in each year (Appendix Table \ref{tab:long_run_toprankgov_add} Panel (a)). This reflects that the total number of exam passers was largely determined by government demand-side factors (e.g., the capacity of each ministry) rather than supply-side factors (e.g., the quality of exam takers). 

Second, recall that we define our cohort by the year of entering a higher school or its equivalent (and not by the year of becoming top-ranking officials). Within each cohort, the number of years taken from entering a higher school to the appointment for the first top-ranking position varied widely from 20 to 30 years (Supplementary Materials Section C.2 and Figure D.2). 
Individuals in a given cohort were therefore promoted to top-ranking positions in different years. This limits the concern that a potential correlation between the number of top-ranking positions and the lagged periods of centralized admissions may drive our results.

Another potential threat is that some officials might have died prematurely in wars before rising to prominence. This may cause a bias in estimates if the number of such officials are not balanced for decentralization and centralization cohorts.
However, we show that the numbers of HCSE passers and top-ranking officials who died in wars change little between these cohorts (Supplementary Materials Table D.6).

\subsubsection*{Effect on All Occupational Elites}

Finally, we use PIR data to investigate ifmeritocratic centralization produced more occupational elites in general, not just government officials.
Starting with a graphical illustration, in Appendix Figure \ref{fig:careerelites_total}, we plot the number of all occupational elites 
for each of the 1934 and 1939 editions of the PIR by cohort. 
There may be age-specific trends in the number of elites listed in the PIR because individuals in their 50s are more likely to hold prominent positions. To distinguish cohort effects from age effects, we use the two editions to observe the same centralization cohorts in their late 40s in 1934 as well as in their early 50s in 1939. 
In both editions, the number of elites is greater for centralization cohorts even after age trends are taken into consideration.

More formally, in Table \ref{tab:careerelites_total}, we present regression results for all occupational elites, as well as major subgroups of elites for which we have a sufficient number of observations (i.e., top income earners, medal recipients, and corporate executives).
We pool data of the two editions of the PIR and count the number of elites aged 40--69 by cohort in each edition to obtain the cohort-edition level data.
In Columns (1)--(3), we regress this count on the binary indicator of centralization cohorts. Alternatively, in Columns (4)--(6), we use the continuous measure of the cohort's exposure to centralization  (as in Table \ref{tab:long_run} Panel D). We control for non-linear age trends (defined by either quadratic age trends, quartic age trends, or edition-specific quadratic age trends) as well as the edition fixed effect. 

Panel A shows that the number of all occupational elites is 5--13\% higher among centralization cohorts. 
In Panels B--D, we find that the number of the top 0.05\% income earners, medal recipients, and corporate executives increased by 7--9\%, 13--24\%, and 2--11\% among centralization cohorts, respectively. 
In all specifications, almost all of the estimates are statistically significant based on robust standard errors and standard errors clustered at the cohort level. 

To summarize, the centralized admission system appears more productively efficient not only for producing top bureaucrats, but also for producing other types of individuals in leadership positions. 

\section{Conclusion}\label{section:direction}


The design of selective school admissions persistently impacts the production and distribution of society's leaders. 
We reveal this fact by examining the world's first recorded use of nationally centralized admissions and its subsequent abolitions in the early twentieth century. 
While centralization was designed to make the school seat allocations more meritocratic, there turns out to be a tradeoff between meritocracy and equal access to selective higher education and career advancement.  Meritocratic centralization led students to apply to more selective schools and make more inter-regional applications. 
As high-achieving students were located disproportionately in urban areas, however, centralization caused urban high achievers to crowd out rural applicants from advancing to higher education. 

Importantly, these impacts persist: Several decades later, the centralized system produced more individuals in leadership positions, especially top-ranking government officials. 
The distributional effect also persists: Meritocratic centralization increased the number of high income earners, medal recipients, and other occupational elites coming from urban areas relative to those from rural areas. 

Even though our study uses the admission reforms unique to our historical setting, the implications of our study may have broader relevance. 
For instance, the distributional consequences of centralized meritocratic admissions may be a reason why many countries combine centralized admissions with regional quotas and affirmative action policies or continue to use seemingly inefficient decentralized college admissions. 
While this paper focuses on the particular decentralized system, the takeaway from our results would be relevant for other types of decentralized admissions as long as costly applications make applicants self-select into a limited number of schools. For any such decentralized system, the meritocracy-equity tradeoff is an important concern. 

Methodologically, the use of natural experiments in history may also be valuable for studying the long-run effects of market designs in other areas, such as housing, labor, and health markets. 
The disadvantage of using historical events is the limited availability of data.
The ideal way to alleviate the data concerns would be to use modern administrative data.
One may imagine linking administrative tax return data and school district data to measure the long-run effects of school choice reforms around the world in the past few decades.
Such an effort would be a fruitful complement to our historical study.

\newpage

\bibliographystyle{aea}
{\footnotesize
\bibliography{reference.bib}}

\newpage

 \newgeometry{right=1in,left=1in,top=1.2in,bottom=1.2in}

 \begin{landscape}
\begin{table}[h!]\vspace{-0.2in}
        \caption{Short-run Effects of Centralization on Enrollment}\label{tab:mobility1}
\begin{center}
\vspace{-0.25in}
\subfloat[Centralization Broke Local Monopoly and Increased Regional Mobility across the Country]{
\scalebox{0.76}{
\begin{tabular}{lcccccccc} \hline \hline
 & (1) & (2) & (3) & (4) & (5) & (6) & (7) & (8) \\
Dependent variable = No. of entrants to: & Sch. 1 & Sch. 2 & Sch. 3 & Sch. 4 & Sch. 5 & Sch. 6 & Sch. 7 & Sch. 8 \\ \hline
 &  &  &  &  &  &  &  &  \\
Centralized x Born in school's prefecture & -26.70 & -18.60 & -15.67 & -23.45 & -27.86 & -22.72 & -48.25 & -13.79 \\
 & (0.000)*** & (0.000)*** & (0.000)*** & (0.000)*** & (0.000)*** & (0.000)*** & (0.000)*** & (0.000)*** \\
  & [0.003]*** & [0.022]** & [0.073]* & [0.005]*** & [0.001]*** & [0.073]* & [0.001]*** & [0.375] \\
Centralized x Born near school's prefecture (1--100 km) & 0.13 & -2.97 & -4.09 & -9.41 & -11.60 & -2.10 & -1.86 & 0.60 \\
 & (0.852) & (0.265) & (0.056)* & (0.004)*** & (0.001)*** & (0.109) & (0.000)*** & (0.523) \\
  & [0.794] & [0.037]** & [0.000]*** & [0.001]*** & [0.001]*** & [0.021]** & [0.456] & [0.838] \\
 &  &  &  &  &  &  &  &  \\
Observations & 1,457 & 1,457 & 1,410 & 1,457 & 1,410 & 1,410 & 1,269 & 1,081 \\
Year FE, Prefecture FE & Yes & Yes & Yes & Yes & Yes & Yes & Yes & Yes \\ 
Mean dep var & 7.95 & 5.55 & 6.23 & 5.69 & 6.32 & 5.23 & 5.02 & 5.74 \\
Mean dep var (School's pref. under Decentralization) & 104.70 & 62.86 & 56.15 & 60.33 & 73.38 & 76.35 & 95.94 & 77.00 \\
 Mean dep var (Within 1--100km under Decentralization) & 9.12 & 20.90 & 17.87 & 27.07 & 34.96 & 8.37 & 8.56 & 15.53 \\ 
 \hline \hline
\end{tabular}

}} \\ \vspace{0.05in}
\subfloat[Centralization Increased Urban-born Entrants to Schools 2--8]{
\scalebox{0.76}{
\begin{tabular}{lccccccccc} \hline \hline
 & (1) & (2) & (3) & (4) & (5) & (6) & (7) & (8)  \\
Dependent variable = No. of entrants to:  & All schools & Sch. 2 & Sch. 3 & Sch. 4 & Sch. 5 & Sch. 6 & Sch. 7 & Sch. 8 \\ \hline
 &  &   &  &  &  &  &  &  \\
Centralized x Born in Tokyo prefecture & 25.04 &  0.99 & 3.56 & 6.34 & 3.80 & 11.67 & 6.27 & 19.56 \\
 & (0.000)*** &  (0.000)*** & (0.000)*** & (0.000)*** & (0.000)*** & (0.000)*** & (0.000)*** & (0.000)*** \\
  & [0.215] &  [0.760] & [0.255] & [0.019]** & [0.028]** & [0.000]*** & [0.049]** & [0.000]*** \\
Centralized x Born near Tokyo prefecture (1--100 km) & 11.81 &  0.17 & 0.76 & 1.95 & 0.55 & 1.11 & 0.53 & 0.57 \\
 & (0.000)*** &  (0.675) & (0.016)** & (0.000)*** & (0.192) & (0.020)** & (0.290) & (0.365) \\
  & [0.001]*** &  [0.722] & [0.001]*** & [0.000]*** & [0.092]* & [0.009]*** & [0.307] & [0.079]* \\
 &   &  &  &  &  &  &  \\
Observations & 1,457 & 1,457 & 1,410 & 1,457 & 1,410 & 1,410 & 1,269 & 1,081 \\
Year FE, Prefecture FE & Yes & Yes & Yes & Yes & Yes & Yes & Yes & Yes \\ 
Mean dep var & 45.43 & 5.55 & 6.23 & 5.69 & 6.32 & 5.23 & 5.02 & 5.74 \\
Mean dep var (Tokyo pref. under Decentralization) & 201.10 & 27.52 & 10.85 & 14.48 & 6.10 & 9.200 & 11.72 & 20.21 \\
 Mean dep var (Within 1--100km from Tokyo pref. & 26.23 & 6.74 & 1.21 & 2.87 & 0.75 & 1.24 & 1.55 & 3.23 \\
  under Decentralization)  \\ \hline \hline
\end{tabular}

}} 
\end{center}
{\scriptsize \textit{Notes}:  
Using the prefecture-year level data in 1900--1930, we define the dependent variable as the number of entrants who were born in the prefecture and entered the school indicated in the column in each year. 
In both panels, we control for year fixed effects, prefecture fixed effects, the number of middle school graduates in the prefecture, and the number of higher schools other than Schools 1--8 in the prefecture. In Panel (b), we additionally control for ``Born in school's prefecture'' and ``Born near school's prefecture (1--100 km).''
``Mean dep var" shows the mean of the dependent variable during decentralization for all prefecture-year observations. 
``Mean dep var (school's pref. under decentralization)'' shows the mean number of entrants to the school under the decentralized system, restricted to those born in the prefecture where the school is located. 
``Mean dep var (within 1--100km under decentralization)'' shows the mean number of entrants to the school during decentralization, restricted to those born in the prefectures within 100 km (excluding the prefecture where the school is located).
Parentheses contain p-values based on standard errors clustered at the prefecture level. Square brackets contain p-values based on standard errors clustered at the year level.
***, **, and * mean significance at the 1\%, 5\%, and 10\% levels, respectively.
See Sections \ref{section:mobility} and  \ref{section:equality} for discussions about these tables. 
See Supplementary Materials Section B.1 for details of the data used in these tables.
\par}
\end{table} 
 \end{landscape}

  \restoregeometry

 \begin{landscape}  
\begin{table}[h!]   \vspace{-0.5in}  
 \caption{Long-run Impacts of Centralization: Difference-in-Differences Estimates}\label{tab:long_run}   \vspace{-0.2in}  
\begin{center}
\scalebox{0.76}{
\begin{tabular}{lcccccccc}
\toprule
\toprule
                    &\multicolumn{1}{c}{(1)}&\multicolumn{1}{c}{(2)}&\multicolumn{1}{c}{(3)}&\multicolumn{1}{c}{(4)}&\multicolumn{1}{c}{(5)}&\multicolumn{1}{c}{(6)}&\multicolumn{1}{c}{(7)}&\multicolumn{1}{c}{(8)}\\
                    &\multicolumn{1}{c}{\shortstack{Imperial\\Univ.\\grads}}&\multicolumn{1}{c}{\shortstack{Top 0.01\% \\income\\earners}}&\multicolumn{1}{c}{\shortstack{Top 0.05\% \\income\\earners}}&\multicolumn{1}{c}{\shortstack{Medal\\recipients}}&\multicolumn{1}{c}{\shortstack{Corporate\\executives}}&\multicolumn{1}{c}{\shortstack{Top\\politicians \& \\bureaucrats}}&\multicolumn{1}{c}{\shortstack{Imperial\\Univ.\\professors}}&\multicolumn{1}{c}{\shortstack{All\\occupational\\elites}}\\
\hline &  \\ & \multicolumn{8}{c}{Panel A: Baseline Specification} \\ 
Age 17 under centralization&        2.36   &        0.47   &        1.56   &        2.13   &        1.41   &        0.84   &        0.35   &        6.68   \\
                    &     (0.018)** &     (0.007)***&     (0.057)*  &     (0.014)** &     (0.072)*  &     (0.000)***&     (0.048)** &     (0.113)   \\
                    &     [0.000]***   &     [0.117]   &     [0.007]***   &     [0.000]***   &     [0.070]*   &     [0.043]**   &     [0.155]   &     [0.001]***   \\
Observations        &         658   &         658   &         658   &         658   &         658   &         658   &         658   &         658   \\

\\ & \multicolumn{8}{c}{Panel B: With Control Variables} \\ 
Age 17 under centralization&        1.54   &        0.46   &        1.48   &        1.80   &        1.21   &        0.71   &        0.33   &        4.78   \\
                    &     (0.003)***&     (0.013)** &     (0.050)*  &     (0.007)***&     (0.070)*  &     (0.000)***&     (0.062)*  &     (0.079)*  \\
                    &     [0.011]**   &     [0.083]*   &     [0.001]***   &     [0.000]***   &     [0.074]*   &     [0.080]*   &     [0.163]   &     [0.005]***   \\
Observations        &         658   &         658   &         658   &         658   &         658   &         658   &         658   &         658   \\

\\ & \multicolumn{8}{c}{Panel C: Bidirectional Specification with Control Variables} \\ 
Age$\le$17 in 1902  &        1.20   &        0.69   &        1.39   &        1.79   &        2.23   &        0.48   &        0.25   &        4.60   \\
                    &     (0.045)** &     (0.010)***&     (0.001)***&     (0.002)***&     (0.010)** &     (0.018)** &     (0.245)   &     (0.073)*  \\
                    &     [0.056]*   &     [0.013]**   &     [0.002]***   &     [0.001]***   &     [0.030]**   &     [0.252]   &     [0.496]   &     [0.033]**   \\
Age$\le$17 in 1908  &       -1.83   &       -0.27   &       -1.56   &       -1.81   &       -0.36   &       -0.89   &       -0.39   &       -4.92   \\
                    &     (0.005)***&     (0.105)   &     (0.216)   &     (0.026)** &     (0.587)   &     (0.001)***&     (0.044)** &     (0.105)   \\
                    &     [0.001]***   &     [0.379]   &     [0.003]***   &     [0.018]**   &     [0.643]   &     [0.064]*   &     [0.072]*   &     [0.021]**   \\
Observations        &         658   &         658   &         658   &         658   &         658   &         658   &         658   &         658   \\

\\ & \multicolumn{8}{c}{Panel D: Centralization Exposure with Control Variables} \\ 
Cohort's exposure to centralization&        1.43   &        0.50   &        1.56   &        1.75   &        1.14   &        0.60   &        0.30   &        4.71   \\
                    &     (0.005)***&     (0.020)** &     (0.079)*  &     (0.009)***&     (0.050)*  &     (0.011)** &     (0.102)   &     (0.075)*  \\
                    &     [0.014]**   &     [0.105]   &     [0.000]***   &     [0.000]***   &     [0.152]   &     [0.081]*   &     [0.256]   &     [0.006]***   \\
                    &               &               &               &               &               &               &               &               \\
Observations        &         705   &         705   &         705   &         705   &         705   &         705   &         705   &         705   \\
Cohort FE, Birth pref. FE&         Yes   &         Yes   &         Yes   &         Yes   &         Yes   &         Yes   &         Yes   &         Yes   \\
Mean dep var        &        8.02   &        1.27   &        5.33   &        5.92   &        7.67   &        1.77   &        0.79   &       33.15   \\
Mean dep var (Tokyo area&        7.90   &        1.50   &        7.05   &        5.24   &        9.48   &        1.35   &        0.79   &       37.95   \\
under decentralization)&               &               &               &               &               &               &               &               \\
\hline \hline
\end{tabular}

}
\end{center}
{\scriptsize \textit{Notes}: 
This table shows difference-in-differences estimates of the long-run effects of the centralized admissions on the geographical origins of elites. 
We construct prefecture-cohort level data by counting the number of specified individuals listed in PIR (1934, 1939) by birth prefecture by birth cohort (born in 1880--1894) and dividing this count by 10,000 male births in each prefecture. 
``Age 17 under centralization" in Panels A and B is the indicator variable that takes 1 if the cohort turned age 17 (minimum application age) under the centralized admissions in 1902--1907.
``Age$\le$17 in 1902 (or 1908)" in Panel C is the indicator variable that takes 1 if the cohort turned age 17 in 1902 (or 1908) or later. 
``Cohort's exposure to centralization" in Panel D is a continuous measure of the cohort's intensity of exposure to the centralized admissions in 1902--1907.
In Panels A, B, and C, we drop the cohort who turned age 17 in 1901 from the sample. 
In Panels B, C, and D, we control for time- and cohort-varying prefecture characteristics, i.e., the number of primary schools in the prefecture in the year when the cohort turned eligible age, the number of middle-school graduates in the prefecture in the year when the cohort turned age 17, and log of GDP of the prefecture when the cohort turned age 20. 
``Mean dep var" is the mean of the dependent variable for all prefecture-cohort observations, and
``Mean dep var (Tokyo area under decentralization)" is that for the Tokyo area under the decentralized admissions. 
Parentheses contain p-values based on standard errors clustered at the prefecture level. Square brackets contain wild cluster bootstrap p-values based on standard errors clustered at the cohort level. 
***, **, and * mean significance at the 1\%, 5\%, and 10\% levels, respectively. 
See Section \ref{section:long_run_analysis1} for discussions about this table. \par}
\end{table} 

  \end{landscape}

\newgeometry{right=0.6in,left=0.6in,top=1.5in,bottom=1.5in}  
\begin{table}[h!] \vspace{-0.4in}
        \caption{Long-run Impacts of Centralization: Destinations of Elites}\label{tab:long_run_destination}
\begin{center}
\scalebox{0.76}{
\begin{tabular}{lcccccccc}
\toprule
\toprule
                    &\multicolumn{1}{c}{(1)}&\multicolumn{1}{c}{(2)}&\multicolumn{1}{c}{(3)}&\multicolumn{1}{c}{(4)}&\multicolumn{1}{c}{(5)}&\multicolumn{1}{c}{(6)}&\multicolumn{1}{c}{(7)}&\multicolumn{1}{c}{(8)}\\
                    &\multicolumn{1}{c}{\shortstack{Imperial\\Univ.\\grads}}&\multicolumn{1}{c}{\shortstack{Top 0.01\% \\income\\earners}}&\multicolumn{1}{c}{\shortstack{Top 0.05\% \\income\\earners}}&\multicolumn{1}{c}{\shortstack{Medal\\recipients}}&\multicolumn{1}{c}{\shortstack{Corporate\\executives}}&\multicolumn{1}{c}{\shortstack{Top\\politicians \& \\bureaucrats}}&\multicolumn{1}{c}{\shortstack{Imperial\\Univ.\\professors}}&\multicolumn{1}{c}{\shortstack{All\\occupational\\elites}}\\
\hline &  \\ & \multicolumn{8}{c}{Panel A: Baseline Specification} \\ 
Age 17 under centralization&        2.93   &        0.36   &        1.62   &        2.10   &        2.16   &        0.59   &        0.48   &        9.83   \\
                    &     (0.027)** &     (0.089)*  &     (0.304)   &     (0.037)** &     (0.203)   &     (0.001)***&     (0.131)   &     (0.157)   \\
                    &     [0.002]***   &     [0.259]   &     [0.164]   &     [0.005]***   &     [0.051]*   &     [0.055]*   &     [0.135]   &     [0.003]***   \\

\\ & \multicolumn{8}{c}{Panel B: Adding Control Variables} \\ 
Age 17 under centralization&        2.35   &        0.65   &        2.22   &        1.98   &        2.63   &        0.48   &        0.39   &        8.82   \\
                    &     (0.008)***&     (0.152)   &     (0.271)   &     (0.029)** &     (0.216)   &     (0.005)***&     (0.047)** &     (0.136)   \\
                    &     [0.019]**   &     [0.001]***   &     [0.019]**   &     [0.004]***   &     [0.011]**   &     [0.085]*   &     [0.150]   &     [0.007]***   \\
                    &               &               &               &               &               &               &               &               \\
Observations        &         656   &         656   &         656   &         656   &         656   &         656   &         656   &         656   \\
Cohort FE, Birth pref. FE&         Yes   &         Yes   &         Yes   &         Yes   &         Yes   &         Yes   &         Yes   &         Yes   \\
Mean dep var        &        7.33   &        1.13   &        4.84   &        5.35   &        6.96   &        1.58   &        0.76   &       30.77   \\
Mean dep var (Tokyo area&       10.44   &        2.08   &        9.19   &        6.92   &       12.40   &        2.19   &        0.98   &       46.08   \\
under decentralization)&               &               &               &               &               &               &               &               \\
\hline \hline
\end{tabular}

}
\end{center}
{\scriptsize \textit{Notes}:     
This table shows difference-in-differences estimates of the long-run effects of the centralized admission system on the geographical destinations of elites. 
By birth cohort (born in 1880--1894), we count  the number of specified elites who reside in each prefecture as adults in 1934 or 1939 and divide this count by 10,000 male births in each prefecture. 
Unlike the previous tables, all outcome variables are measured at the prefecture of residence. 
In Panel B, we control for time- and cohort-varying prefecture characteristics, i.e., the number of primary schools in the prefecture in the year when the cohort turned age 17, the number of middle-school graduates in the prefecture in the year when the cohort turned age 17, log GDP of the prefecture when the cohort turned age 20, and birth population of the cohort in the prefecture. 
We exclude the cohort who turned age 17 in 1901. 
Parentheses contain p-values based on standard errors clustered at the prefecture level. 
Square brackets contain wild cluster bootstrap p-values based on standard errors clustered at the cohort level.
***, **, and * mean significance at the 1\%, 5\%, and 10\% levels, respectively.
See Section \ref{section:long_run_analysis1} for discussions about this table. \par}
\end{table}

\begin{table}[h!]
        \caption{Long-run Impacts of Centralization: National Production of Top Government Officials}\label{tab:long_run_toprankgov} \vspace{-0.15in}
\begin{center}
\scalebox{0.76}{
\begin{tabular}{lcccccccc}
\hline
\hline
                    &\multicolumn{2}{c}{\shortstack{Top-ranking\\ officials}}&\multicolumn{2}{c}{\shortstack{Top-ranking\\ officials\\graduated from\\Schools 1--8}}&\multicolumn{2}{c}{\shortstack{Top-ranking\\ officials\\not graduated from\\Schools 1--8}}\\
                    \cmidrule(lr){2-3}\cmidrule(lr){4-5}\cmidrule(lr){6-7}
                    &\multicolumn{1}{c}{(1)}   &\multicolumn{1}{c}{(2)}   &\multicolumn{1}{c}{(3)}   &\multicolumn{1}{c}{(4)}   &\multicolumn{1}{c}{(5)}   &\multicolumn{1}{c}{(6)}   \\
\hline
Centralized         &        4.40   &        4.86   &        4.85   &        5.60   &       -0.44   &       -0.74   \\
                    &     (0.004)***&     (0.011)** &     (0.003)***&     (0.000)***&     (0.660)   &     (0.570)   \\
                    &               &               &               &               &               &               \\
\hline
Observations        &          33   &          33   &          33   &          33   &          33   &          33   \\
Time trend          &   Quadratic   &   6th order   &   Quadratic   &   6th order   &   Quadratic   &   6th order   \\
Control exam passers&         Yes   &         Yes   &         Yes   &         Yes   &         Yes   &         Yes   \\
Mean dep var        &       28.55   &       28.55   &       19.52   &       19.52   &        9.03   &        9.03   \\
(decentralization)  &               &               &               &               &               &               \\
\hline \hline
\end{tabular}

} \vspace{0.4cm} 
\end{center}
{\scriptsize \textit{Notes}:     
This table shows OLS estimates of the effects of the centralized admissions on the number of top-ranking higher civil officials.
The estimates are based on the cohort level data (1898--1930), where cohort is defined by the year of entering a higher school or its equivalent.
The data is compiled from the complete list of individuals who passed the administrative HCSE in 1894--1941 and their biographical information.
``Top-ranking officials'' is the number of top-ranking officials in cohort $t$ (i.e., the number of individuals who entered a higher school or its equivalent in year $t$, passed the administrative HCSE, and were internally promoted to the top three ranks of higher civil service in their lifetime).
``Centralized'' is the indicator variable that takes 1 if cohort $t$ entered a higher school or its equivalent under the centralized admissions in 1902--07, 1917--18, and 1926--27.
``Mean dep var (decentralization)'' is the mean of the dependent variable for the cohorts who entered a higher school or its equivalent under the decentralized admissions.
In all regressions, we control for either quadratic time trends or 6th order polynomial time trends. We also control for the number of exam passers (individuals in cohort $t$ who passed the administrative HCSE). 
Parentheses contain P values based on Newey-West standard errors with the maximum lag order of 3.
***, **, and * mean significance at the 1\%, 5\%, and 10\% levels, respectively.
See Section \ref{section:long_run_analysis2} for discussions about this table.
 \par} 
\end{table} 

\newgeometry{right=0.8in,left=0.8in,top=0.9in,bottom=0.9in}   
\begin{table}[h!]
\caption{Long-run Impacts of Centralization: National Production of Occupational Elites}\label{tab:careerelites_total}
\begin{center}
\scalebox{0.76}{
\begin{tabular}{lcccccc}
\toprule
\toprule
                    &\multicolumn{6}{c}{\shortstack{No. of PIR-listed individuals in each category \\ (No. of PIR-listed individuals aged 50 in each category = 100)}}\\\cmidrule(lr){2-7}
                    &\multicolumn{1}{c}{(1)}   &\multicolumn{1}{c}{(2)}   &\multicolumn{1}{c}{(3)}   &\multicolumn{1}{c}{(4)}   &\multicolumn{1}{c}{(5)}   &\multicolumn{1}{c}{(6)}   \\
 \hline \\ & \multicolumn{6}{c}{Panel A: The number of all occupational elites} \\\\[-1ex]
Age 17 under centralization&        9.92   &        4.75   &       11.63   &               &               &               \\
                    &     (0.000)***&     (0.024)** &     (0.000)***&               &               &               \\
                    &     [0.001]***   &     [0.086]*   &     [0.001]***   &               &               &               \\
Cohort's exposure to centralization&               &               &               &       11.28   &        5.36   &       13.18   \\
                    &               &               &               &     (0.000)***&     (0.020)** &     (0.000)***\\
                    &               &               &               &     [0.002]***   &     [0.092]*   &     [0.001]***   \\
R-squared           &        0.92   &        0.97   &        0.93   &        0.93   &        0.97   &        0.93   \\

 \\ & \multicolumn{6}{c}{Panel B: The number of top 0.05\% income earners} \\\\[-1ex]
Age 17 under centralization&        7.59   &        7.10   &        7.77   &               &               &               \\
                    &     (0.031)** &     (0.014)** &     (0.045)** &               &               &               \\
                    &     [0.057]*   &     [0.017]**   &     [0.058]*   &               &               &               \\
Cohort's exposure to centralization&               &               &               &        9.22   &        8.29   &        9.48   \\
                    &               &               &               &     (0.014)** &     (0.011)** &     (0.020)** \\
                    &               &               &               &     [0.047]**   &     [0.017]**   &     [0.049]**   \\
R-squared           &        0.89   &        0.95   &        0.89   &        0.90   &        0.95   &        0.90   \\

\\ & \multicolumn{6}{c}{Panel C: The number of medal recipients} \\\\[-1ex]
Age 17 under centralization&       21.29   &       13.47   &       19.24   &               &               &               \\
                    &     (0.000)***&     (0.003)***&     (0.000)***&               &               &               \\
                    &     [0.006]***   &     [0.035]**   &     [0.007]***   &               &               &               \\
Cohort's exposure to centralization&               &               &               &       23.68   &       14.65   &       21.41   \\
                    &               &               &               &     (0.000)***&     (0.002)***&     (0.000)***\\
                    &               &               &               &     [0.002]***   &     [0.047]**   &     [0.006]***   \\
R-squared           &        0.85   &        0.91   &        0.86   &        0.85   &        0.91   &        0.86   \\

\\ & \multicolumn{6}{c}{Panel D: The number of corporate executives} \\\\[-1ex]
Age 17 under centralization&        6.45   &        2.77   &        8.65   &               &               &               \\
                    &     (0.050)*  &     (0.281)   &     (0.008)***&               &               &               \\
                    &     [0.052]*   &     [0.227]   &     [0.017]**   &               &               &               \\
Cohort's exposure to centralization&               &               &               &        8.09   &        3.81   &       10.56   \\
                    &               &               &               &     (0.014)** &     (0.157)   &     (0.001)***\\
                    &               &               &               &     [0.012]**   &     [0.090]*   &     [0.003]***   \\
R-squared           &        0.91   &        0.95   &        0.93   &        0.92   &        0.96   &        0.93   \\
                    &               &               &               &               &               &               \\
Observations        &          58   &          58   &          58   &          60   &          60   &          60   \\
Age control         &   Quadratic   &     Quartic   &Edition-specific   &   Quadratic   &     Quartic   &Edition-specific   \\
                    &               &               &   quadratic   &               &               &   quadratic   \\
PIR edition FE      &         Yes   &         Yes   &         Yes   &         Yes   &         Yes   &         Yes   \\
\hline \hline
\end{tabular}

}
\end{center} \vspace{-0.1in}
{\scriptsize \textit{Notes}: 
This table shows the long-run effects of the centralized admission system on the number of all occupational elites (Panel A), the top 0.05\% income earners (Panel B), medal recipients (Panel C), and corporate managers (Panel D), using PIR data. 
To distinguish cohort effects from age effects, we use two editions of the PIR published in 1934 and 1939, count the number of specified individuals (aged 40--69) by cohort in each edition, 
and pool the resulting cohort-edition level data.
We standardize the dependent variable by setting the number of specified individuals at age 50 in each edition to be 100. 
In Columns (1)-(3), ``Age 17 under centralization" takes 1 if the cohort turned age 17 in 1902--1907, and takes 0 otherwise, and the cohort who turned age 17 in 1901 is dropped from the sample. 
In Columns (4)-(6), ``Cohort's exposure to centralization" is the continuous measure used in Table \ref{tab:long_run} Panel D. 
In addition to the edition fixed effect, we control for quadratic age trends in Columns (1) and (4), quartic age trends in Columns (2) and (5), and edition-specific quadratic age trends in Columns (3) and (6).  
Parentheses contain p-values based on robust standard errors.
Square brackets contain wild cluster bootstrap p-values based on standard errors clustered at the cohort level.
***, **, and * mean significance at the 1\%, 5\%, and 10\% levels, respectively.
See Section \ref{section:long_run_analysis2} for discussions about this table.
 \par} 
\end{table}
\restoregeometry

\clearpage 
\appendix

\renewcommand\thefigure{\thesection.\arabic{figure}}
\setcounter{figure}{0}  
\renewcommand\thetable{\thesection.\arabic{table}}
\setcounter{table}{0}  


\pagenumbering{arabic}
\renewcommand*{\thepage}{A-\arabic{page}}

\onehalfspacing

\section{Online Appendix}

\subsection{The Evolution of the Admission System}
\label{appendix:admission_system}

Table \ref{tab:chronological_table2} shows changes in the admission system of Schools 1--8 (\textit{Kanritsu Koutou Gakkou} in Japanese) from 1900 to 1930. 
Despite the population increase and the growing demand for higher education, the number of National Higher Schools increased only slightly from 6 to 8 in 1900--1918 due to tight fiscal constraints.
With the economic boom of WWI, the government expanded the higher education system and increased the number of Schools from 8 in 1918 to 25 in 1925.\footnote{In the late 1920s, in addition to 25 national higher schools, 3 local public higher schools, 4 private higher schools, and one colonial higher school were established.}
Such expansions notwithstanding, Schools 1--8 remained the most distinguished among all higher schools throughout the pre-WWII period.

\subsection{Relation to Historical Literature} 
\label{appendix:historical_literature}
  
The repeated reforms of the admission system of Schools 1--8 have been examined by historians of Japanese education, most notably by \citeappendix{yoshino, yoshino2}, \citeappendix{takeuchi}, and \citeappendix{amano2, amano, amano3}. 
The preceding studies are mostly descriptive and qualitative in nature, providing institutional and historical details of the reforms. Among them, \citeappendix{yoshino, yoshino2} presents the most comprehensive historical accounts and basic statistics (such as the number of applicants and enrollment) combining a variety of historical documents. In this study, we reproduced and improved his data, using the same documents and additional sources as described in Supplementary Materials Section B.1.
The most closely related research is the study by \citeappendix{sono2, sono1}. She investigates regional variations in access to higher schools by comparing the number of students per population across prefectures across years. While her research remains descriptive, we provide quasi-experimental research designs to understand not only the short-run but also the long-run impacts of the admission reforms.

Our study is also related to historical studies of elites in pre-WWII Japan. \citeappendix{aso} uses the \textit{Personnel Inquiry Records} (PIR) to examine occupational, educational, and regional compositions of elites and their evolution over time. His research is descriptive and based on a small sample of PIR-listed individuals, while we use a complete sample to examine the impact of school admission reforms on elite formation. As more closely related research, \citeappendix{ichimura} use the complete sample of the PIR to examine the long-run impact of middle school expansion on elite formation.

 \newgeometry{right=0.65in,left=0.65in,top=1.2in,bottom=1in} 
\subsection{Additional Tables and Figures}

\begin{figure}[h!]
        \caption{College Admissions around the World Today}\label{tab:world}
        \includegraphics[width=1\textwidth]{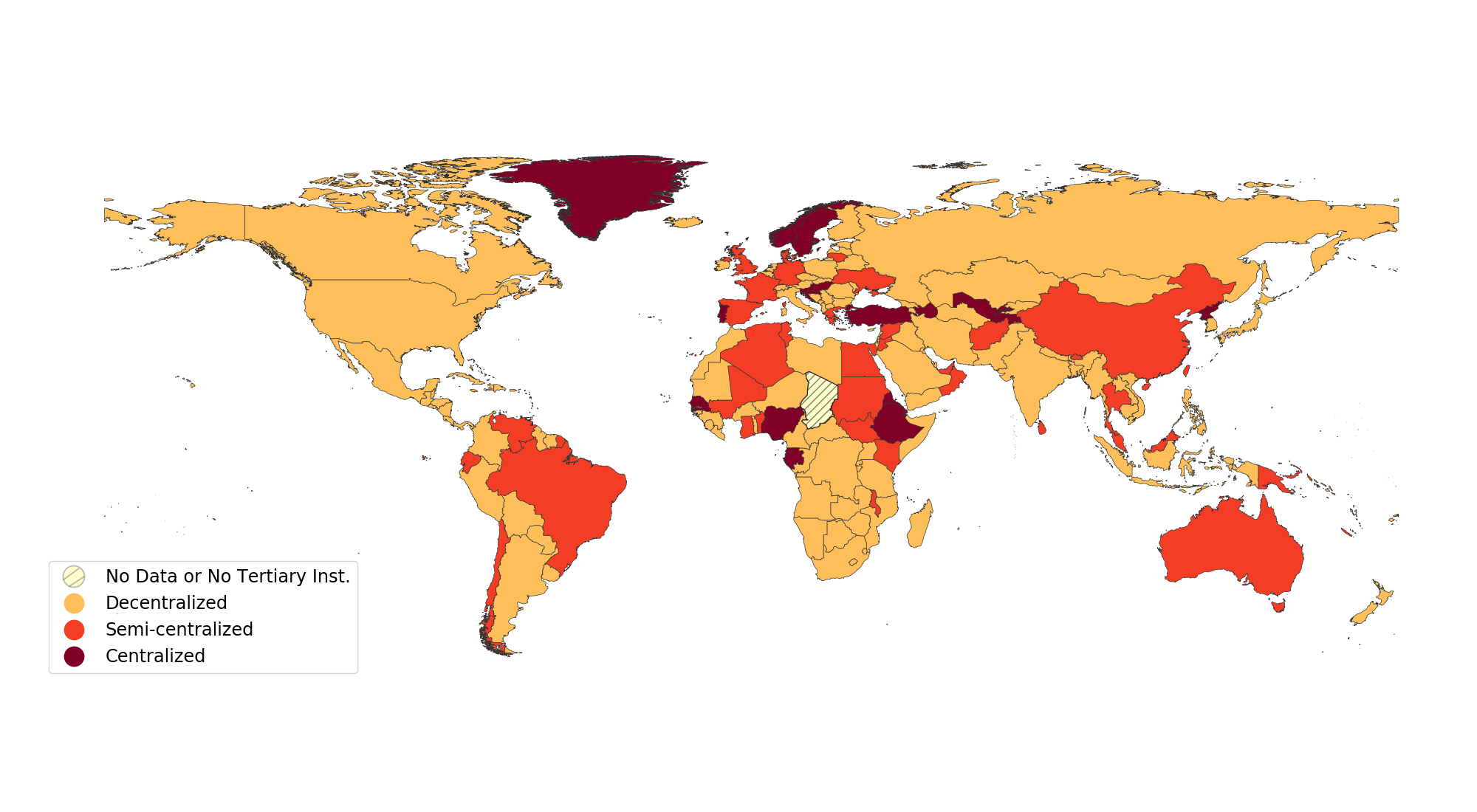}  
        {\footnotesize \textit{Notes}: 
This figure summarizes each country and territory's college admission system today. Dark red color (e.g. Norway): Regionally- or nationally-centralized college admissions where a single-application, single-offer assignment algorithm (well-defined rule) is used to make admissions to both public and private universities. Medium orange color (e.g. Brazil): Semi-centralized college admissions defined as either (1) there is a centralized system, but not all universities (such as private universities) are included in the single-application, single-offer system or (2) students submit a single application and receive multiple offers. Light orange color (e.g. U.S.): Decentralized college admissions where each college defines its own admission standards and rules. Yellow with diagonal lines (e.g. Chad): Not enough information available or if the country or territory does not have tertiary institutions. We summarize the information sources at \url{https://www.scribd.com/document/437545135/Online-Appendix191018}. See Section \ref{section:background} for discussions about this figure.  \par}
\end{figure} 
\restoregeometry
\clearpage
\newpage
 
\begin{figure}[h!]
        \caption{Map of Schools 1--8 and Definition of the Tokyo Area}\label{fig:map}
                \begin{center}
         \includegraphics[width=0.8\textwidth]{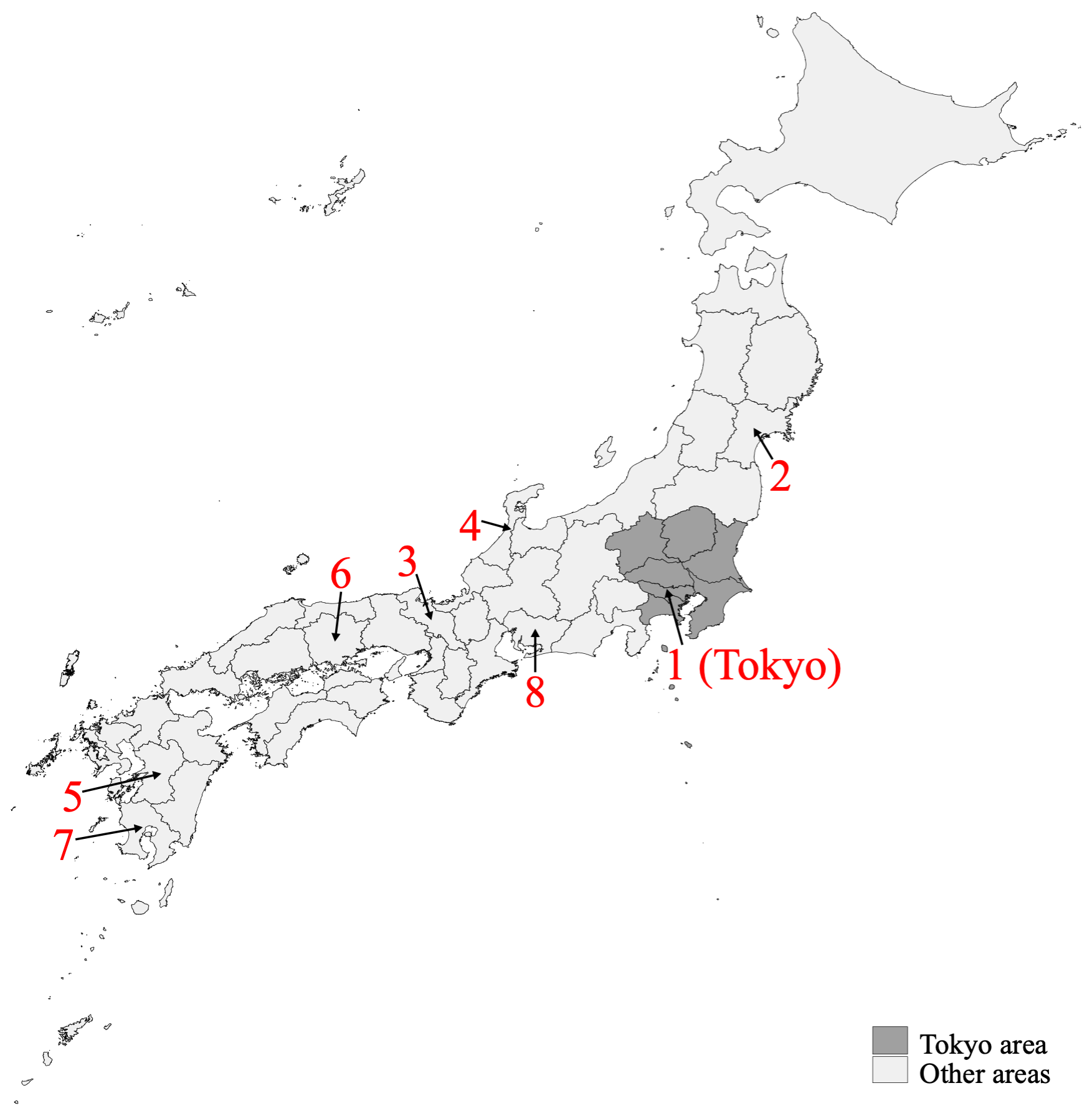}
        \end{center}
        {\footnotesize \textit{Notes}: 
This figure shows the locations of Schools 1--8 and the location of the Tokyo area (in dark gray color) defined as a set of 7 prefectures that are within 100 km from Tokyo (i.e., Tokyo, Chiba, Kanagawa, Saitama, Ibaraki, Tochigi, and Gunma prefectures). 
See Sections \ref{section:background} and \ref{section:equality} for discussions about this figure. \par}
\end{figure}

 \newgeometry{right=0.65in,left=0.65in,top=1.2in,bottom=1in}   
\begin{table}[h!]
        \caption{The Evolution of the Admission System}\label{tab:chronological_table2}
                \begin{center}
         \includegraphics[width=1\textwidth]{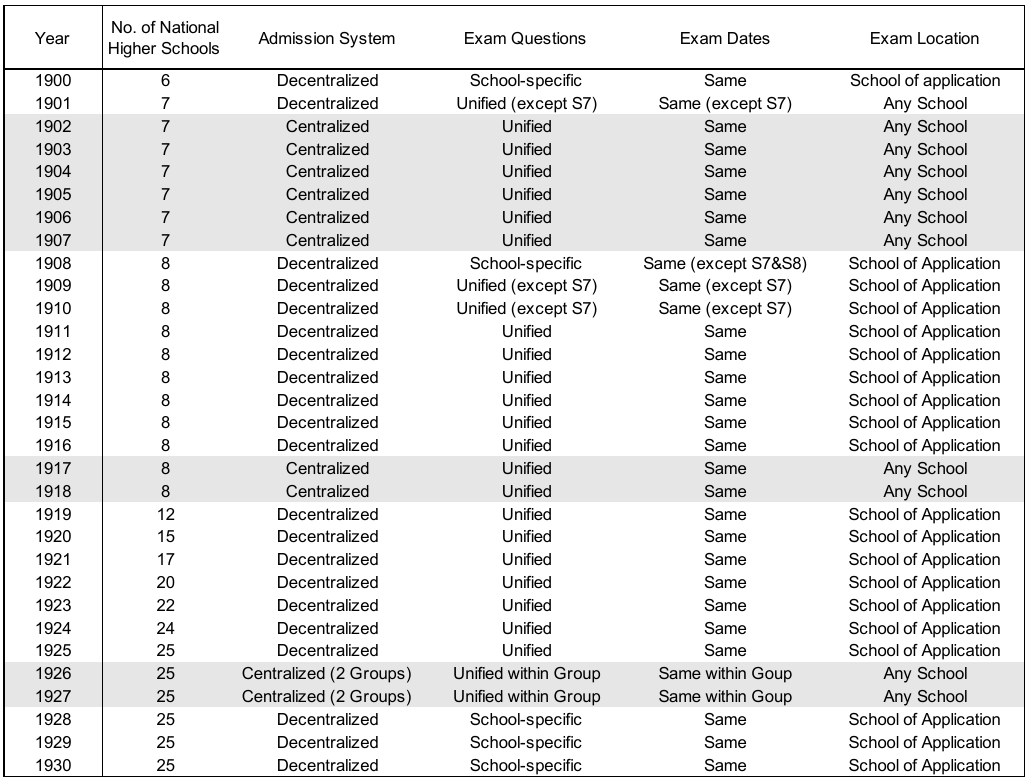}
        \end{center}
        {\footnotesize \textit{Notes}: 
This table shows changes in the admission system of the National Higher Schools (including Schools 1--8) from 1900 to 1930. 
In 1901, all schools held their exams on the same date, with the exception of newly established School 7. 
In 1908, all schools held their exams on the same date, with the exception of Schools 7 and 8. 
See \citeappendix{moriguchi2021higher} for historical details and the sources of information. 
See Section \ref{section:admission_reforms} and Appendix Section \ref{appendix:admission_system} for discussions about this table. \par}
\end{table}
\restoregeometry

\begin{figure}[h!]
        \caption{Centralized Assignment Rule}\label{fig:algo}
 \begin{center}
 \includegraphics[scale=0.5]{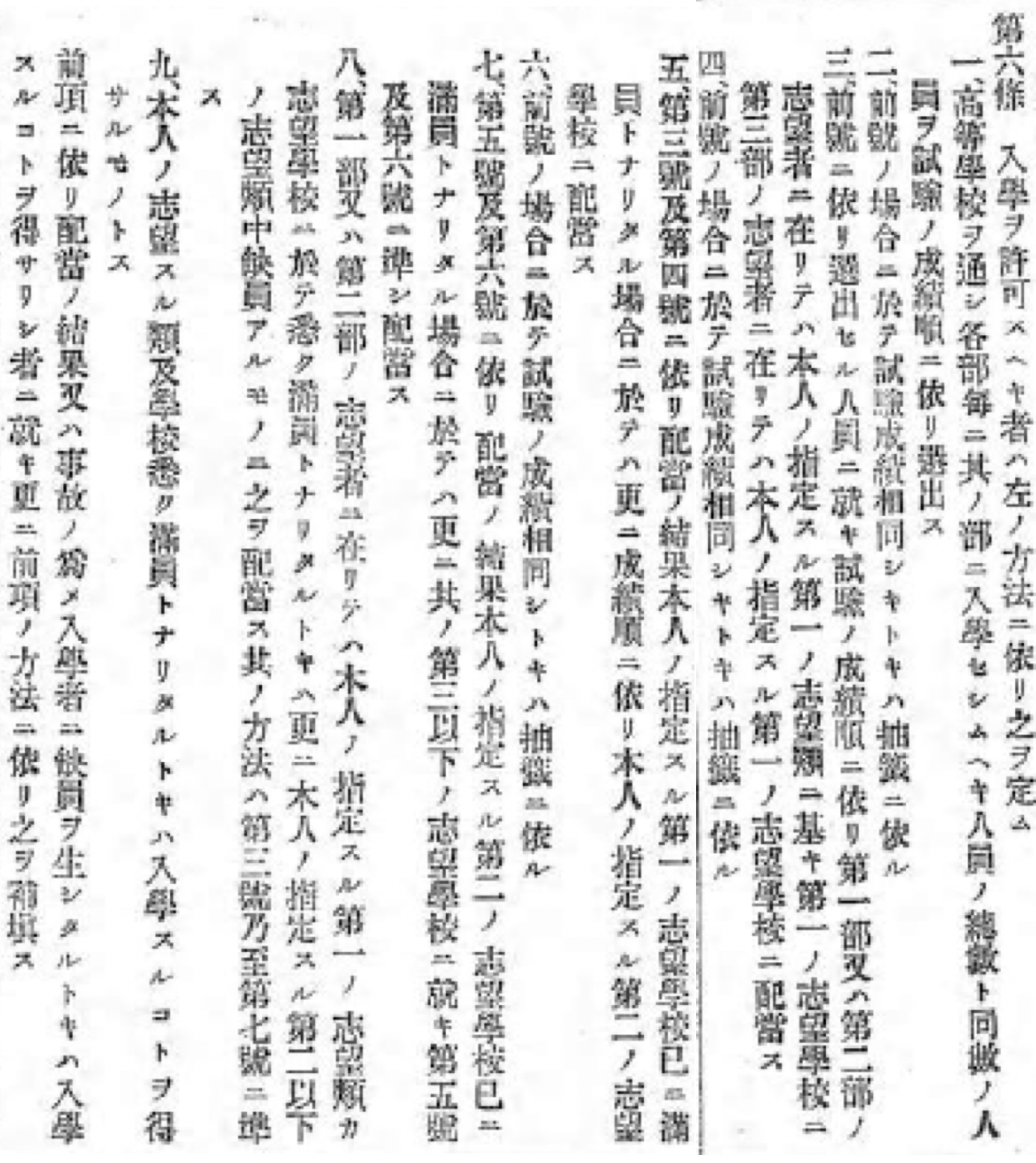}\\
 \end{center}
           {\footnotesize \textit{Notes}: 
This figure is a reprint of the assignment algorithm of the centralized admission system stated in the Ordinance of the Ministry of Education No.4 published in \textit{Government Gazette} No.1419, pp.580-581, on April 27, 1917. 
See Sections \ref{section:background} and \ref{section:mobility} for an English translation and discussions. \par}
\end{figure}

 \newgeometry{right=0.65in,left=0.65in,top=0.8in,bottom=1in}   
\begin{table}[h!]
        \caption{Summary Statistics}\label{tab:statistics} 
        \begin{center}
\scalebox{0.76}{
\begin{tabular}{l c c c c c c}\hline\hline
\multicolumn{1}{c}{\textbf{Variable}} & \textbf{Mean}
 & \textbf{Std. Dev.}& \textbf{Median} & \textbf{N}\\ \hline
 \multicolumn{5}{l}{\textbf{Year level data on short-run outcomes, 1900--1930}} \\ 
No. of applicants to Schools 1--8  & 10777 & 4122 & 10187 & 27\\
Share of applicants choosing School 1 as their first choice  & 0.29 & 0.11 & 0.24 & 27\\
No. of entrants to Schools 1--8  & 2007 & 333 & 2147 & 31\\ \hline 
\multicolumn{5}{l}{\textbf{Applicant level data on short-run outcomes, 1916--1917}} \\ 
Distance between middle-school prefecture and the first-choice school (km) & 225 & 272 & 117  & 20913\\
Applying to School 1 as first choice & 0.33 & 0.47 & 0  & 20913\\ \hline 
\multicolumn{5}{l}{\textbf{Entrant level data on short-run outcomes, 1900--1930}} \\ 
Distance between birth prefecture and the school entered (km)  & 224 & 255 & 139 & 65251\\
Entering the nearest school from birth prefecture & 0.49 & 0.5 & 0 & 66193\\
Born in Tokyo prefecture & 0.09 & 0.29 & 0 & 66193\\
Born in the Tokyo area (7 prefectures within 100 km from Tokyo) & 0.17 & 0.38 & 0 & 66193\\ \hline 
 \multicolumn{5}{l}{\textbf{Prefecture-year level data on short-run outcomes, 1900--1930}} \\ 
No. of entrants to Schools 1-8  & 45.06 & 37.45 & 34 & 1469\\
No. of entrants to School 1 & 7.88 & 14.11 & 5& 1469\\
No. of entrants to School 2 & 5.50 & 10.40 & 2 & 1469\\
No. of entrants to School 3 & 6.19 & 10.34 & 3 & 1421\\
No. of entrants to School 4 & 5.64 & 9.91 & 3 & 1469\\
No. of entrants to School 5 & 6.27 & 14.20 & 1 & 1422\\
No. of entrants to School 6 & 5.19 & 11.80 & 2 & 1421\\
No. of entrants to School 7 & 5.03 & 12.99 & 2 & 1328\\
No. of entrants to School 8 & 5.67 & 12.45 &2  & 1093\\
No. of middle-school graduates &  533 & 618 & 359 & 1457 \\
No. of national higher schools other than Schools 1--8 & 0.14 & 0.43 & 0& 1469\\ \hline 
 \multicolumn{5}{l}{\textbf{Prefecture-cohort level data on long-run outcomes, individuals listed in PIR (1934, 1939) and born in 1880--1894}} \\ 
No. of all Imperial University graduates (in 10,000 male births, the same hereafter) & 8.02 & 5.32 & 7.13  & 705\\
No. of individuals in the top 0.01\% income group & 1.27 & 1.84 & 0.84 & 705\\
No. of individuals in the top 0.05\% income group & 5.33 & 6.83 & 3.50 & 705\\
No. of civilians receiving medal of the Order of Fifth Class and above & 5.92 & 3.84 & 5.19 & 705\\
No. of corporate executives with a positive amount of tax payment & 7.67 & 6.79 & 6.23 & 705\\
No. of top politicians and high-ranking bureaucrats & 1.77 & 1.59 & 1.48 & 705\\
No. of Imperial University professors & 0.79 & 1.04 & 0.58 & 705\\
No. of all occupational elites & 33.15 & 22.42 & 28.87 & 705\\
 \hline 
 \multicolumn{5}{l}{\textbf{Cohort level data on long-run outcomes, government officials entering higher school or equivalent in 1898--1930}} \\ 
No. of top-ranking officials (internally promoted to top three ranks) &  29.76 & 8.40 &  29.48 & 33 \\
No. of top-ranking official who are Schools 1--8  graduates &  21.30 & 7.76 &  21 & 33 \\
No. of top-ranking officials who are not Schools 1--8 graduates & 8.46 & 5.69 & 6.77 & 33\\
 \hline \hline 
\end{tabular}
}
\end{center}
{\footnotesize 
\par}
\end{table} 
\restoregeometry

 \newgeometry{right=0.65in,left=0.65in,top=0.8in,bottom=1in}   
\begin{table}[h!]
        \caption{Centralization Caused Applicants Across the Country to Apply More Aggressively}\label{tab:application}
\begin{center}
\scalebox{0.76}{
\begin{tabular}{l c c c c c c c c c c c}\hline\hline
 & \multicolumn{9}{c}{A. Selecting School 1 as First Choice}  \\ \hline
 &  &  &  &  &  &  &  &  &  \\
Centralized & 0.159 & 0.192 & 0.151 & 0.146 & 0.128 & 0.168 & 0.180 & 0.166 & 0.114 \\
 & (0.000)*** & (0.000)*** & (0.003)*** & (0.001)*** & (0.142) & (0.001)*** & (0.001)*** & (0.007)*** & (0.001)*** \\
Constant & 0.248 & 0.494 & 0.169 & 0.0892 & 0.178 & 0.107 & 0.184 & 0.0813 & 0.127 \\
 & (0.001)*** & (0.000)*** & (0.002)*** & (0.003)*** & (0.018)** & (0.002)*** & (0.000)*** & (0.015)** & (0.088)* \\
&  &  &  &  &  &  \\
Sample region & All & S1 Region & S2 Region & S3 Region & S4 Region & S5 Region & S6 Region & S7 Region & S8 Region \\
Observations & 20,913 & 6,505 & 2,555 & 3,248 & 1,266 & 2,730 & 2,276 & 615 & 1,718 \\
\end{tabular}
}

\scalebox{0.76}{
\begin{tabular}{l c c c c c c c c c c c}\hline\hline
 & \multicolumn{9}{c}{B. First Choice Application Distance}  \\ \hline 
 &  &  &    \\
Centralized & -2.534 & -92.88 & 10.95 & 2.080 & -15.74 & 128.0 & 46.52 & 145.4 & -25.57 \\
 & (0.914) & (0.000)*** & (0.670) & (0.720) & (0.541) & (0.003)*** & (0.012)** & (0.021)** & (0.264) \\
Constant & 226.2 & 231.7 & 289.7 & 158.8 & 166.7 & 252.6 & 294.1 & 218.0 & 154.2 \\
 & (0.000)*** & (0.000)*** & (0.008)*** & (0.002)*** & (0.061)* & (0.002)*** & (0.001)*** & (0.092)* & (0.051)* \\ 
&  &  &  &  &  &  \\
Sample region & All & S1 Region & S2 Region & S3 Region & S4 Region & S5 Region & S6 Region & S7 Region & S8 Region \\  
Observations & 20,913 & 6,505 & 2,555 & 3,248 & 1,266 & 2,730 & 2,276 & 615 & 1,718 \\  \hline \hline
\end{tabular}
}
\end{center}
{\footnotesize \textit{Notes}: 
In Panel A, we estimate the effects of the centralized admissions on the propensity of an applicant to select the most prestigious and selective school (School 1 in Tokyo) as the first choice, using the applicant-level data in 1916 (under the decentralized system) and 1917 (under the centralized system). 
The prefecture-level application data is available only for these two years. We estimate the following regression: $Y_{it} = \alpha +\beta \times Centralized_{t} + \epsilon_{it},$ where $Y_{it}$ is the indicator that applicant $i$ in year $t$ selects School 1 as the first choice. $Centralized_t$ is the indicator that year $t$ is under the centralized system. 
        To observe regional variation, we estimate the equation separately by region of the applicant's middle school. More specifically, we group applicants into ``school regions" based on which school (among Schools 1--8) is closest to the applicant's middle school in 1916. 
        The following map shows the locations of the eight school regions. 
        In Panel B, we estimate the effects of centralization on the application distance defined by the distance between an applicant's first-choice school and middle school. 
Parentheses contain p-values based on standard errors clustered at the prefecture level. 
        ***, **, and * mean significance at the 1\%, 5\%, and 10\% levels, respectively.
See Section \ref{section:application} for discussions about this table. }
                \begin{center}
 \includegraphics[width=80mm]{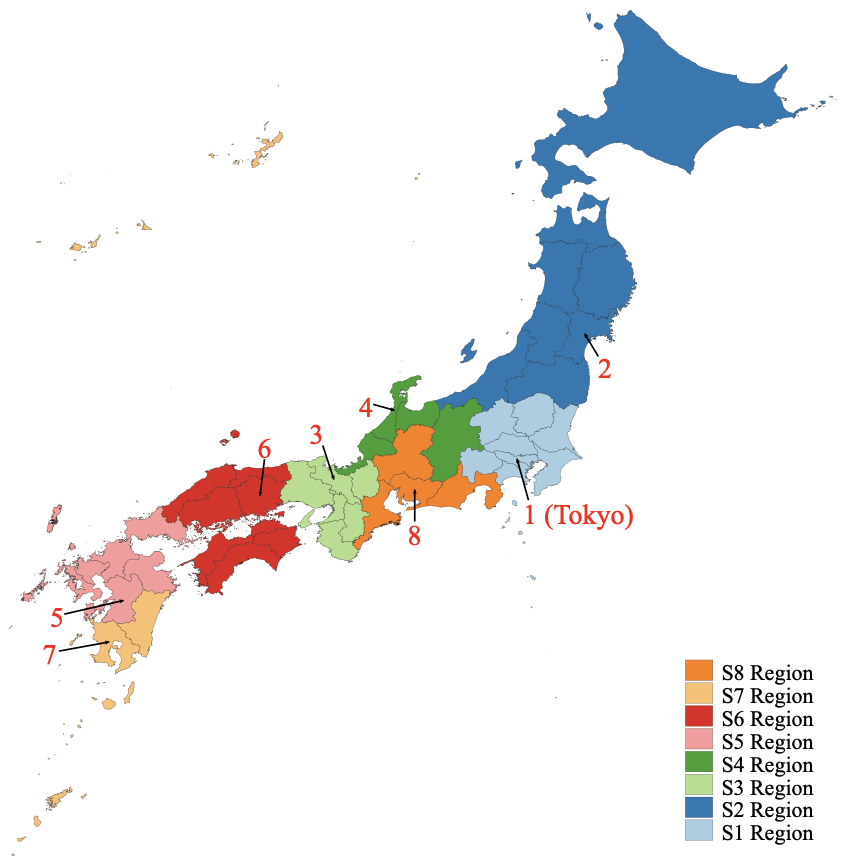}
 \end{center} 
\end{table} 
\restoregeometry

\begin{table}[h!]
    \caption{Characteristics of the Tokyo Area}    \label{tab:tokyo_area}
    \begin{center}
    \scalebox{0.76}{
        \begin{tabular}{lccccc} \hline \hline
      &  Mean & Std. Dev. & Min & Max  & N \\ \hline
     \underline{Tokyo Area} \\
\ \ \ Population in prefecture (in million) & 1.497 & 0.866 & 0.820 & 5.437 & 217\\
\ \ \ GDP per capita in prefecture (in 1,000 yen) & 0.209 & 0.095 & 0.108 & 0.469 & 217\\
\ \ \ Middle-school graduates in prefecture & 0.871 & 1.172 & 0.011 & 6.427 & 217\\
\ \ \ Middle-school graduates in nearby prefectures (1-100 km) & 4.604 & 2.150 & 0.395 & 12.212 & 217\\
 \hline
     \underline{Other Areas} \\ 
\ \ \ Population in prefecture (in million) & 1.072 & 0.522 & 0.414 & 3.568 & 1240\\
\ \ \ GDP per capita in prefecture (in 1,000 yen) & 0.173 & 0.058 & 0.097 & 0.494 & 1240\\
\ \ \ Middle-school graduates in prefecture & 0.474 & 0.431 & 0.029 & 3.307 & 1240\\
\ \ \ Middle-school graduates in nearby prefectures (1-100 km) & 1.299 & 1.594 & 0.000 & 11.729 & 1240\\
     \hline \hline    
    \end{tabular} 
    }
    \end{center}
    {\footnotesize \textit{Notes}: This table shows the characteristics of the Tokyo area (defined as 7 prefectures within 100 km of Tokyo as in the map in Appendix Figure \ref{fig:map}) compared to other areas. All numbers are the prefecture-level average in 1900--1930. GDP per capita is in real terms expressed in 1934--1936 prices. \par}
\end{table}

\begin{table}[hbp]\vspace{-0.7in}
\caption{Replacing the Tokyo Area Indicator by Urban Characteristics}\label{tab:whytokyoarea}\vspace{-0.2in}
\begin{center}
\scalebox{0.76}{
\begin{tabular}{lccccc} \hline \hline
 & (1) & (2) & (3)  \\
 & \multicolumn{3}{c}{Entrants to Schools 1-8}  \\ \hline
 &  &  &      \\
 &  &  &    \\
Centralized $\times$& 2.63 &  &    \\
 Population in prefecture  & (0.440) &  &   \\
  & [0.183] &  &   \\
Centralized $\times$  &  & 3.16 &    \\
GDP per capita in prefecture &  & (0.274) &  \\
  &  & [0.224] &    \\
Centralized $\times$  &  &  & 3.73  \\
Middle-school graduates in prefecture &  &  & (0.003)***  \\
  &  &  & [0.171]  \\
Centralized $\times$  &  &  & 4.32  \\
 Middle-school graduates in nearby prefectures &  &  & (0.001)***  \\
  &  &  & [0.019]**  \\
Population in prefecture & 11.74 &  &   \\
 & (0.061)* &  &    \\
  & [0.000]*** &  &    \\
GDP per capita in prefecture &  & 17.30 &    \\
 &  & (0.099)* &    \\
  &  & [0.001]*** &    \\
Middle-school graduates in prefecture & 10.70 & 13.64 & 15.65  \\
 & (0.004)*** & (0.001)*** & (0.000)*** \\
  & [0.000]*** & [0.000]*** & [0.000]***  \\
Middle-school graduates in nearby prefectures &  &  & -0.73   \\
 &  &  & (0.765)   \\
  &  &  & [0.322]   \\
 &  &  &   \\
Observations & 1,457 & 1,457 & 1,457  \\
Year FE, Prefecture FE & Yes & Yes & Yes \\
 Mean dep var & 45.43 & 45.43 & 45.43  \\ \hline \hline 
\end{tabular}
 
}
\end{center}
{\footnotesize \textit{Notes}:  
This table uses the prefecture-year level data in 1900--1930.  
The dependent variable is the number of students from birth prefecture $p$ who entered one of Schools 1--8 in year $t$.  
``Population in prefecture'' is population in prefecture $p$ in year $t$.
``GDP per capita in prefecture'' is real gross value-added per capita in prefecture $p$ in year $t$. 
``Middle-school graduates in prefecture'' is the number of students who graduated from middle schools in prefecture $p$ in year $t$.
``Middle-school graduates in nearby prefectures'' is the number of students who graduated from middle schools in the prefectures within 100 km from prefecture $p$ (excluding prefecture $p$) in year $t$. 
All variables interacted with ``Centralized'' are standardized to be mean 0 and standard deviation 1. 
We control for year fixed effects, prefecture fixed effects, and the number of higher schools other than Schools 1--8 in prefecture $p$ in year $t$. We also control for ``Born in school's prefecture," ``Born near school's prefecture (1--100 km)," and ``Born near school's prefecture (100-300 km)" as in Table \ref{tab:mobility1}. 
Parentheses contain p-values based on standard errors clustered at the prefecture level. Square brackets contain p-values based on standard errors clustered at the year level.
***, **, and * mean significance at the 1\%, 5\%, and 10\% levels, respectively.
See Section \ref{section:equality} for discussions about this table. \par} 
\end{table} 
 
\begin{landscape}
\begin{table}[h!]
        \caption{Testing Exogeneity of Centralization}\label{tab:other}
\begin{center}
\subfloat[Placebo Outcomes]{
\scalebox{0.76}{
\begin{tabular}{lccccccc}
\toprule
\toprule
                    &\multicolumn{1}{c}{(1)}&\multicolumn{1}{c}{(2)}&\multicolumn{1}{c}{(3)}&\multicolumn{1}{c}{(4)}&\multicolumn{1}{c}{(5)}&\multicolumn{1}{c}{(6)}&\multicolumn{1}{c}{(7)}\\
                    &\multicolumn{1}{c}{\shortstack{No. middle\\school\\graduates}}&\multicolumn{1}{c}{\shortstack{No. entrants\\to Schools 1--8}}&\multicolumn{1}{c}{\shortstack{No.  entrants\\to School 1}}&\multicolumn{1}{c}{\shortstack{No. applicants\\to Schools 1--8}}&\multicolumn{1}{c}{\shortstack{Ratio of\\entrants to\\applicants}}&\multicolumn{1}{c}{\shortstack{Mean\\ age of\\ entrants}}&\multicolumn{1}{c}{\shortstack{Government\\expenditure\\for\\ Higher Schools}}\\
\hline
Centralized         &        2.06   &       -0.11   &       -0.01   &       -0.34   &        0.01   &       -0.03   &        0.85   \\
                    &     (0.182)   &     (0.103)   &     (0.132)   &     (0.620)   &     (0.577)   &     (0.691)   &     (0.674)   \\
\hline
Observations        &          31   &          31   &          31   &          27   &          27   &          26   &          31   \\
Mean dep. var       &       26.42   &        2.10   &        0.36   &       10.78   &        0.28   &       19.03   &       15.17   \\
\hline \hline
\end{tabular}

}} \vspace{0.4cm} \\ 
\subfloat[Main Outcomes]{
\scalebox{0.76}{
\begin{tabular}{lccccc}
\toprule
\toprule
                    &\multicolumn{1}{c}{(1)}&\multicolumn{1}{c}{(2)}&\multicolumn{1}{c}{(3)}&\multicolumn{1}{c}{(4)}&\multicolumn{1}{c}{(5)}\\
                    &\multicolumn{1}{c}{\shortstack{No. applicants\\ to School 1}}&\multicolumn{1}{c}{\shortstack{Share of\\ applicants\\ to School 1}}&\multicolumn{1}{c}{\shortstack{Enrollment\\ distance}}&\multicolumn{1}{c}{\shortstack{No. entrants\\ born in\\ Tokyo area}}&\multicolumn{1}{c}{\shortstack{Share of\\ entrants\\ born in\\ Tokyo area}}\\
\hline
Centralized         &        1.16   &        0.17   &       59.36   &        0.05   &        0.04   \\
                    &     (0.000)***&     (0.001)***&     (0.000)***&     (0.074)*  &     (0.001)***\\
\hline
Observations        &          27   &          27   &          31   &          31   &          31   \\
Mean dep. var       &        2.73   &        0.25   &      205.27   &        0.36   &        0.16   \\
\hline \hline
\end{tabular}

}} 
\end{center}
{\footnotesize \textit{Notes}: Panel (a) tests if important institutional variables are correlated with the timing of centralization using year-level data. 
Panel (b) does the same for our main short-run outcomes using year-level data. 
All numbers are at the national-level from 1900 to 1930. 
The numbers of middle school graduates, entrants, and applicants are denominated by 1,000. 
In all regressions, quadratic time trends (i.e. trend and trend squared, where the trend is defined by ``year - 1899'') and the number of middle school graduates (except for Column 1) are controlled. 
Parentheses contain p-values based on Newey-West standard errors with the maximum lag order of 3. 
See Section \ref{section:robustness} for discussions about this table. 
\par}
\end{table} 
\end{landscape}

\begin{table}[h!]
        \caption{Occupational Distributions of PIR-listed Individuals} \label{tab:occupation_PIR}
\begin{center}
\vspace{-0.25in}
\subfloat[1934 PIR]{
\scalebox{0.76}{
\begin{tabular}{lcccc} \hline\hline
                &\multicolumn{1}{c}{\shortstack{(1)\\All individuals\\listed}}&\multicolumn{1}{c}{\shortstack{(2)\\Sampled\\cohorts}}&\multicolumn{1}{c}{\shortstack{(3)\\Top 0.05\% \\income earners}}&\multicolumn{1}{c}{\shortstack{(4)\\Medal\\recipients}}\\
                &\multicolumn{1}{c}{}     &\multicolumn{1}{c}{}     &\multicolumn{1}{c}{}     &\multicolumn{1}{c}{}     \\
Corporate executives and managers&                    0.594&                    0.596&                    0.723&                    0.301\\
Politicians and bureaucrats&                    0.095&                    0.125&                    0.040&                    0.314\\
Scholars, lawyers, and artists&                    0.134&                    0.161&                    0.046&                    0.439\\
Engineers and physicians&                    0.079&                    0.111&                    0.032&                    0.274\\
Military personnels&                    0.052&                    0.051&                    0.015&                    0.000\\
Landlords       &                    0.087&                    0.063&                    0.124&                    0.001\\
Imperial and peerage family members&                    0.015&                    0.012&                    0.008&                    0.027\\
None of the above&                    0.130&                    0.108&                    0.123&                    0.077\\
\\
No. of observations&                26,177&                12,041&                 2,781&                 3,171\\

\hline\hline
\end{tabular}

}} \\ \vspace{0.15in}
\subfloat[1939 PIR]{
\scalebox{0.76}{
\begin{tabular}{lcccc} \hline\hline
                &\multicolumn{1}{c}{\shortstack{(1)\\All individuals\\listed}}&\multicolumn{1}{c}{\shortstack{(2)\\Sampled\\cohorts}}&\multicolumn{1}{c}{\shortstack{(3)\\Top 0.05\% \\income earners}}&\multicolumn{1}{c}{\shortstack{(4)\\Medal\\recipients}}\\
                &\multicolumn{1}{c}{}     &\multicolumn{1}{c}{}     &\multicolumn{1}{c}{}     &\multicolumn{1}{c}{}     \\
Corporate executives and managers&                    0.660&                    0.672&                    0.811&                    0.320\\
Politicians and bureaucrats&                    0.075&                    0.070&                    0.030&                    0.213\\
Scholars, lawyers, and artists&                    0.119&                    0.142&                    0.047&                    0.473\\
Engineers and physicians&                    0.090&                    0.110&                    0.046&                    0.259\\
Military personnels&                    0.045&                    0.047&                    0.020&                    0.000\\
Landlords       &                    0.048&                    0.033&                    0.060&                    0.001\\
Imperial and peerage family members&                    0.006&                    0.004&                    0.004&                    0.016\\
None of the above&                    0.130&                    0.109&                    0.094&                    0.087\\
\\
No. of observations&                55,742&                28,423&                 3,550&                 4,740\\

\hline\hline
\end{tabular}

}} 

\end{center}
{\footnotesize \textit{Notes}: 
Panels (a) and (b) show the occupational distributions of individuals listed in the 1934 and 1939 editions of the PIR, respectively. Column (1) shows the share of each occupational category in all PIR-listed individuals. Column (2) restricts the sample to cohorts who were born in 1880--1894. Columns (3) and (4) further restrict the sample to the top 0.05\% income earners and civilian medal recipients (as defined in Section \ref{section:long_run_analysis1}), respectively. Occupational categories are not mutually exclusive, except for ``None of the above'' which is defined as individuals not included in any of the specified categories. See Section \ref{section:long_run_analysis1} for discussions about this table. 
\par}
\end{table}



 \newpage 
 \begin{figure}[h!]
\caption{Sampling Rates of High Income Earners}\label{fig:sampling_rate1} \vspace{-0.2in}
 \begin{center}
          \begin{subfigure}[b]{0.45\textwidth}
                \caption{PIR (1934)}
             \includegraphics[width=1\textwidth]{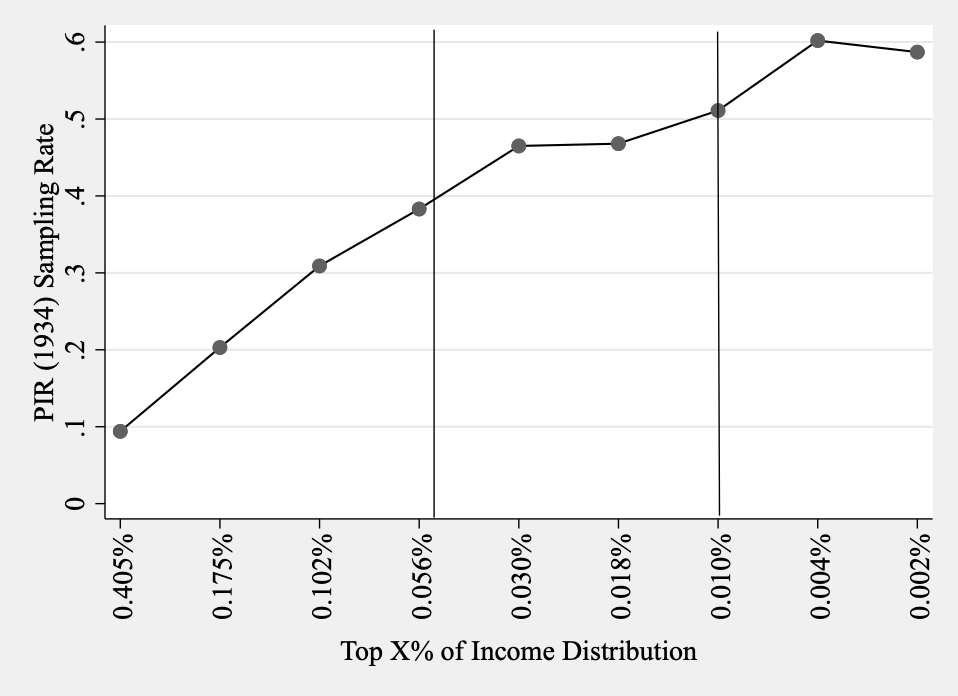}
        \end{subfigure}
                ~ 
        \begin{subfigure}[b]{0.45\textwidth}
                \caption{PIR (1939)}
               \includegraphics[width=1\textwidth]{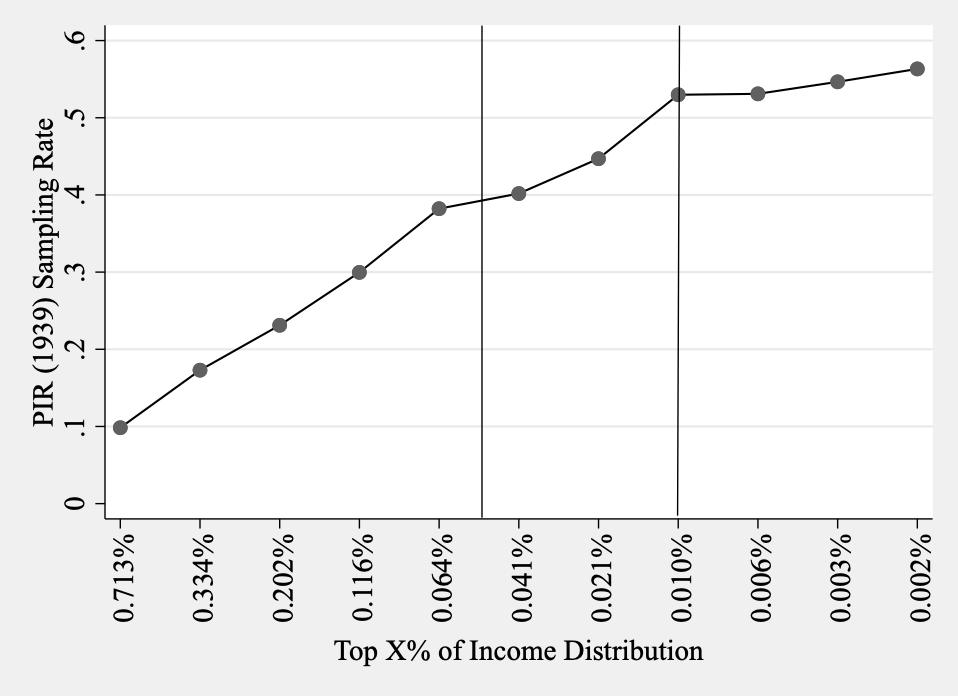} 
        \end{subfigure}
 \end{center}
           {\footnotesize \textit{Notes}: 
This figure plots the sampling rate of the high income earners listed in the 1934 and 1939 editions of the PIR by the income level expressed as a top percentile of the national income distribution.
The sampling rates and the top income percentiles are computed from the complete count data in the National Tax Bureau Yearbook. The vertical lines indicate the top 0.05\% and 0.01\% thresholds used in our analysis. 
See Section \ref{section:long_run_analysis1} for discussions about this figure.\par}
\end{figure}    

 \begin{figure}[h!]
        \caption{Sampling Rates of Top Income Earners across Prefectures}\label{fig:sampling_rate2}
 \begin{center}
        \begin{subfigure}[b]{0.45\textwidth}
                \caption{Top 0.01\% income (PIR 1934)}
                \includegraphics[width=\textwidth]{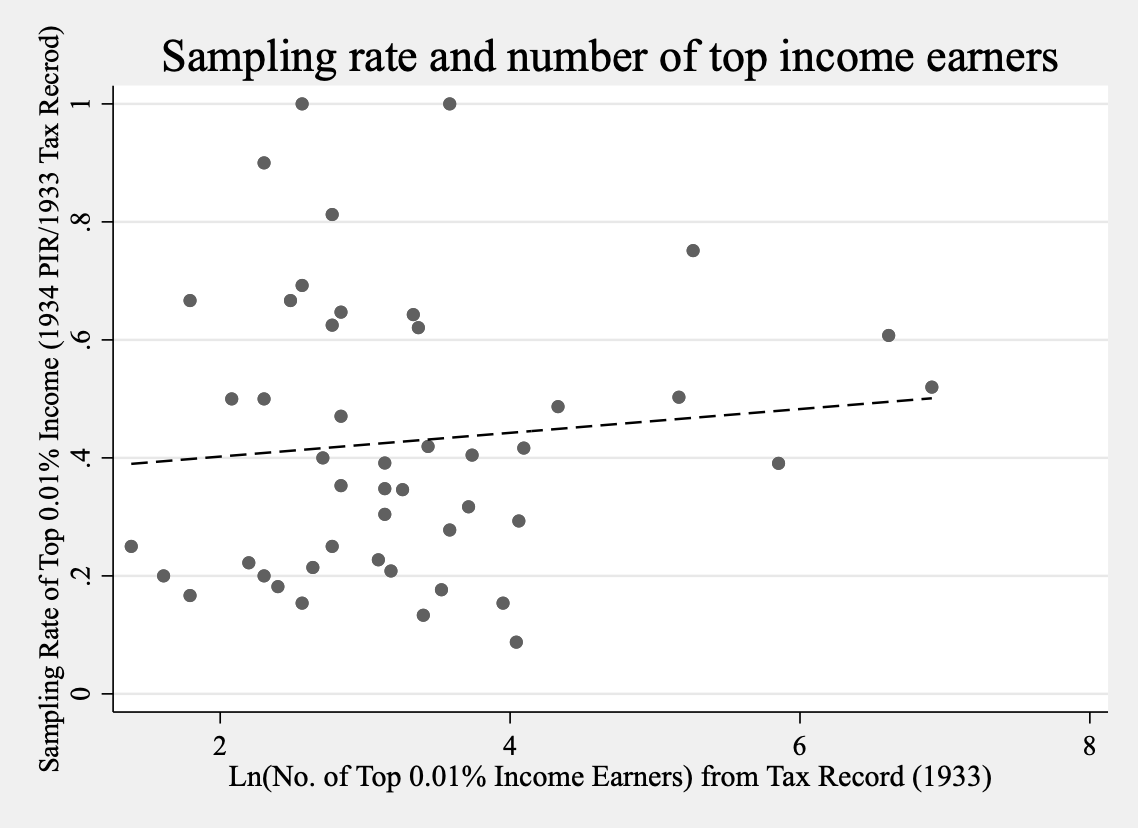}   
        \end{subfigure}  
                ~ 
        \begin{subfigure}[b]{0.45\textwidth}
                \caption{Top 0.01\% income (PIR 1939)}
                \includegraphics[width=\textwidth]{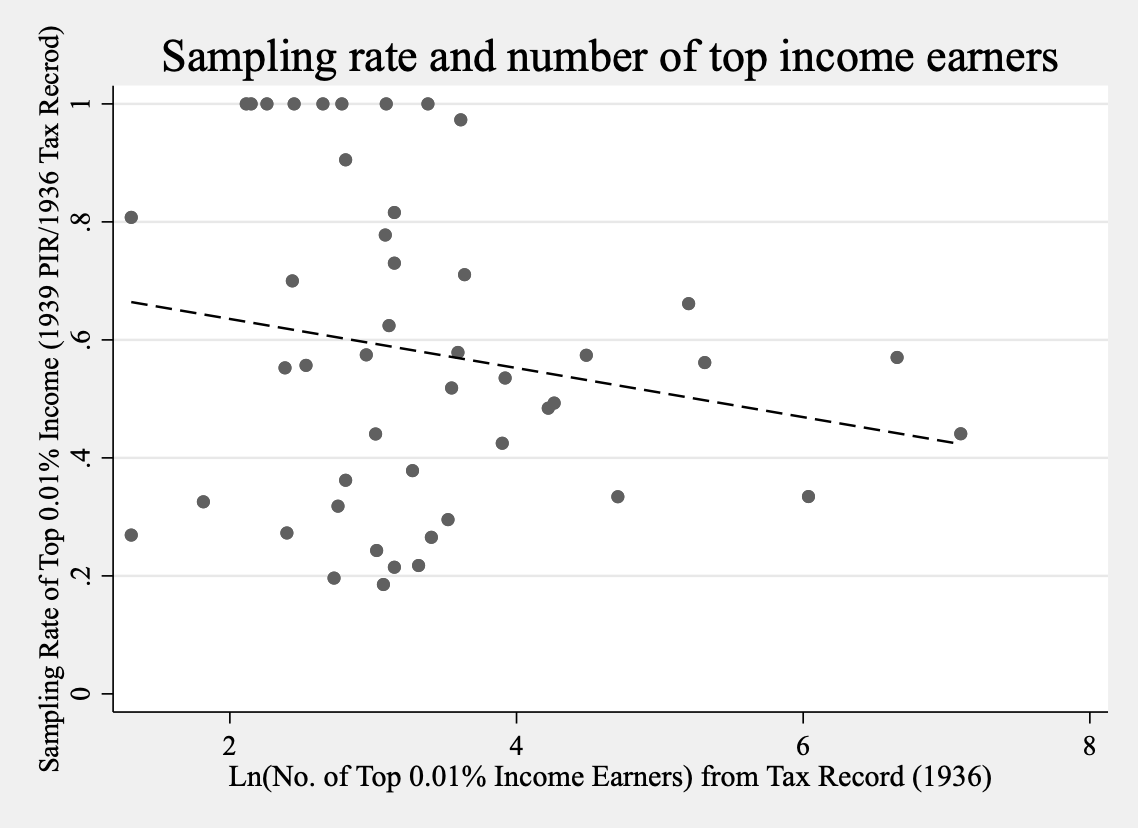}   
        \end{subfigure}   
        ~
          \begin{subfigure}[b]{0.45\textwidth}
          \vspace{0.2in}
                \caption{Schools 1--8 entrants (PIR 1934) }
                 \includegraphics[width=\textwidth]{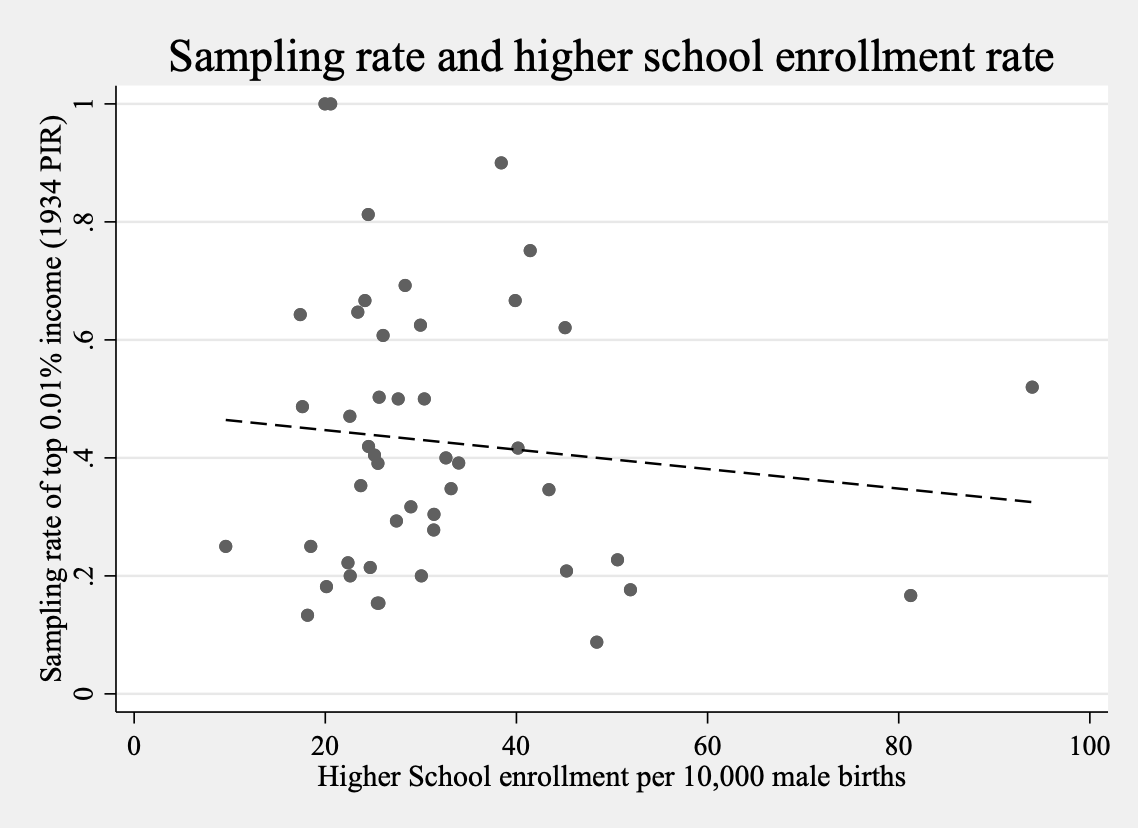}
        \end{subfigure}
                ~ 
        \begin{subfigure}[b]{0.45\textwidth}
                \caption{Schools 1--8 entrants (PIR 1939) }
                \includegraphics[width=\textwidth]{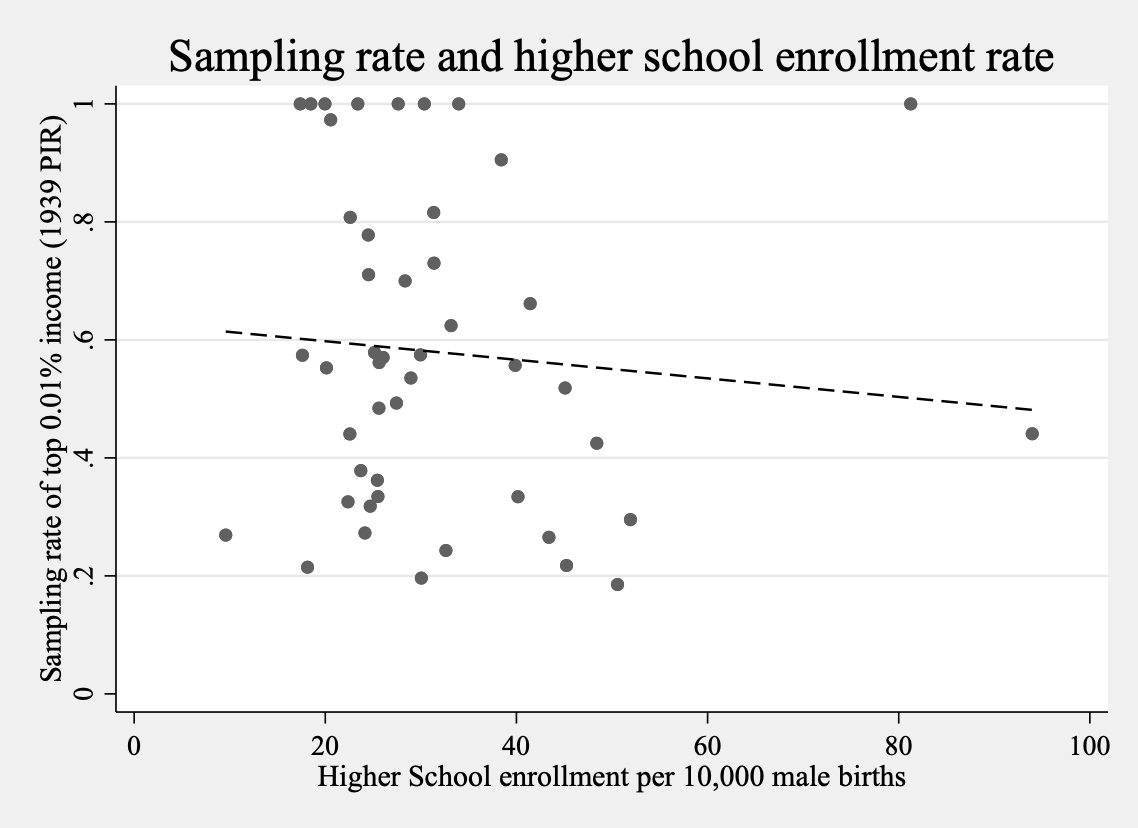}   
        \end{subfigure}
 \end{center}
{\footnotesize \textit{Notes}: These figures plot the prefecture-level sampling rates of PIR (1934) and PIR (1939) for the top 0.01\% income earners against (1) the complete counts of the top 0.01\% income earners (Panels (a) and (b)) and (2) the number of entrants to Schools 1--8 per birth population (Panels (c) and (d)), respectively.
The vertical axis shows the sampling rate of the PIR for the top 0.01\% income earners, defined as the number of PIR-listed top 0.01\% income earners residing in the prefecture divided by the complete count of the top 0.01\% income earners residing in the prefecture. We calculate the prefecture-level complete counts using income tax statistics from the National Tax Bureau Statistical Yearbooks (see Supplementary Materials Section B.2.1 for details).
In Panels (a) and (b), the horizontal axis shows the log of the complete counts of the top 0.01\% income earners.
In Panels (c) and (d), the horizontal axis shows the number of entrants to Schools 1--8 who were born in the prefecture (per 10,000 male births) for cohorts born in 1880--1894. Tokyo prefecture has the highest number  in both the complete count of the top 0.01\% income earners and the school enrollment rate.
The dashed lines are linear regression estimates.
See Section \ref{section:long_run_analysis1} for further discussion of these figures.
 \par}
\end{figure}

   \begin{figure}[hbp] \vspace{-0.2in}
        \caption{Long-run Impacts of Centralization: Geographical Origins of Elites (Other Subgroups)}\label{fig:long-run_additional} 
         \begin{center} \vspace{-0.2in}
        \begin{subfigure}[b]{0.43\textwidth}
                \caption{}
                \includegraphics[width=\textwidth]{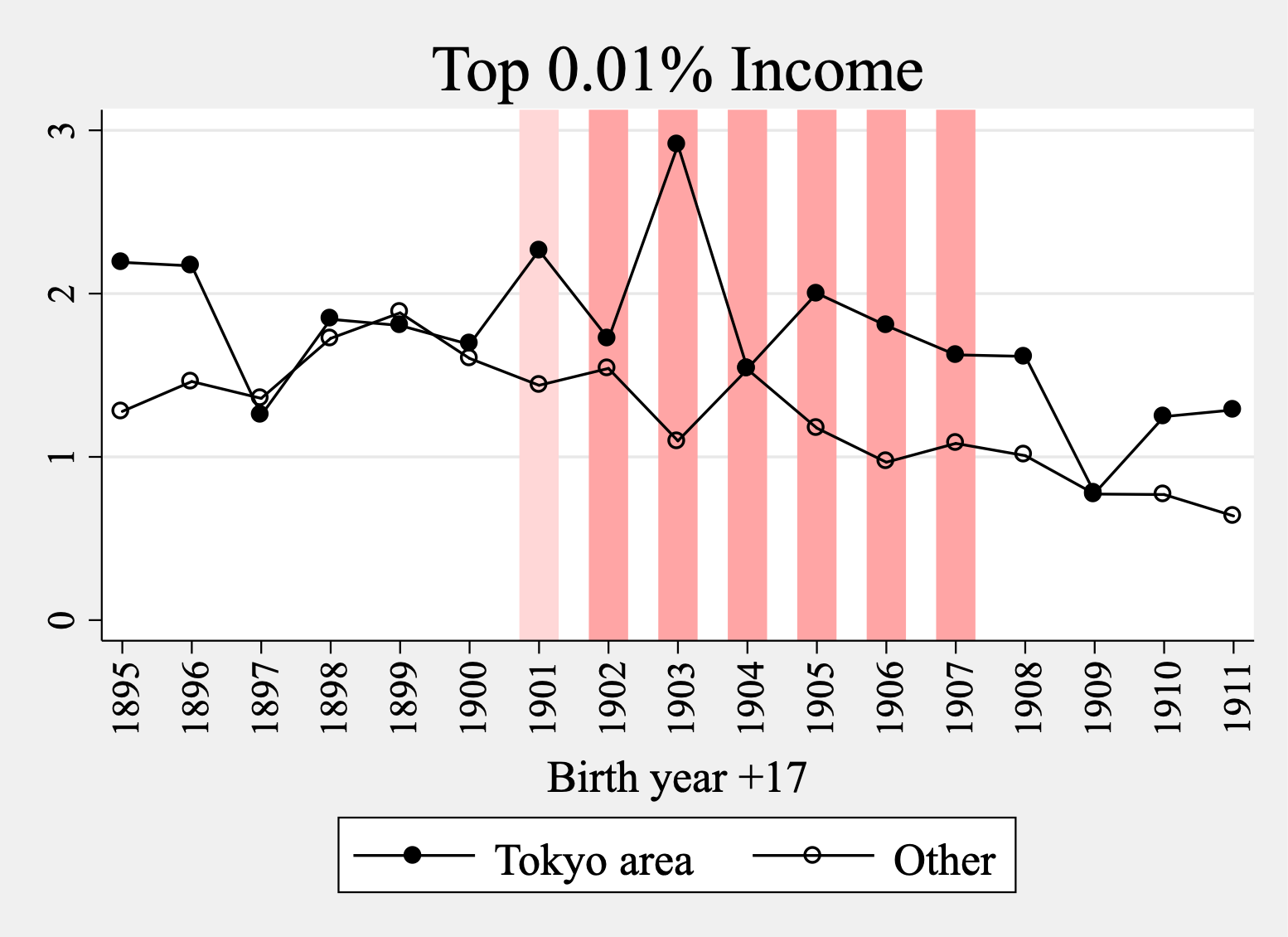}
        \end{subfigure}
                ~ 
        \begin{subfigure}[b]{0.43\textwidth}
                \caption{}
                \includegraphics[width=\textwidth]{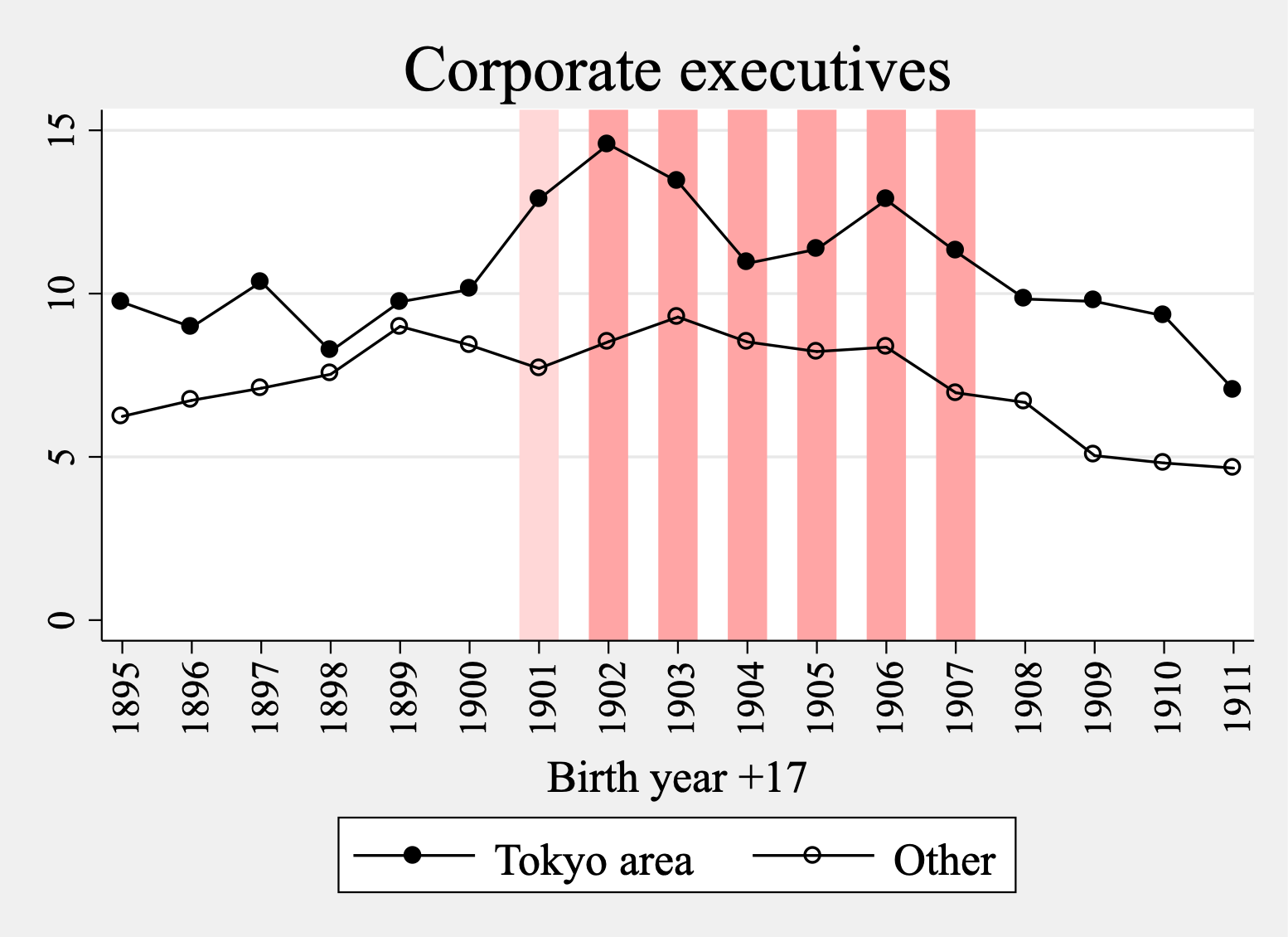}            
        \end{subfigure}
                ~
        \begin{subfigure}[b]{0.43\textwidth}
                \caption{}
                \includegraphics[width=\textwidth]{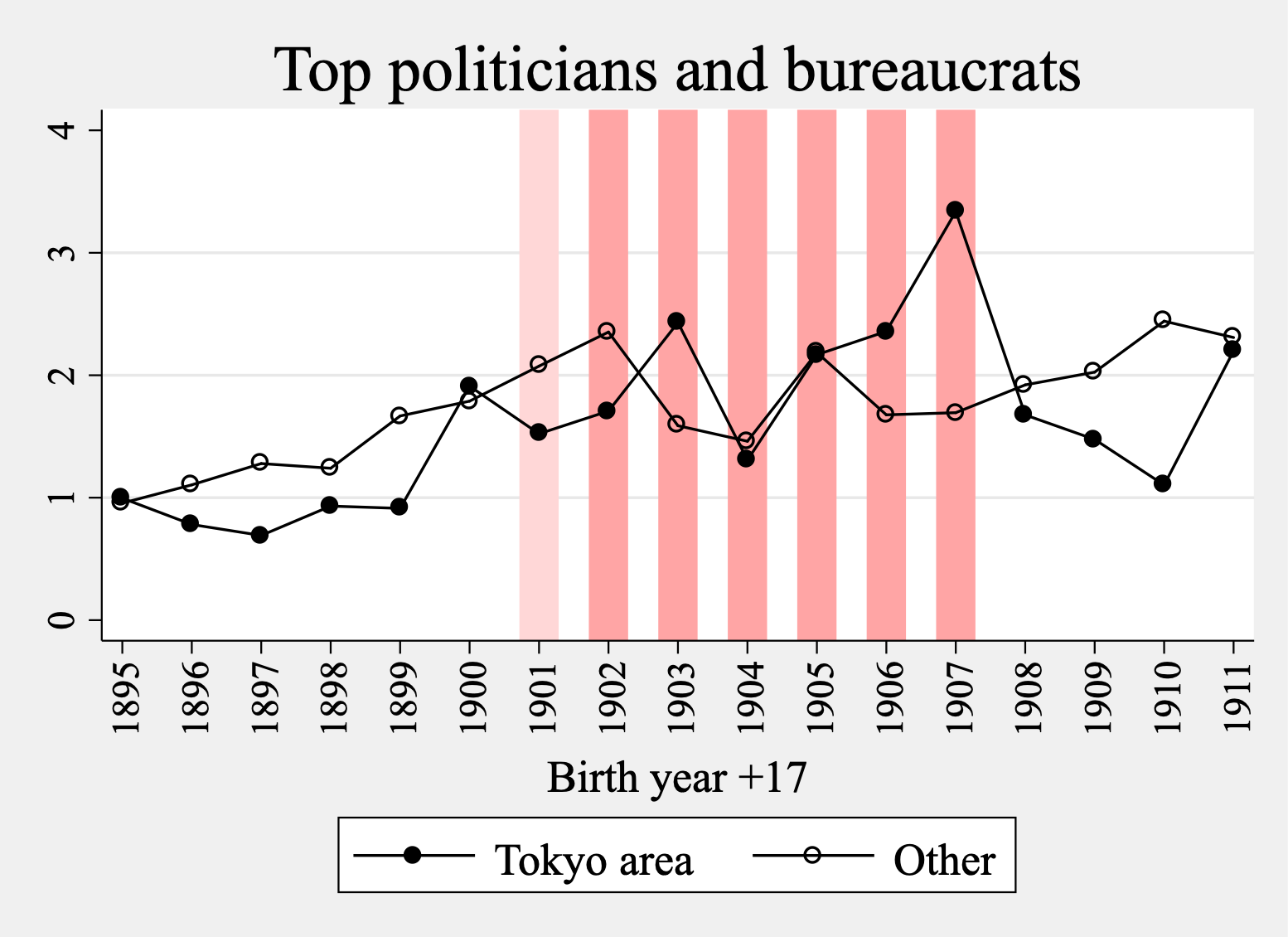}
        \end{subfigure}
                ~ 
        \begin{subfigure}[b]{0.43\textwidth}
                \caption{}
                \includegraphics[width=\textwidth]{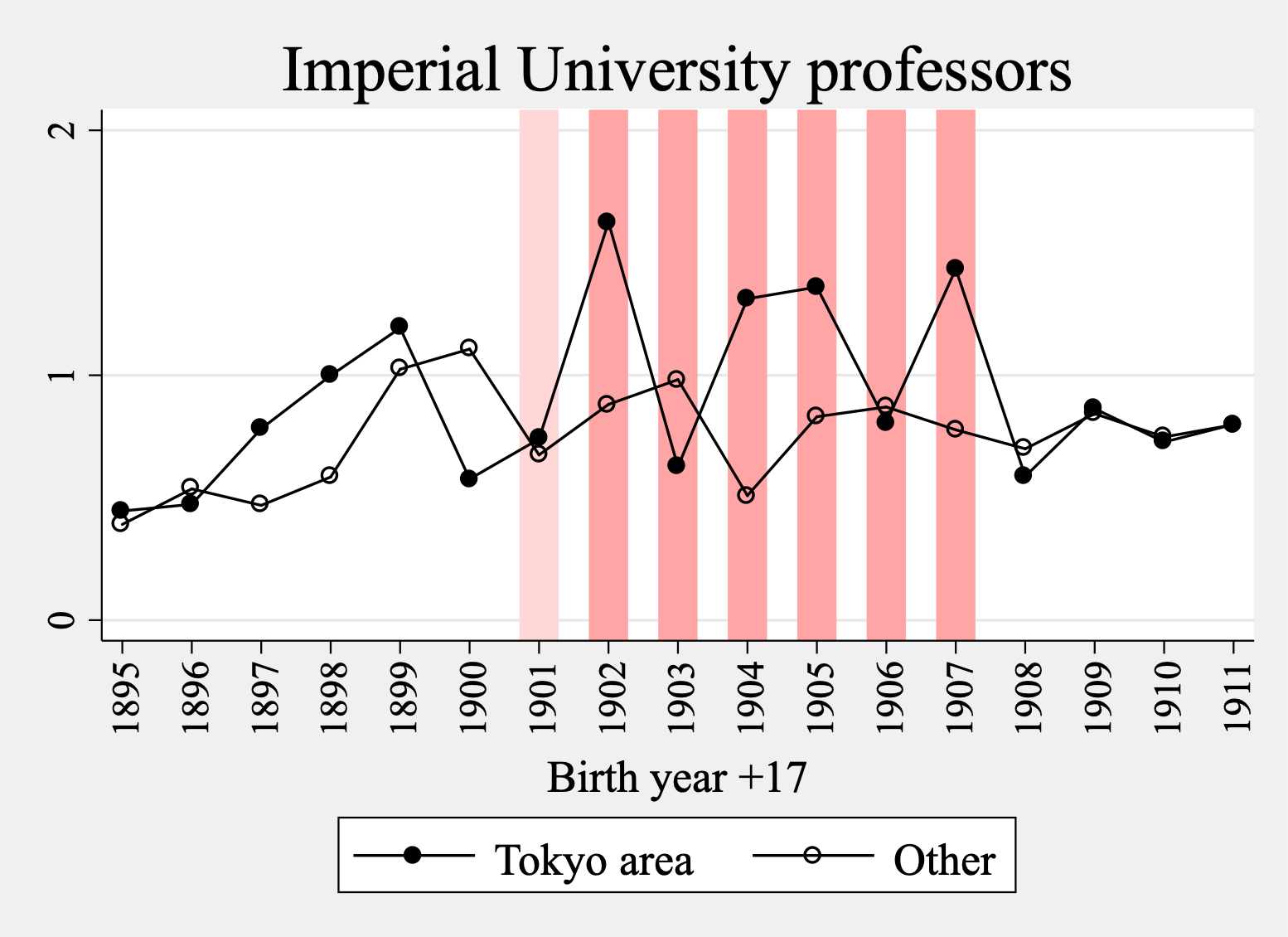}         
        \end{subfigure}  
        \end{center}
        {\footnotesize \textit{Notes}: 
This figure shows additional difference-in-differences plots that compare the average number of elites (per 10,000 male births in the prefecture) born in prefectures inside and outside the Tokyo area by cohort (see Figure \ref{fig:long-run1}).
The plots are based on the data from PIR (1934, 1939), which covers cohorts who were born in 1878--1894 and turned age 17 (minimum application age) in 1895--1911.
The cohorts turning age 17 in 1902--1907 (under centralization) are colored in dark pink. 
The cohort turning age 17 in 1901 (under decentralization) and age 18 in 1902 (under centralization) is colored in pale pink.
See Section \ref{section:long_run_analysis1} for discussions about this figure. \par}
\end{figure}

 \newgeometry{right=0.6in,left=0.6in,top=1in,bottom=1in}   
\begin{table}[h!]
\caption{Long-run Impacts: Pre-event Trends Are Parallel}\label{tab:pretrend}
\begin{center}
\scalebox{0.7}{
\begin{tabular}{lcccccccc}
\hline
\hline
                    &\multicolumn{1}{c}{(1)}&\multicolumn{1}{c}{(2)}&\multicolumn{1}{c}{(3)}&\multicolumn{1}{c}{(4)}&\multicolumn{1}{c}{(5)}&\multicolumn{1}{c}{(6)}&\multicolumn{1}{c}{(7)}&\multicolumn{1}{c}{(8)}\\
                    &\multicolumn{1}{c}{\shortstack{Imperial\\Univ.\\grads}}&\multicolumn{1}{c}{\shortstack{Top 0.01\% \\income\\earners}}&\multicolumn{1}{c}{\shortstack{Top 0.05\% \\income\\earners}}&\multicolumn{1}{c}{\shortstack{Medal\\recipients}}&\multicolumn{1}{c}{\shortstack{Corporate\\executives}}&\multicolumn{1}{c}{\shortstack{Top\\politicians \& \\bureaucrats}}&\multicolumn{1}{c}{\shortstack{Imperial\\Univ.\\professors}}&\multicolumn{1}{c}{\shortstack{All\\occupational\\elites}}\\
\\ \hline \\ & \multicolumn{8}{c}{Panel A: Coefficient for each year} \\ 
Tokyo area $\times$ Cohort 1895 &        0.93   &        0.80   &        0.44   &        0.32   &        1.73   &       -0.11   &        0.56   &        1.26   \\
                    &     (0.750)   &     (0.117)   &     (0.823)   &     (0.876)   &     (0.394)   &     (0.916)   &     (0.113)   &     (0.849)   \\
Tokyo area $\times$ Cohort 1896 &        0.87   &        0.62   &        0.17   &        0.76   &        0.54   &       -0.46   &        0.45   &       -0.80   \\
                    &     (0.746)   &     (0.299)   &     (0.854)   &     (0.713)   &     (0.726)   &     (0.643)   &     (0.112)   &     (0.877)   \\
Tokyo area $\times$ Cohort 1897 &       -0.39   &       -0.20   &        0.62   &       -0.15   &        1.49   &       -0.72   &        0.81   &        3.85   \\
                    &     (0.870)   &     (0.672)   &     (0.640)   &     (0.925)   &     (0.231)   &     (0.563)   &     (0.077)*  &     (0.344)   \\
Tokyo area $\times$ Cohort 1898 &       -0.39   &        0.02   &       -0.05   &       -0.21   &       -0.97   &       -0.43   &        0.91   &       -1.17   \\
                    &     (0.831)   &     (0.974)   &     (0.980)   &     (0.872)   &     (0.598)   &     (0.636)   &     (0.028)** &     (0.845)   \\
Tokyo area $\times$ Cohort 1899 &        0.91   &       -0.15   &        0.17   &        0.44   &       -0.89   &       -0.86   &        0.67   &        1.86   \\
                    &     (0.571)   &     (0.755)   &     (0.877)   &     (0.798)   &     (0.542)   &     (0.335)   &     (0.236)   &     (0.414)   \\
F-statistic for joint significance&        1.23   &        0.79   &        0.29   &        0.33   &        1.93   &        0.57   &        1.84   &        1.43   \\
p-value for joint significance&       0.310   &       0.560   &       0.916   &       0.895   &       0.108   &       0.720   &       0.125   &       0.232   \\

\\ & \multicolumn{8}{c}{Panel B: Linear Specification} \\ 
Tokyo area ($<$ 100km) $\times$ Time trend&       -0.13   &       -0.17   &       -0.08   &       -0.07   &       -0.44   &       -0.01   &       -0.06   &       -0.09   \\
                    &     (0.822)   &     (0.117)   &     (0.766)   &     (0.823)   &     (0.129)   &     (0.952)   &     (0.247)   &     (0.934)   \\

\\ & \multicolumn{8}{c}{Panel C: Linear with Control Variables} \\ 
Tokyo area ($<$ 100km) $\times$ Time trend&       -0.70   &       -0.18   &       -0.28   &       -0.27   &       -0.63   &       -0.05   &       -0.06   &       -0.81   \\
                    &     (0.002)***&     (0.107)   &     (0.130)   &     (0.243)   &     (0.021)** &     (0.656)   &     (0.353)   &     (0.152)   \\

\\ & \multicolumn{8}{c}{Panel D: Squared Trend Term} \\ 
Tokyo area ($<$ 100km) $\times$ Time trend&       -1.24   &       -1.82   &        2.10   &        1.15   &       -2.04   &       -0.30   &        1.96   &        8.84   \\
                    &     (0.497)   &     (0.335)   &     (0.355)   &     (0.630)   &     (0.615)   &     (0.875)   &     (0.044)** &     (0.207)   \\
Tokyo area ($<$ 100km) $\times$ Time trend$^2$&        0.03   &        0.08   &       -0.11   &       -0.07   &        0.07   &        0.01   &       -0.10   &       -0.46   \\
                    &     (0.771)   &     (0.372)   &     (0.292)   &     (0.569)   &     (0.724)   &     (0.899)   &     (0.036)** &     (0.172)   \\
                    &               &               &               &               &               &               &               &               \\
Observations        &         278   &         278   &         278   &         278   &         278   &         278   &         278   &         278   \\
Cohort FE, Birth pref. FE&         Yes   &         Yes   &         Yes   &         Yes   &         Yes   &         Yes   &         Yes   &         Yes   \\
\hline \hline
\end{tabular}

}
\end{center}
          {\footnotesize \textit{Notes}: 
 This table tests if there are differences in pre-event trends between urban and rural areas in the difference-in-differences analysis in Table \ref{tab:long_run}. 
We use the sample of cohorts who were born in 1878--1883 and thus turned age 17 in 1895--1900. In Panel A,  we run the following regression in Panel A: 
$$Y_{pt} = \sum_t \beta_t \times Cohort_t \times Tokyo\_area_p + \alpha_p+\alpha_t+ \epsilon_{pt},$$
\noindent where $Cohort_t$ is an indicator for cohort $t$. The cohort who turned age 17 in 1900 is the base cohort. ``F-statistic (p-value) for joint significance'' shows the F-statistic (and its p value) for the null hypothesis that all of the pre-event coefficients ($\beta_t$) are zero. 

\noindent In Panels B and C: we run the following regression:
$$Y_{pt} =\beta \times Trend_t \times Tokyo\_area_p + \alpha_p+\alpha_t+ \epsilon_{pt},$$

\noindent where $Trend_t$ is defined as the cohort's birth year minus 1870 (the linear time trend). 
In Panel D, we run the following regression:
$$Y_{pt} =\beta_1 \times Trend_t \times Tokyo\_area_p + \beta_2 \times (Trend_t)^2 \times Tokyo\_area_p + \alpha_p+\alpha_t+ \epsilon_{pt},$$

In Panels C and D, we add for the same control variables as Panel B of Table \ref{tab:long_run}.
All the other variables are defined in the same way as in Table \ref{tab:long_run}. Parentheses contain p-values based on standard errors clustered at the prefecture level. 
See Section \ref{section:long_run_analysis1} for discussions about this table. \par}
\end{table} 
\restoregeometry

   \begin{figure}[hbp] \vspace{-0.4in}
        \caption{Long-run Impacts of Centralization: Sensitivity Analysis}\label{fig:long-run_sensitivity} 
         \begin{center} \vspace{-0.2in}
        \begin{subfigure}[b]{0.6\textwidth}
                \caption{}
                \includegraphics[width=\textwidth]{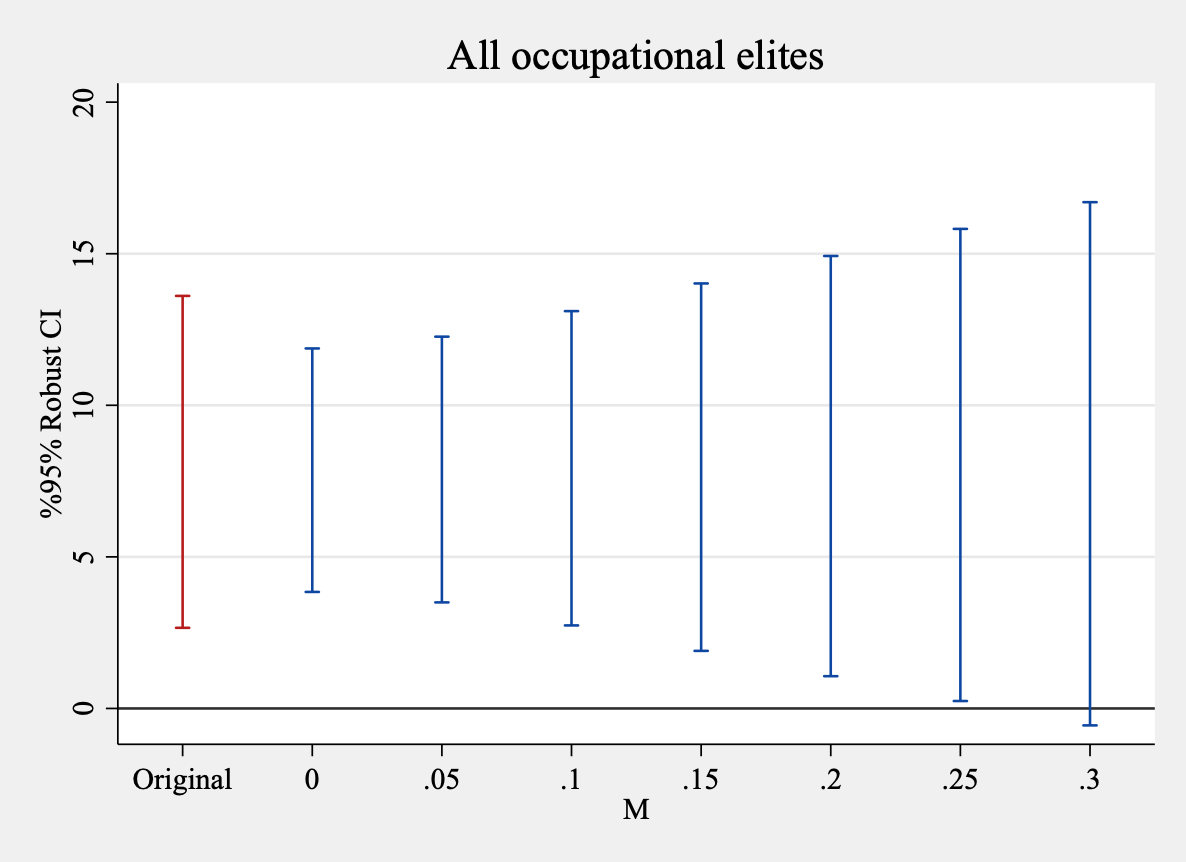}
        \end{subfigure}
                ~ 
        \begin{subfigure}[b]{0.6\textwidth}
                \caption{}
                \includegraphics[width=\textwidth]{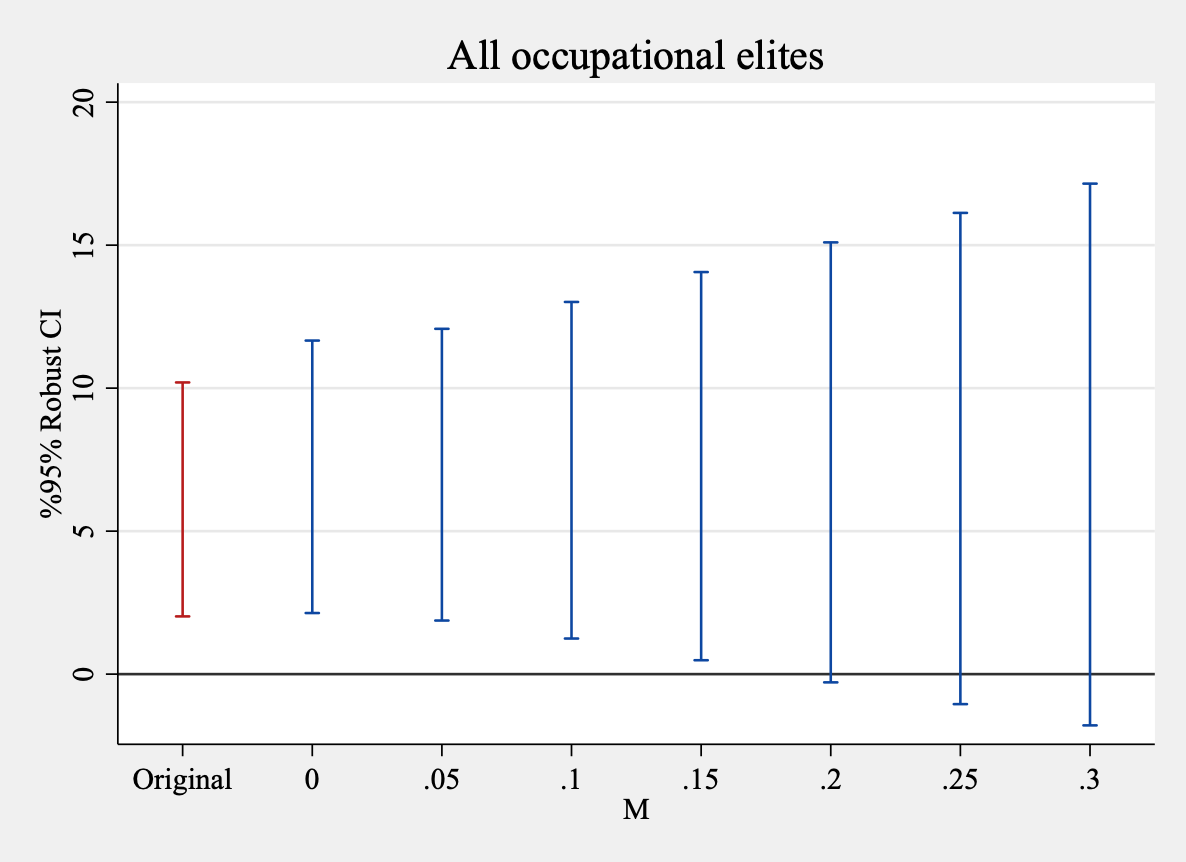}            
        \end{subfigure}  
        \end{center}
        {\footnotesize \textit{Notes}: This figure shows the sensitivity of the difference-in-difference estimates for all occupational elites to possible violations of the parallel trend assumption, following \citeappendix{Rambachan_Roth2023}'s method.         
        The horizontal axis measures the potential degree of differential trends. 
        Panel (a) shows the baseline estimates with prefecture and cohort fixed effects without control variables, and panel (b) adds control variables. 
        The figures thus show that our estimates are robust to bounded differential pre-trends of up to 0.15--0.2 percentage points per year. 
        As the method of \citeappendix{Rambachan_Roth2023} does not accommodate bidirectional events, we here exclude the cohorts who turned age 17 in 1908--1911 from the sample used for estimating Table \ref{tab:long_run} Panels A--C. See Section \ref{section:long_run_analysis1} for discussions about this figure. \par}
\end{figure}

 \newgeometry{right=0.6in,left=0.6in,top=1.5in,bottom=1.5in}   
\begin{table}[h!]
\caption{Long-run Impacts: Difference-in-Differences Estimates (Using PIR (1934) and PIR (1939) Separately)}\label{tab:JPIR1934}
\begin{center}
\scalebox{0.76}{
\begin{tabular}{lcccccccc}
\toprule
\toprule
                    &\multicolumn{1}{c}{(1)}&\multicolumn{1}{c}{(2)}&\multicolumn{1}{c}{(3)}&\multicolumn{1}{c}{(4)}&\multicolumn{1}{c}{(5)}&\multicolumn{1}{c}{(6)}&\multicolumn{1}{c}{(7)}&\multicolumn{1}{c}{(8)}\\
                    &\multicolumn{1}{c}{\shortstack{Imperial\\Univ.\\grads}}&\multicolumn{1}{c}{\shortstack{Top 0.01\% \\income\\earners}}&\multicolumn{1}{c}{\shortstack{Top 0.05\% \\income\\earners}}&\multicolumn{1}{c}{\shortstack{Medal\\recipients}}&\multicolumn{1}{c}{\shortstack{Corporate\\executives}}&\multicolumn{1}{c}{\shortstack{Top\\politicians \& \\bureaucrats}}&\multicolumn{1}{c}{\shortstack{Imperial\\Univ.\\professors}}&\multicolumn{1}{c}{\shortstack{All\\occupational\\elites}}\\
\hline &  \\ & \multicolumn{8}{c}{Panel A: Baseline Specification (PIR 1934)} \\ 
Age 17 under centralization&        1.62   &        0.27   &        1.08   &        1.32   &        0.64   &        0.40   &        0.01   &        3.20   \\
                    &     (0.032)** &     (0.097)*  &     (0.042)** &     (0.071)*  &     (0.054)*  &     (0.006)***&     (0.969)   &     (0.157)   \\
                    &     [0.001]***   &     [0.018]**   &     [0.019]**   &     [0.000]***   &     [0.053]*   &     [0.159]   &     [0.982]   &     [0.005]***   \\
Observations        &         658   &         658   &         658   &         658   &         658   &         658   &         658   &         658   \\

\\ & \multicolumn{8}{c}{Panel B: With Control Variables (PIR 1934)} \\ 
Age 17 under centralization&        1.28   &        0.30   &        1.19   &        1.00   &        0.71   &        0.28   &        0.02   &        2.94   \\
                    &     (0.021)** &     (0.050)*  &     (0.043)** &     (0.055)*  &     (0.062)*  &     (0.071)*  &     (0.871)   &     (0.143)   \\
                    &     [0.001]***   &     [0.015]**   &     [0.002]***   &     [0.002]***   &     [0.002]***  &     [0.297]   &     [0.890]   &     [0.001]***   \\
Observations        &         658   &         658   &         658   &         658   &         658   &         658   &         658   &         658   \\

\\ & \multicolumn{8}{c}{Panel C: Baseline Specification (PIR 1939)} \\ 
Age 17 under centralization&        2.26   &        0.47   &        0.97   &        1.94   &        1.24   &        0.62   &        0.33   &        5.97   \\
                    &     (0.007)***&     (0.004)***&     (0.074)*  &     (0.007)***&     (0.042)** &     (0.003)***&     (0.014)** &     (0.076)*  \\
                    &     [0.000]***   &     [0.036]**   &     [0.015]**   &     [0.000]***   &     [0.072]*   &     [0.022]**   &     [0.123]   &     [0.000]***   \\
Observations        &         658   &         658   &         658   &         658   &         658   &         658   &         658   &         658   \\

\\ & \multicolumn{8}{c}{Panel D: With Control Variables (PIR 1939)} \\ 
Age 17 under centralization&        1.49   &        0.40   &        0.83   &        1.67   &        0.97   &        0.52   &        0.32   &        4.20   \\
                    &     (0.000)***&     (0.007)***&     (0.084)*  &     (0.004)***&     (0.037)** &     (0.003)***&     (0.022)** &     (0.042)** \\
                    &     [0.015]**   &     [0.042]**   &     [0.020]**   &     [0.002]***   &     [0.098]*   &     [0.085]*   &     [0.117]   &     [0.004]***   \\
                    &               &               &               &               &               &               &               &               \\
Observations        &         658   &         658   &         658   &         658   &         658   &         658   &         658   &         658   \\
Cohort FE, Birth pref. FE&         Yes   &         Yes   &         Yes   &         Yes   &         Yes   &         Yes   &         Yes   &         Yes   \\
Mean dep var        &        7.32   &        0.92   &        3.54   &        5.24   &        6.55   &        1.19   &        0.70   &       30.52   \\
\hline \hline
\end{tabular}

}
\end{center}
{\footnotesize \textit{Notes}: In this table, we repeat the same difference-in-differences analysis as in Table \ref{tab:long_run} Panels A and B, but using the observations from the two editions of the PIR in 1934 and 1939 separately.
In PIR (1934), we observe the cohorts born in 1880--1894 when they are 40 to 54 years old. 
In PIR (1939), we observe the cohorts born in 1880--1894 when they are 45 to 59 years old. 
All the variables are defined as in Table \ref{tab:long_run}. Parentheses contain p-values based on standard errors clustered at the prefecture level. Square brackets contain wild cluster bootstrap p-values based on standard errors clustered at the cohort level. 
See Section \ref{section:long_run_analysis1} for discussions about this table. \par} 
\end{table} 
\restoregeometry

 \begin{table}[h!]
        \caption{Long-run Impacts: Difference-in-Difference Estimates (Adjusted by PIR Sampling Rates)}\label{tab:long_run_inversesample}
\begin{center}
\scalebox{0.76}{
\begin{tabular}{lcccc}
\toprule
\toprule
                    &\multicolumn{1}{c}{(1)}&\multicolumn{1}{c}{(2)}&\multicolumn{1}{c}{(3)}&\multicolumn{1}{c}{(4)}\\
                    &\multicolumn{1}{c}{\shortstack{Top 0.01\% \\PIR (1934)}}&\multicolumn{1}{c}{\shortstack{Top 0.05\% \\PIR (1934)}}&\multicolumn{1}{c}{\shortstack{Top 0.01\% \\PIR (1939)}}&\multicolumn{1}{c}{\shortstack{Top 0.05\% \\PIR (1939)}}\\
\hline &  \\ & \multicolumn{4}{c}{Panel A: Baseline Specification} \\ 
Age 17 under centralization&        0.57   &        3.78   &        1.08   &        2.05   \\
                    &     (0.148)   &     (0.024)** &     (0.009)***&     (0.197)   \\
                    &     [0.046]**   &     [0.018]**   &     [0.039]**   &     [0.059]*   \\

\\ & \multicolumn{4}{c}{Panel B: With Control Variables} \\ 
Age 17 under centralization&        0.77   &        4.02   &        0.97   &        1.52   \\
                    &     (0.074)*  &     (0.006)***&     (0.009)***&     (0.263)   \\
                    &     [0.092]*   &     [0.024]**  &     [0.023]**   &     [0.118]   \\
                    &               &               &               &               \\
Observations        &         658   &         658   &         658   &         658   \\
Cohort FE, Birth pref. FE&         Yes   &         Yes   &         Yes   &         Yes   \\
Mean dep var        &        1.26   &       11.37   &        1.73   &       10.80   \\
Mean dep var &        1.36   &       10.77   &        1.63   &       10.96   \\
(Tokyo area under decentralization) \\ 
\hline \hline
\end{tabular}

}
\end{center}
{\footnotesize \textit{Notes}: 
This table shows difference-in-differences estimates of the long-run effects of the centralized admissions on the geographical origins of elites. 
The dependent variable is the number of top 0.01\% (or 0.05\%) income earners listed in PIR (1934) (or PIR (1939)) by birth prefecture and birth cohort (for cohorts born in 1880--1894), multiplied by the inverse of sampling rate of the top income earners in the prefecture. 
The prefecture-level sampling rate is defined by the number of the PIR-listed top income earners residing in the prefecture divided by the complete count of the top income earners in the prefecture. The prefecture-level complete counts of the top income earners are calculated using income tax statistics from the National Tax Bureau Statistical Yearbooks (see Supplementary Materials Section B.2.1 for details).
The rest of the specification is the same as Appendix Table \ref{tab:JPIR1934}. 
See Section \ref{section:long_run_analysis1} for discussions about this table. \par}
\end{table} 

 \begin{table}[h!]
        \caption{Long-run Impacts: Difference-in-Difference Estimates (Subsample of Prefectures with High Sampling Rates)}\label{tab:long_run_high_sample_rate}
\begin{center}
\scalebox{0.76}{
\begin{tabular}{lcccc}
\toprule
\toprule
                    &\multicolumn{1}{c}{(1)}&\multicolumn{1}{c}{(2)}&\multicolumn{1}{c}{(3)}&\multicolumn{1}{c}{(4)}\\
                    &\multicolumn{1}{c}{\shortstack{Top 0.01\% \\PIR (1934)}}&\multicolumn{1}{c}{\shortstack{Top 0.05\% \\PIR (1934)}}&\multicolumn{1}{c}{\shortstack{Top 0.01\% \\PIR (1939)}}&\multicolumn{1}{c}{\shortstack{Top 0.05\% \\PIR (1939)}}\\
\hline &  \\ & \multicolumn{4}{c}{Panel A: Baseline Specification} \\ 
Age 17 under centralization&        0.30   &        1.14   &        0.40   &        0.93   \\
                    &     (0.148)   &     (0.086)*  &     (0.050)*  &     (0.130)   \\
                    &     [0.200]   &     [0.022]**   &     [0.177]   &     [0.040]**   \\

\\ & \multicolumn{4}{c}{Panel B: With Control Variables} \\ 
Age 17 under centralization&        0.31   &        1.35   &        0.22   &        0.74   \\
                    &     (0.136)   &     (0.098)*  &     (0.273)   &     (0.167)   \\
                    &     [0.112]   &     [0.004]***   &     [0.432]   &     [0.114]   \\
                    &               &               &               &               \\
Observations        &         364   &         364   &         364   &         364   \\
Cohort FE, Birth pref. FE&         Yes   &         Yes   &         Yes   &         Yes   \\
Mean dep var        &        0.80   &        3.50   &        1.21   &        4.52   \\
Mean dep var &        0.78   &        4.36   &        1.10   &        4.51   \\
 (Tokyo area under decentralization) \\ 
\hline \hline
\end{tabular}

}
\end{center}
{\footnotesize \textit{Notes}: 
This table shows difference-in-differences estimates of the long-run effects of the centralized admissions on the geographical origins of elites. 
The dependent variable is the number of top 0.01\% (or 0.05\%) income earners listed in PIR (1934) (or PIR (1939)) by birth prefecture and birth cohort (for cohorts born in 1880--1894). For a robustness check, we use a subsample of 26 prefectures whose average sampling rates of top 0.01\% income earners are equal to or higher than the average sampling rate of Tokyo prefecture in PIR (1934) and PIR (1939).
The prefecture-level sampling rate is defined by the number of the PIR-listed top income earners residing in the prefecture divided by the complete count of the top income earners in the prefecture. The prefecture-level complete counts of the top income earners are calculated using income tax statistics from the National Tax Bureau Statistical Yearbooks (see Supplementary Materials Section B.2.1 for details).
The rest of the specification is the same as Appendix Table \ref{tab:JPIR1934}. 
See Section \ref{section:long_run_analysis1} for discussions about this table. \par}
\end{table}

\begin{table}[h!]
        \caption{Long-run Impacts of Centralization: Placebo Tests and Pathways}\label{tab:long_run_placebo}
\begin{center}
\scalebox{0.76}{
\begin{tabular}{lcccc}
\toprule
\toprule
                    &\multicolumn{1}{c}{(1)}&\multicolumn{1}{c}{(2)}&\multicolumn{1}{c}{(3)}&\multicolumn{1}{c}{(4)}\\
                    &\multicolumn{1}{c}{\shortstack{Placebo:\\Population}}&\multicolumn{1}{c}{\shortstack{Placebo: \\Landlords}}&\multicolumn{1}{c}{\shortstack{Pathway: \\Fraction moved \\in the long-run}}&\multicolumn{1}{c}{\shortstack{Pathway: \\Distance moved \\in the long-run}}\\
\hline &  \\ & \multicolumn{4}{c}{Panel A: Baseline Specification} \\ 
Age 17 under centralization&        0.03   &       -0.14   &       -0.02   &      -15.62   \\
                    &     (0.321)   &     (0.575)   &     (0.334)   &     (0.199)   \\
                    &     [0.283]   &     [0.445]   &     [0.274]   &     [0.352]   \\

\\ & \multicolumn{4}{c}{Panel B: Adding Control Variables} \\ 
Age 17 under centralization&       -0.01   &        0.05   &       -0.02   &      -21.59   \\
                    &     (0.789)   &     (0.689)   &     (0.228)   &     (0.148)   \\
                    &     [0.666]   &     [0.731]   &     [0.206]   &     [0.242]   \\
                    &               &               &               &               \\
Observations        &         658   &         658   &         658   &         658   \\
Cohort FE, Birth pref. FE&         Yes   &         Yes   &         Yes   &         Yes   \\
Mean dep var        &        1.17   &        0.60   &        0.30   &      298.51   \\
Mean dep var (Tokyo area&        1.33   &        1.94   &        0.38   &      467.32   \\
under decentralization)&               &               &               &               \\
\hline \hline
\end{tabular}

}
\end{center}
          {\footnotesize \textit{Notes}: 
This table provides placebo tests and explores pathways of the long-run effects. 
We construct the prefecture-cohort level data focusing on the birth cohort born in 1880--1894.
In (1), ``Population'' is the cohort's male birth population in the birth prefecture (where the unit is 10000 persons). 
In (2), ``Landlords" is defined as individuals listed in the PIR whose occupations include landlords, but excluding the top 0.05\% income earners, medal recipients, corporate executives, top politicians and bureaucrats, and Imperial University professors. We count the number of landlords by birth prefecture and birth cohort dividing them by average birth population in the prefecture and multiplying them by 10000. 
In (3), ``Fraction moved" is defined as the fraction of individuals listed in PIR (1934, 1939) whose prefecture of residence is different from his birth prefecture. 
In (4), ``Distance moved" is defined as the average distance between the birth prefecture and the prefecture of residence among individuals listed in PIR (1934, 1939). 
``Age 17 under centralization" is the indicator variable that takes 1 if the cohort became age 17 under the centralized admissions in 1902--07.
``Mean dep var" shows the mean of the dependent variable for all prefecture-cohort observations. 
``Mean dep var (Tokyo area under decentralization)" shows the mean of the dependent variable in the Tokyo area under the decentralized admissions. 
In Panel B, we control for time- and cohort-varying prefecture characteristics, i.e., the number of primary schools in the prefecture in the year when the cohort turned eligible age, the number of middle-school graduates in the prefecture in the year when the cohort turned age 17, log GDP of the prefecture when the cohort turned age 20, and the birth population of the cohort in the prefecture (except for Column (1)). 
Parentheses contain p-values based on standard errors clustered at the prefecture level. Square brackets contain wild cluster bootstrap p-values based on standard errors clustered at the cohort level. ***, **, and * mean significance at the 1\%, 5\%, and 10\% levels, respectively.
See Section \ref{section:long_run_analysis1} for discussions about this table. \par}
\end{table}

\begin{table}[h!]
        \caption{Long-run Impacts of Centralization: National Production of Top Government Officials (Additional Results)}\label{tab:long_run_toprankgov_add} \vspace{-0.15in}
\begin{center}
\subfloat[Passers of the Higher Civil Service Exams]{
\scalebox{0.76}{
\begin{tabular}{lcccccccc}
\hline
\hline
                    &\multicolumn{2}{c}{\shortstack{Exam passers\\graduated from\\School 1--8}}&\multicolumn{2}{c}{\shortstack{Exam passers\\not graduated from\\Schools 1--8}}&\multicolumn{2}{c}{\shortstack{Exam passers}}\\\cmidrule(lr){2-3}\cmidrule(lr){4-5}\cmidrule(lr){6-7}
                    &\multicolumn{1}{c}{(1)}   &\multicolumn{1}{c}{(2)}   &\multicolumn{1}{c}{(3)}   &\multicolumn{1}{c}{(4)}   &\multicolumn{1}{c}{(5)}   &\multicolumn{1}{c}{(6)}   \\
\hline
Centralized         &       12.47   &       18.31   &      -12.47   &      -18.31   &        7.18   &       -6.65   \\
                    &     (0.042)** &     (0.000)***&     (0.042)** &     (0.000)***&     (0.763)   &     (0.410)   \\
                    &               &               &               &               &               &               \\
\hline
Observations        &          33   &          33   &          33   &          33   &          33   &          33   \\
Time trend          &   Quadratic   &   6th order   &   Quadratic   &   6th order   &   Quadratic   &   6th order   \\
Control exam passers&         Yes   &         Yes   &         Yes   &         Yes   &          No   &          No   \\
Mean dep var        &      101.92   &      101.92   &       92.07   &       92.07   &      194.00   &      194.00   \\
(decentralization)  &               &               &               &               &               &               \\
\hline \hline
\end{tabular}

}} \\ 
 \vspace{0.4cm} 
 \subfloat[Top-Ranking Higher Civil Officials Controlling for Exam Passers in Each Group]{
\scalebox{0.76}{
\begin{tabular}{lcccc}
\hline
\hline
                    &\multicolumn{2}{c}{\shortstack{Top-ranking\\ officials\\graduated from\\Schools 1--8}}&\multicolumn{2}{c}{\shortstack{Top-ranking\\ officials\\not graduated from\\Schools 1--8}}\\\cmidrule(lr){2-3}\cmidrule(lr){4-5}
                    &\multicolumn{1}{c}{(1)}   &\multicolumn{1}{c}{(2)}   &\multicolumn{1}{c}{(3)}   &\multicolumn{1}{c}{(4)}   \\
\hline
Centralized         &        2.51   &        2.44   &        0.20   &        1.87   \\
                    &     (0.043)** &     (0.079)*  &     (0.805)   &     (0.169)   \\
                    &               &               &               &               \\
\hline
Observations        &          33   &          33   &          33   &          33   \\
Time trend          &   Quadratic   &   6th order   &   Quadratic   &   6th order   \\
Control exam passers&         Yes   &         Yes   &         Yes   &         Yes   \\
Mean dep var        &       19.52   &       19.52   &        9.03   &        9.03   \\
(decentralization)  &               &               &               &               \\
\hline \hline
\end{tabular}

}} 
 \vspace{0.2cm} 
\end{center}
{\scriptsize \textit{Notes}:     
Panel (a) shows OLS estimates of the effects of the centralized admissions on the number of individuals who passed the administrative division of the Higher Civil Service Exams (administrative HCSE).
Panel (b) shows OLS estimates of the effects of the centralized admissions on the number of top-ranking higher civil officials, controlling for the number of exam passers in the specified group.
The estimates are based on the cohort level data (1898--1930), where cohort is defined by the year of entering a higher school or its equivalent.
The data is compiled from the complete list of individuals who passed the administrative HCSE in 1894--1941 and their biographical information.
``Exam passers'' is the number of individuals in cohort $t$ who passed the administrative HCSE.
``Top-ranking officials'' is the number of top-ranking officials in cohort $t$ (i.e., the number of individuals who entered a higher school or its equivalent in year $t$, passed the administrative HCSE, and were internally promoted to the top three ranks of higher civil service in their lifetime).
``Centralized'' is the indicator variable that takes 1 if cohort $t$ entered a higher school or its equivalent under the centralized admissions in 1902--07, 1917--18, and 1926--27.
``Mean dep var (decentralization)'' is the mean of the dependent variable for the cohorts who entered a higher school or its equivalent under the decentralized admissions.
In all regressions, we control for either quadratic time trends or 6th order polynomial time trends. We also control for the number of HCSE exam passers in Panel (a) Columns (3)--(8) and Panel (b). 
Parentheses contain P values based on Newey-West standard errors with the maximum lag order of 3.
***, **, and * mean significance at the 1\%, 5\%, and 10\% levels, respectively.
See Section \ref{section:long_run_analysis2} for discussions about this table.
 \par} 
\end{table}

\begin{figure}[h!]
\caption{Total Number of All Occupational Elites by Cohort in PIR (1934) and PIR (1939) }\label{fig:careerelites_total}
\begin{center}
\includegraphics[width=90mm]{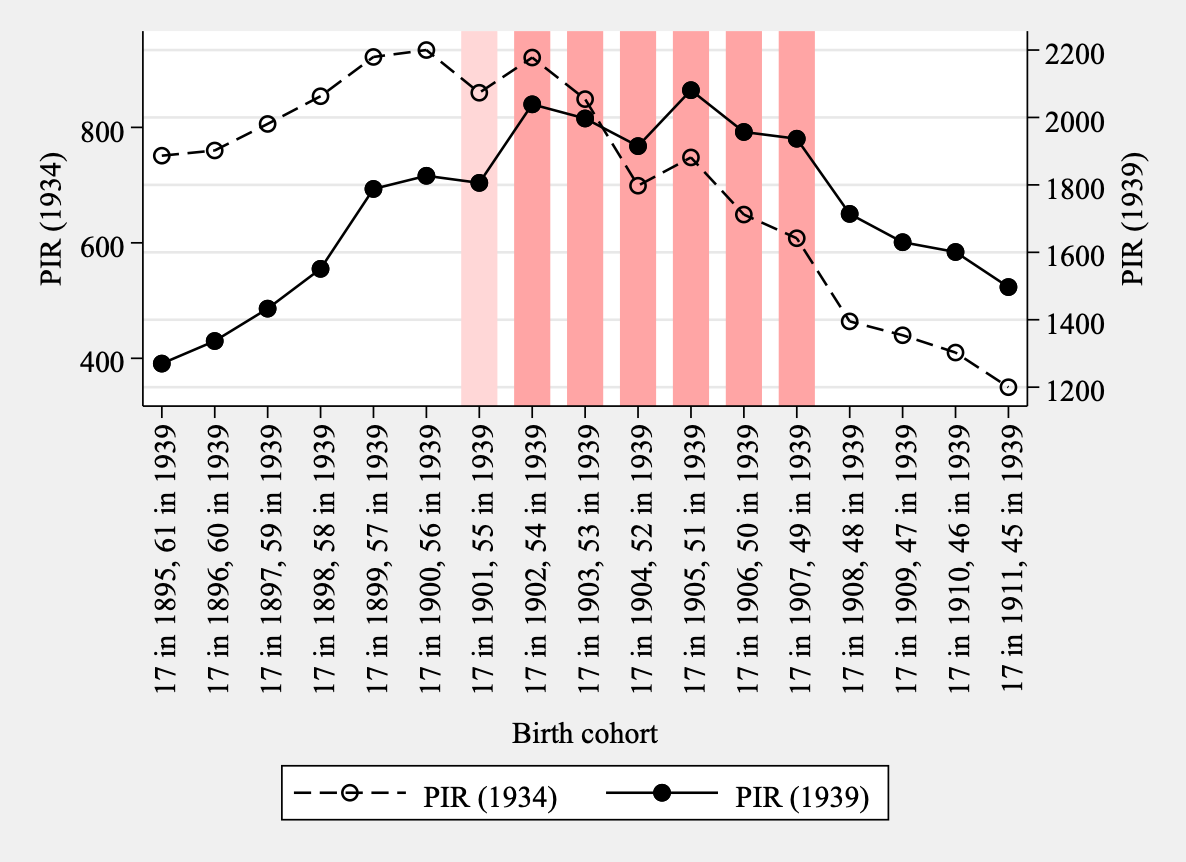}
\end{center}
{\footnotesize \textit{Notes}:
This figure plots the total number of all occupational elites listed in  each of the 1934 and 1939 editions of the PIR by cohort.
Cohorts born in 1878--1894 and turning age 17 in 1895--1911 are observed when they are 40 to 56 years old in PIR (1934) and when they are 45 to 61 years old in PIR (1939).
The cohorts turning age 17 in 1902--1907 (under centralization) are colored in dark pink. 
The cohort turning age 17 in 1901 (under decentralization) and age 18 in 1902 (under centralization) is colored in pale pink.
In both of the editions, centralization cohorts tend to have a greater number of listed individuals compared to decentralization cohorts. See Table \ref{tab:careerelites_total} for a statistical analysis and Section \ref{section:long_run_analysis2} for discussions about this figure.
\par}
\end{figure}

\clearpage
\bibliographystyleappendix{aea}
{\footnotesize\bibliographyappendix{reference}}

\end{document}